\def\FullVersion{} %
\def\ShowAuthor{} %

\ifdefined\LLNCS
\documentclass{llncs}
\renewcommand{\paragraph}{\subsubsection}
\else
\documentclass[12pt]{article}
\usepackage{amsthm}
\usepackage[margin=1in]{geometry}
\fi

\usepackage{graphicx,color,url,booktabs,comment}
\usepackage{tikz}
\usetikzlibrary {chains, fit, shapes.geometric, arrows.meta}
\usepackage{authblk}
\usepackage{amsmath, amssymb,amsbsy}
\usepackage{multicol}
\usepackage{framed}
\usepackage{enumitem}
\usepackage{bbm}
\usepackage{mathrsfs}
\usepackage{qcircuit}
\usepackage{braket}
\usepackage{algorithm}
\usepackage{algpseudocode}
\usepackage{xcolor,float}
\usepackage{pagecolor}
\usepackage[utf8]{inputenc}
\usepackage{tabularx}
\usepackage[pdfstartview=FitH,colorlinks,linkcolor=blue,filecolor=blue,citecolor=blue,urlcolor=blue]{hyperref}
\MakeRobust{\Call}
\usepackage{xspace}
\usepackage{cleveref} 
\usepackage{mdframed}
\usepackage{amsfonts}
\usepackage{tikz}
\usepackage{tikz-cd}
\usepackage{tikz-qtree}
\usepackage{adjustbox}
\usepackage{blindtext}
\usepackage{makecell}
\usetikzlibrary{positioning}
\ifdefined\LLNCS
\else
\usepackage[margin=1in]{geometry}
\fi

\usepackage{setspace}
\newcounter{protocol}
\newenvironment{protocol}[1]
{
\refstepcounter{protocol}
\par\setlength{\parindent}{0pt}
\rule{\textwidth}{0.3mm}
\textbf{Protocol~\theprotocol} #1 
\vspace{-2.6mm}

\hrulefill \break
}
{
\hrulefill \break
\par
}

\newcounter{securityGame}

\newcounter{balgorithm}

\newcounter{numConstruction}
\newenvironment{Construction}[1]
{
\refstepcounter{numConstruction}
\par\setlength{\parindent}{0pt}
\rule{\textwidth}{0.3mm}
\vspace{-2.6mm}\textbf{Construction~\thenumConstruction} #1 

\hrulefill \break
}
{
\hrulefill \break
\par
}

\newcounter{simulator}
\newenvironment{simulator}[1]
{
\refstepcounter{simulator}
\par\setlength{\parindent}{0pt}
\rule{\textwidth}{0.3mm}
\vspace{-2.6mm}\textbf{Simulator~\thesimulator} #1 

\hrulefill 
}
{
\hrulefill \break
\par
}

\newenvironment{idealmodel}[1]
{
\begin{mdframed}
    \begin{center}\textbf{#1}\end{center}
}
{
\end{mdframed}
}

\ifdefined\LLNCS
\newenvironment{thm}{\begin{theorem}}{\end{theorem}}

\newenvironment{rmk}{\begin{remark}}{\end{remark}}
\newenvironment{lem}{\begin{lemma}}{\end{lemma}}
\newenvironment{cor}{\begin{corollary}}{\end{corollary}}
\newtheorem{dfn}[definition]{Definition}
\else
\newtheorem{thm}{Theorem}[section]

\newtheorem{lemma}[thm]{Lemma}
\newtheorem{claim}[thm]{Claim}

\theoremstyle{definition}
\newtheorem{dfn}[thm]{Definition}
\fi

\ifdefined\LLNCS
\else
\newtheorem*{thm*}{Theorem}
\newenvironment{theorem}{\begin{thm}}{\end{thm}}
\makeatletter
\newtheorem*{rep@theorem}{\rep@title}
\newcommand{\newreptheorem}[2]{%
\newenvironment{rep#1}[1]{%
 \def\rep@title{#2 \ref{##1}}%
 \begin{rep@theorem}}%
 {\end{rep@theorem}}}
\makeatother

\newreptheorem{thm}{Theorem}
\newreptheorem{lem}{Lemma}
\fi

\ifdefined\LLNCS
\newaliascnt{claiml}{theorem}
\newtheorem{claiml}[claiml]{Claim}
\aliascntresetthe{claiml}

\crefname{claiml}{Claim}{Claims}
\else\fi

\crefname{lemma}{Lemma}{Lemmata}
\crefname{claim}{Claim}{Claims}
\crefname{figure}{Figure}{Figures}
\crefname{corollary}{Corollary}{Corollaries}
\crefname{proposition}{Proposition}{Propositions}
\crefname{conjecture}{Conjecture}{Conjectures}
\crefname{definition}{Definition}{Definitions}
\crefname{remark}{Remark}{Remarks}
\crefname{example}{Example}{Examples}
\crefname{algorithm}{Algorithm}{Algorithms}
\crefname{bAlgorithm}{Algorithm}{Algorithms}
\crefname{protocol}{Protocol}{Protocols}
\Crefformat{algorithm}{\textup{Algorithm~#2#1#3}}

\renewcommand{\cref}{\Cref} 

\ifdefined\LLNCS
\newenvironment{proofof}[1]{\begin{proof}[of~#1]}{\end{proof}}
\newenvironment{proofsketch}{\begin{trivlist} \item {\it Proof sketch.}} {\qed\end{trivlist}}
\newenvironment{proofsketchof}[1]{\begin{proofsketch}[of~#1]}{\end{proofsketch}}
\else

\newenvironment{proofsketch}{\begin{trivlist} \item {\it Proof sketch.}} {\qed\end{trivlist}}

\fi

\let\mathbb\relax 
\DeclareMathAlphabet{\mathbb}{U}{msb}{m}{n}

\def\ot{\otimes}

\newcommand{\ie}{{i.e.,\ }}
\newcommand{\eg}{{e.g.,\ }}

\def\eps{\epsilon}

\def\cD{\mathcal{D}}

\def\cT{{\cal T}}

\def\Dec{\mathsf{Dec}}

\def\bbC{\mathbb{C}}

\def\bbN{\mathbb{N}}

\def\bbZ{\mathbb{Z}}

\newcommand{\zo}{\{0,1\}}

\newcommand{\dyad}[1]{\ket{#1}\!\bra{#1}}

\newcommand{\onreg}[2]{\ensuremath{{#1}^{\gray{#2}}}}

\DeclareMathOperator{\poly}{poly}
\DeclareMathOperator{\tr}{tr}

\newcommand{\qpt}{{\sc qpt}\xspace}
\newcommand{\secparam}{\lambda}

\newcommand{\given}{\ensuremath{\;\middle|\;}}
\newcommand{\randsample}{\from}

\newcommand{\inner}[1]{\langle{#1}\rangle}

\newcommand{\of}[1]{\left(#1\right)}

\newcommand{\ketbra}[1]{\dyad{#1}}

\newcommand{\mixednospace}[1]{\mathsf{Mixed}\left[#1\right]}
\newcommand{\mixed}[1]{\text{ }\mixednospace{#1}}

\DeclareMathOperator*{\E}{\mathbb{E}}

\newcommand{\negl}{\operatorname{negl}}
\newcommand{\Tr}{\tr}

\newcommand{\Range}{\textsc{Range}}

\def\real{{\mathsf{Real}}}
\def\ideal{{\mathsf{Ideal}}}
\def\acc{{\mathsf{Acc}}}
\def\rej{{\mathsf{Rej}}}

\def\MSP{{\mathsf{MSP}}}

\def\BoBW{{\mathsf{BoBW}}}
\def\MPQC{{\mathsf{MPQC}}}

\newcommand{\hybrid}{\mathcal{H}}

\def\benum{\begin{enumerate}}
\def\eenum{\end{enumerate}}
\def\bit{\begin{itemize}}
\def\eit{\end{itemize}}
\def\bdesc{\begin{description}}
\def\edesc{\end{description}}

\newcommand{\act}{\cdot}
\newcommand{\measurement}{\mathsf{\Pi}}
\newcommand{\twirl}{\mathcal{T}}

\newcommand{\CliffordGroup}{\mathscr{C}}
\newcommand{\PauliGroup}{\mathscr{P}}

\newcommand{\identity}{\mathbbm{1}}
\newcommand{\GL}{\operatorname{GL}}

\newcommand{\Ig}{I}
\newcommand{\Tg}{\mathsf{T}}

\newcommand{\Hg}{\mathsf{H}}
\newcommand{\Xg}{\mathsf{X}} 

\newcommand{\Zg}{\mathsf{Z}}
\newcommand{\Sg}{\mathsf{S}}
\newcommand{\Pg}{\mathsf{S}}

\newcommand{\Setup}{\mathsf{Setup}}
\newcommand{\Enc}{\mathsf{Enc}}
\newcommand{\Send}{\mathsf{Send}}
\newcommand{\Audit}{\mathsf{Audit}}
\newcommand{\Recv}{\mathsf{Recv}}
\newcommand{\sk}{\mathsf{sk}}
\newcommand{\dk}{\mathsf{dk}}
\newcommand{\pf}{\mathsf{pf}}
\newcommand{\EncGate}{\mathsf{EncG}}
\newcommand{\Gen}{\mathsf{Gen}}

\newcommand{\TP}{\mathsf{TP}}
\newcommand{\TPSend}{\mathsf{TP.Send}}
\newcommand{\TPReceive}{\mathsf{TP.Recv}}

\newcommand{\AQA}{\mathsf{AQA}}

\newcommand{\PVIA}{\mathsf{PVIA}}
\newcommand{\SWIA}{\mathsf{SWIA}}

\newcommand{\hyb}{\mathsf{hybrid}}
\newcommand{\EPRS}{\mathcal{S}}
\newcommand{\EPRR}{\mathcal{R}}
\newcommand{\EPRC}{\mathcal{C}}

\newcommand{\CX}{\mathsf{CX}}

\newcommand{\numinputs}{\ell} 

\newcommand{\Pc}{\mathsf{P}}

\newcommand{\adv}{{\mathcal{A}}}

\newcommand{\cMPC}{{\mathsf{cMPC}}}
\newcommand{\trustp}{\mathsf{T}}
\newcommand{\abort}{\normalfont{\texttt{abort}}}

\newcommand{\aux}{{\mathsf{aux}}}

\newcommand{\anc}{{\mathsf{anc}}}

\newcommand{\pvia}{secure with publicly verifiable identifiable abort\xspace}
\newcommand{\bobw}{best-of-both-worlds\xspace}

\newcommand{\total}{\operatorname{total}}

\newcommand{\out}{\operatorname{out}}

\newcommand{\circuit}{{C}}

\newcommand{\from}{\leftarrow}

\newcommand{\CAuth}{\mathsf{CliffordAuth}}

\newcommand{\QG}{\mathsf{QG}}
\newcommand{\QGarble}{\mathsf{QG.Garble}}
\newcommand{\QGEval}{\mathsf{QG.Eval}}
\newcommand{\QGSim}{\mathsf{QG.Sim}}

\newcommand{\numtraps}{\secparam}

\newcommand{\Sender}{\mathsf{P_{s}}}
\newcommand{\Receiver}{\mathsf{P_{r}}}
\newcommand{\Observer}{\mathsf{O}}

\newcommand{\gray}[1]{\textcolor{gray}{#1}}
\newcommand{\AR}{\mathsf{AR}}

\newcommand{\QECC}{\mathsf{QECC}}
\newcommand{\QEnc}{{\mathsf{QECC.Enc}}}
\newcommand{\QDec}{\mathsf{QECC.Dec}}
\newcommand{\Qnum}{n} 

\newcommand{\thres}{t}

\newcommand{\Sim}{\mathsf{Sim}}

\newcommand{\Den}[1]{\cD^{#1}}

\makeatletter
\newcommand{\pushright}[1]{\ifmeasuring@#1\else\omit\hfill$\displaystyle#1$\fi\ignorespaces}
\newcommand{\pushleft}[1]{\ifmeasuring@#1\else\omit$\displaystyle#1$\hfill\fi\ignorespaces}
\makeatother

\floatstyle{boxed} 
\restylefloat{figure}

\title{Best-of-Both-Worlds Multiparty Quantum Computation with Publicly Verifiable Identifiable Abort}

\ifdefined\ShowAuthor
\author[1]{Kai-Min Chung}
\author[2]{Mi-Ying (Miryam) Huang}
\author[1,3]{Er-Cheng Tang}
\author[2]{Jiapeng Zhang}
\affil[1]{Academia Sinica, Taiwan}
\affil[2]{University of Southern California, United States}
\affil[3]{University of Washington, United States}

\else
\author{}
\date{}
\fi

\begin{document}

\maketitle
\begin{abstract}
Alon et al. (CRYPTO 2021) introduced a multiparty quantum computation protocol that is secure with identifiable abort (MPQC-SWIA). However, their protocol allows only inside MPQC parties to know the identity of malicious players. This becomes problematic when two groups of people disagree and need a third party, like a jury, to verify who the malicious party is. This issue takes on heightened significance in the quantum setting, given that quantum states may exist in only a single copy.
Thus, we emphasize the necessity of a protocol with \emph{publicly verifiable identifiable abort} (PVIA), enabling outside observers with only classical computational power to agree on the identity of the malicious party in case of an abort. However, achieving MPQC with PVIA poses significant challenges due to the no-cloning theorem, and previous works proposed by Mahadev (STOC 2018) and Chung et al. (Eurocrypt 2022) for classical verification of quantum computation fall short. 

In this paper, we obtain the first MPQC-PVIA protocol assuming post-quantum oblivious transfer and a classical broadcast channel. The core component of our construction is a new authentication primitive called \emph{auditable quantum authentication} (AQA) that identifies the malicious sender with overwhelming probability. Additionally, we provide the first MPQC protocol with best-of-both-worlds (BoBW) security, which guarantees output delivery with an honest majority and remains secure with abort even if the majority is dishonest. Our \bobw MPQC protocol also satisfies PVIA upon abort.

\end{abstract}

\ifdefined\FullVersion
\newpage
\tableofcontents
\clearpage
\fi

\section{Introduction}
Secure multiparty computation (MPC) allows two or more parties to compute a function on their joint private inputs securely \cite{Yao86}. Most of the MPC literature studies classical functionality over classical inputs with different notions of security, such as \emph{full security}, \emph{security with abort}, and \emph{security with identifiable abort} \cite{RB89,GMW87,IOZ14}. 

Recently, secure multiparty quantum computation (MPQC) has raised research interest. Most of the works consider the fully quantum setting \ie the functionality, including inputs and outputs, is quantum. Like in the classical setting, it is known that an honest majority is both sufficient \cite{CGS02,BCG06}, and necessary \cite{ABDR04} to achieve full security, which guarantees output delivery for everyone. In light of this, the study of MPQC protocols in the dishonest majority setting has focused on the weaker notion of \emph{security with abort} \cite{DNS12,DGJ+20,BCKM21}, which allows all honest parties to abort when they detect an attack. However, such a notion is vulnerable to a denial-of-service attack because an attacker can repeatedly induce aborts. For this reason, a more recent work \cite{ACC+21} has proposed an MPQC protocol with identifiable abort (MPQC-SWIA) that allows all honest parties to agree on the identity of a corrupted party in case of an abort. Regrettably, the identification mechanism of \cite{ACC+21} only allows the participants of the protocol to identify a malicious party. This is unsatisfactory in many practical scenarios because during a dispute, external observers are aware that two groups of people are in disagreement, but it is unclear which side is acting maliciously. Consider an instance where a client accuses a tech company of failing to provide a service and, therefore, refuses payment. Conversely, the company asserts that they have indeed provided the service. In such cases, it becomes vital to employ a publicly verifiable protocol to assess their integrity. This is especially important in the quantum setting, where each party may possess only one copy of their quantum input. Once the quantum inputs are ruined, it results in the irreversible loss of inputs for honest parties. Therefore, we consider a notion of security called \emph{publicly verifiable identifiable abort} (PVIA) that allows everyone, including outside observers, to identify the malicious party. We ask:
\begin{quote}
    {\it Is it possible to construct MPQC with publicly verifiable identifiable abort (PVIA)?}
\end{quote}

In the classical setting, one can turn MPC-SWIA into MPC-PVIA almost for free. A publicly verifiable protocol can be obtained by requiring each party to broadcast their messages and proofs to outside observers. Unfortunately, this simple solution does not work in the quantum setting due to the no-cloning theorem. One may be tempted to turn to classical verification of quantum computation (CVQC) \cite{CVQC,CVQCKM} in order to achieve public verifiability. However, this approach is restricted to computation that is performed by a single quantum party with classical outputs, and it is unclear how it can be adapted to fit into the setting of MPQC. 
Furthermore, all existing MPQC protocols face an inherent difficulty in achieving PVIA because their sender-receiver mechanism cannot differentiate a malicious sender from a malicious receiver. To address this issue, we propose a new primitive called Auditable Quantum Authentication (AQA), which subverts the traditional sender-receiver mechanism and holds the sender accountable for his behavior.

While PVIA security can act as insurance for honest parties when a dishonest majority is present, it is desirable to have a stronger security notion, such as full security, if it turns out that the honest parties outnumber the malicious ones. 
An intriguing scenario involves reducing the maximal number of malicious parties allowed for security with abort while conditionally offering full security. Such a notion is called best-of-both-worlds (BoBW) security\footnote{There are different flavors of \bobw security. For example, \cite{BoBW-neg,BoBW-1/p} consider MPC protocols with full security against $\lfloor \frac{n-1}{2} \rfloor$ malicious parties and $(1\slash p)$-security with abort against $n-1$ malicious parties. The notion of $(1\slash p)$-security only requires an inverse polynomial error in distinguishing the real/ideal world.}. In the classical setting, \cite{BoBW-pos} constructs, for every threshold $\thres < \frac{n}{2}$, an MPC protocol that achieves security with abort against $n-1-\thres$ malicious parties and achieves full security tolerating $\thres$ malicious parties. \cite{BoBW-neg} proved that these corruption thresholds are optimal. In the quantum setting, none of the existing MPQC protocols satisfy BoBW security. Therefore, we ask:
\begin{quote}
    {\it Is it possible to construct a single MPQC that achieves full security under an honest majority and is secure with abort under a dishonest majority?}
\end{quote}

\subsection{Our results}
We answer both questions affirmatively in the preprocessing model, which features an offline setup that prepares input-independent auxiliary quantum states. Then, during the online protocol, parties only exchange classical bits. With this approach, the parties can create classical proofs that are accessible to everyone, which in turn facilitates PVIA.
Moreover, combined with quantum error correction code (QECC), the setup can create quantum states that enable distributed computation and ultimately achieve \bobw security. Finally, we show that our offline setup can be instantiated without requiring any trusted third party.

Our first result is an MPQC protocol secure with publicly verifiable identifiable abort (PVIA) under a trusted setup. Similar to existing MPQC works \cite{DNS12,DGJ+20,ACC+21,BCKM21}, we assume that parties have access to an ideal functionality $\cMPC$ for classical MPC (this model is known as the MPC-hybrid model). Here, the classical MPC is assumed to be PVIA-secure, and such an MPC can be based on a post-quantum oblivious transfer (OT) and a classical broadcast channel. 

\begin{theorem}[MPQC-PVIA with trusted setup, informal]\label{thm:informal-PVIA}
There exists a multiparty quantum computation protocol \pvia supporting poly-size quantum circuits in the preprocessing MPC-hybrid model.
\end{theorem}

To achieve \cref{thm:informal-PVIA}, we propose and construct a new primitive called auditable quantum authentication (AQA) that allows a classical auditor to decide the integrity of a quantum message sender. Then, in our MPQC-PVIA protocol, the actions of the trusted auditor will be taken by classical MPC.

Our second result is a \bobw (BoBW) MPQC protocol that achieves full security against $\thres$ corruptions and satisfies security with abort against $n-1-\thres$ corruptions under a trusted setup. We call $\thres$ as the BoBW threshold. 
Here, we assume our underlying classical MPC to be BoBW-secure with threshold $\thres$ as well, which can be based on post-quantum OT for $\thres < \frac{n}{3}$ and additionally requires a classical broadcast channel for $\frac{n}{3} \le t < \frac{n}{2}$.
\begin{theorem}[BoBW-MPQC with trusted setup, informal]\label{thm:informal-BoBW}
There exists a \bobw multiparty quantum computation protocol of threshold $\thres$ supporting poly-size quantum
circuits for any $\thres < \frac{n}{2}$ in the preprocessing MPC-hybrid model.
\end{theorem}

The key to arriving at best-of-both-worlds security is our protocol's compatibility with decentralized quantum computation using QECC. In particular, no single party in our protocol holds all the quantum information of a piece of data during the computation step, as opposed to prior security-with-abort protocols \cite{DGJ+20,ACC+21,BCKM21}.

Combining these two results, we obtain a BoBW-MPQC-PVIA protocol that achieves full security against $\thres$ corruptions and satisfies PVIA security against $n-1-\thres$ corruptions under a trusted setup. The underlying classical MPC should be BoBW-PVIA-secure, which can be based on a post-quantum OT and a classical broadcast channel. 
\begin{theorem}[BoBW-MPQC-PVIA with trusted setup, informal]\label{thm:informal-BoBW-PVIA}
There exists a \bobw multiparty quantum computation protocol secure with publicly verifiable identifiable abort of threshold $\thres$ supporting poly-size quantum circuits for every $\thres < \frac{n}{2}$ in the preprocessing MPC-hybrid model.
\end{theorem}
Furthermore, we can instantiate the setups, thus obtaining the above three results without needing a trusted setup. 
\begin{theorem}[BoBW-MPQC-PVIA without trusted setup, informal]\label{thm:informal-No-Trusted-Setup}
Theorems \ref{thm:informal-PVIA},\ref{thm:informal-BoBW},\ref{thm:informal-BoBW-PVIA} hold in the (standard) MPC-hybrid model.
\end{theorem}
Our main technique for instantiating the setup is to leverage MPQC secure with identifiable abort (SWIA) protocols. Interestingly, the properties of both BoBW and PVIA can be preserved under our instantiation. Note that our instantiation is based on an MPQC-SWIA protocol which, contrasting with the previous result \cite{ACC+21}, only assumes classical MPC.

\begin{table}[H]
\begin{tabular}{ |@{ } c @{ }|@{ } l @{ }|@{ } l @{ }|@{ } l @{ }|} 
  \hline
   \makecell[l]{{}\\{}} & {Dishonest-Majority Regime} & {Honest-Majority Regime} & Assumptions \\ 
  \hline
 \cite{BCG06}  & No Security & Full Security & cMPC\\ 
  \hline
  \makecell[l]{\cite{DGJ+20} \\ \cite{BCKM21}} & Security with Abort $\bigg($\makecell[l]{$\le n-1$ \\ corruptions}$\bigg)$ & Security with Abort & cMPC \\ 
  \hline
  \cite{ACC+21} & Identifiable Abort\;\;\; $\bigg($\makecell[l]{$\le n-1$ \\ corruptions}$\bigg)$ & Identifiable Abort & cMPC+FHE \\
  \hline
  ${\Large \substack{\text{This Work} \\ \scriptscriptstyle{(0 \le \thres < \frac{n}{2}}) }}$ 
  & \makecell[l]{Publicly Verifiable \\ Identifiable Abort} $\bigg($\makecell[l]{$\le n-1-\thres$ \\ corruptions}$\bigg)$ & Full Security $\bigg($\makecell[l]{$\le \thres$ \\ corruptions}$\bigg)$ & cMPC \\
  \hline
\end{tabular}
\centering
\caption{Comparison of MPQC protocols.}
\label{tab:comparison}
\end{table}
\section{Technical Overview}
In this section, we first explain why public verifiability does not follow directly from existing works. Then, we put forth a novel primitive called Auditable Quantum Authentication (AQA), which ensures the secure transmission of quantum outputs and the public identification of malicious identities within a protocol. Following a high-level understanding of AQA, we then incorporate the input encoding and computation steps together to realize MPQC-PVIA. At the end of the section, we discuss the difficulty of achieving \bobw security and elucidate our approach to attaining a BoBW-MPQC protocol.

\subsection{Why is MPQC-PVIA hard to achieve?}

A first observation is that classical techniques for public verifiability cannot apply to their quantum counterparts. Existing methods for classical MPC-PVIA protocols are to commit to classical messages, provide zero-knowledge arguments over the commitments, and let outside observers check whether any party deviates from the protocol. There are several issues when adapting to MPQC in the fully quantum setting. If one considers classical commitments to quantum messages \cite{CVQC}, one cannot fulfill MPQC with purely quantum outputs because such classical commitment schemes always end with measurements. Instead, one may have to consider quantum commitments \cite{gunn2023commitments}. However, quantum commitments are unlikely to be duplicated and broadcast to each party for verification because of the no-cloning theorem. In addition, zero-knowledge arguments for quantum computation (\eg \cite{ZKP-QMA}) only apply to problems with a classical description. Those arguments cannot prove relations involving quantum commitments.

Another difficulty arises because we require the outside observers of MPQC to have only classical computational power. Although there is research on classical verification of quantum computation (CVQC), a seemingly similar task, CVQC needs to be more relaxed because it can only resolve computations with classical outputs conducted by a single quantum prover. The techniques of CVQC fail in the fully quantum setting. Moreover, CVQC already produces an inverse polynomial soundness error when extended from decisional problems \cite{CVQC} to sampling problems \cite{CVQCKM}. Thus, there is little hope that CVQC can aid the construction of MPQC-PVIA.

One may try to upgrade MPQC-SWIA to MPQC-PVIA directly, but there is still a gap between them. The MPQC-SWIA protocol by \cite{ACC+21} is based on a Sequential Authentication primitive that outputs two suspects whenever message tampering is detected. However, it gives no information about the \emph{exact} party that deviates from the protocol. The resulting MPQC-SWIA allows honest parties to agree on the same malicious party when protocol aborts, but an outside observer only sees two groups of people accusing each other. This outcome arises from the conventional utilization of quantum authentication codes\footnote{The prevalent approach in most existing works involves employing authentication codes in this manner \cite{ACC+21,BCKM21,DGJ+20}.}, where the sender sends an authenticated state to the receiver, and the receiver is in charge of measuring the authentication checksum to validate the state.
However, this kind of validation mechanism relies on the synergy of both the sender and the receiver over a single-copy state, which makes it challenging to achieve public verifiability.
To address this, we subvert the old idea and creatively combine quantum authentication codes and quantum teleportation in a white-box manner.

\subsection{Our Solution: Auditable Quantum Authentication (AQA)}

The primary goal of AQA is to establish a mechanism where the sender of an authenticated state is held responsible for his own sending action through a test performed by someone trustworthy.
In a normal quantum authentication scheme, the receiver of an authenticated state runs the decoding algorithm to obtain either the original message or an authentication failure symbol. To learn the true authentication outcome, an outside observer has to trust the party who executes the decoding algorithm. This would require trust in the receiver, who might be malicious. To resolve this issue, we propose an \emph{auditable} quantum authentication scheme that separates the authenticity check from the message decoding process. Importantly, the AQA scheme is equipped with a \emph{classical} auditing algorithm that decides message authenticity and outputs a decoding key for the receiver to recover the message. With AQA, an outside observer can learn the authentication outcome by trusting a classical auditor who executes the auditing algorithm. Later on, we can replace the classical auditor with a publicly verifiable classical MPC (cMPC) to completely remove the need of trust.

AQA is designed to be cooperated by three parties: a sender, a receiver, and a classical auditor. 
We define AQA as consisting of five algorithms: $\Setup, \Enc, \Send, \Audit, \Recv$. In the beginning, $\Setup$ prepares initial states for all the parties, and $\Enc$ produces an authenticated state $\sigma$ for the sender. The sender runs $\Send(\sigma)$ to generate a classical proof $\pf$ showing that the quantum message has been delivered.
Next, the auditor runs $\Audit(\pf)$ to verify the proof and produce a decoding key $\dk$. Afterward, the receiver can run $\Recv(\dk)$ to obtain the quantum message. 
The security of AQA entails that $\Recv(\dk)$ produces the correct quantum message (up to a negligible error) whenever $\Audit(\pf)$ outputs a positive verification outcome.

\paragraph{Constructing a (Simplified) AQA} We will start with a normal quantum authentication scheme ($\Gen$,$\Enc$,$\Dec$), the Clifford code \cite{CliffordCodes} in particular. We aim to keep the encoding procedure $\Enc$ and split its decoding procedure into several parts.
The decoding procedure of Clifford code applies a secret Clifford gate $\onreg{F^\dag}{M,T}$ to an authenticated state $\onreg{\sigma}{M,T}$ and measures $\gray{T}$ register in the computational basis. An authentication failure occurs if the measurement result is not all zeros. Otherwise, the content of the $\gray{M}$ register will be the message state. To make this authentication scheme classically auditable, we consider the following alternative decoding procedure that involves $4$ algorithms. We take $|\gray{M}| = 1$ and $|\gray{T}| = \numtraps$ as an example.

\vspace{2mm}
\noindent
\underline{$\Setup$}:
    \begin{itemize}
    \setlength{\itemsep}{1pt} 
        \item Generate EPR pairs $\{(e_0^i,e_1^i)\}_{i \in [\secparam+1]}$ of length $\secparam+1$. Put $\{e^i_0\}_{i \in [\secparam+1]}$, $\{e_1^i\}_{i \in \{2,\cdots,\secparam+1\}}$, $e_1^1$ into the sending register $\gray{S}$, the checking register $\gray{C}$, and the receiving register $\gray{R}$. 
        \item Apply the secret Clifford gate $F^\dag \randsample \CliffordGroup_{\secparam+1}$ to $\gray{R,C}$.
    \end{itemize}
\underline{$\Send$}:
    \begin{itemize}
    \setlength{\itemsep}{1pt} 
        \item Sending Procedure: Teleport the authenticated state $\sigma$ through the sending register $\gray{S}$.
        \item Proving Procedure: Measure the checking register $\gray{C}$ in the computational basis.
        \item Set the classical proof as the teleportation Pauli $P$ and the measurement result $c$.
    \end{itemize}
\underline{$\Audit$}:
    \begin{itemize}
    \setlength{\itemsep}{1pt} 
        \item Compute the Pauli $F^\dag P F$ and express it as a tensor product of two Pauli gates $\hat{P}_R, \hat{P}_C$ that act on $1, \secparam$ qupits respectively. 
        \item Report an authentication failure if $c \neq x(\hat{P}_C)$. Set the decoding key as the Pauli $\hat{P}_R$.
    \end{itemize}
\underline{$\Recv$}:
    \begin{itemize}
    \setlength{\itemsep}{1pt} 
        \item Apply $\onreg{\hat{P}_R^\dag{}}{R}$ and output the state on the receiving register $\gray{R}$. 
    \end{itemize}
Running these $4$ algorithms in a row is equivalent to running the decoding procedure of Clifford code. This follows almost directly from quantum teleportation: If we denote $F^{\dag} \sigma = (\onreg{\rho}{M}, \onreg{\tau}{T})$ and if teleporting $\sigma$ through $\gray{S}$ during $\Send$ yields teleportation result $P$, then the state in $\gray{R,C}$ would collapse from $F^\dag (\{e_1^i\}_{i\in[\secparam+1]})$ to 
$$F^{\dag} P (\sigma) 
= (F^{\dag} P F) (F^\dag (\sigma)) 
= (\hat{P}_R \ot \hat{P}_C) (\rho, \tau)
= (\hat{P}_R (\rho), \hat{P}_C (\tau))$$
This shows that the measurement result of the checking register $\gray{C}$ equals $x(\hat{P}_C)$ if and only if the measurement result of $\tau$ equals all zeros. 
The current decoding procedure includes a classical algorithm $\Audit$ that determines the authentication outcome, so we can set our simplified AQA as these $4$ algorithms plus the encoding algorithm of Clifford code.

\begin{figure}[H]
\definecolor{myblue}{RGB}{80,80,160}
\definecolor{mygreen}{RGB}{80,160,80}
\scalebox{0.5}{\begin{tikzpicture}[thick,
  every node/.style={draw, circle},
  ssnode/.style={fill=myblue},
  fsnode/.style={fill=mygreen},
  every fit/.style={ellipse, draw, inner sep=-1pt, text width=1.75cm},
  >={Latex[width=3pt, length=4pt]}, shorten >= 2pt, shorten <= 2pt]

\begin{scope}[start chain=going below,node distance=1.5mm]
\foreach \i in {0}
  \node[ssnode,on chain] (s\i) [label=left: $S_{1}$] {};
\foreach \i in {1}
  \node[ssnode,on chain] (s\i) [label=left: $S_{2}$] {};
\foreach \i in {2,3}
  \node[ssnode,on chain] (s\i) [label=left: $\cdot$] {};
\foreach \i in {4}
  \node[ssnode,on chain] (s\i) [label=left: $S_{1+\secparam}$] {};
\end{scope}

\begin{scope}[xshift=3cm, yshift=-0.15cm, start chain=going below, node distance=5mm]
\foreach \i in {0}
    \node[fsnode, on chain] (f\i) [label=right: ${R}$] {};
\foreach \i in {1}
    \node[fsnode, on chain] (f\i) [label=right: ${C_1}$] {};
\foreach \i in {2,3}
  \node[fsnode, on chain] (f\i) [label=right: $\cdot$] {};
\foreach \i in {4}
    \node[fsnode, on chain] (f\i) [label=right: ${C_\secparam}$] {};
\end{scope}

\node [myblue, fit=(s0) (s4), label=above:\itshape, xshift=-0.2cm] {};
\node [myblue, fit=(f1) (f4), label=above:\itshape, xshift=0.3cm, yshift=-.3cm ] {};

\draw[<->, red] (s0) -- (f0);
\draw[<->] (s2) -- (f2);
\draw[<->] (s3) -- (f3);
\draw[<->] (s4) -- (f4);
\draw[<->] (s1) -- (f1);
\end{tikzpicture}}
    \centering
    \caption{AQA Setup: Each edge represents an EPR pair. The nodes on the left contain halves of EPR pairs $\{e^i_0\}_{i \in [\secparam+1]}$, and the nodes on the right contain the other halves of EPR pairs $\{e^i_1\}_{i \in [\secparam+1]}$. The encircled vertices (the sending register $S$ and checking register $C$) are given to the sender. The lonely vertex (receiving register $R$) is given to the receiver.}
    \label{fig:AQA}
\end{figure}
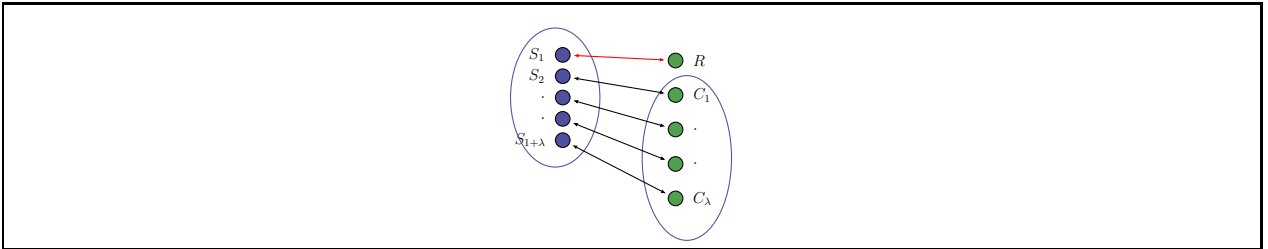

\paragraph{Proving Security of AQA} For the security of AQA, it's crucial for $\Recv$ to recover the original message whenever $\Audit$ doesn't indicate an authentication failure upon receiving proof from an adversarial sender. Take, for instance, a malicious sender who alters the authenticated message $\sigma$ prior to the execution of $\Send$. In such a specific scenario, the security of the simplified AQA is derived from the established equivalence between the processes $(\Send,\Audit,\Recv)$ and $\Dec$ mentioned in the preceding paragraph and grounded on the efficacy of $(\Enc,\Dec)$ as a quantum authentication scheme. However, the simplified AQA is presented mainly as an explanation of how we achieve the audit functionality, and is not yet secure against arbitrary malicious senders. To further protect against adversaries employing arbitrary attacks, our formal AQA in \cref{Sec:AQA} additionally integrates quantum one-time pads. To offer a high-level intuition, the use of quantum one-time pads can split the attack of the malicious sender into a combination of Pauli attacks. Since the malicious sender knows nothing about the random Clifford key $F$, the Clifford twirl will transform Pauli attacks into random Pauli operators distributed across the states, breaking the consistency of $P$ and $c$.

\subsection{From AQA to MPQC-PVIA}
From the previous section, we see that the transmission of quantum information can be audited by a classical party. We now build an MPQC-PVIA protocol with AQA where the auditing is performed by a publicly verifiable classical MPC.

\paragraph{MPQC-PVIA with Setup}
The MPQC-PVIA consists of two phases: an offline setup and an online phase. The offline setup generates EPR pairs that would allow each party to send their input to the server (who is a designated party, say $\Pc_1$), and runs the setup of AQA. During the online phase, every party teleports their input to $\Pc_1$ and $\Pc_1$ only obtains a ciphertext of the joint inputs. Next, $\Pc_1$ performs quantum computation on the ciphertext as instructed by classical MPC. Finally, $\Pc_1$ sends the output ciphertexts to other parties using AQA, which is audited by classical MPC. These three steps in the online phase are called input encoding, computation, and output delivery, respectively.

We now move on to examine security. In our protocol, the parties' inputs are gathered towards $\Pc_1$, and the quantum computation is solely performed by $\Pc_1$. Thus, only $\Pc_1$ can launch an effective attack. The attack would ruin $\Pc_1$'s ciphertext, and $\Pc_1$ would ultimately face an authentication error when transmitting the ciphertext with AQA. In this case, the classical MPC that runs the audit algorithm can publicly output $\Pc_1$ as malicious. As a result, our protocol achieves MPQC while maintaining PVIA security.

\paragraph{Instantiable Setup}\label{InstantiableSetup}
Next, we show how to instantiate our setup with an MPQC protocol secure with identifiable abort (SWIA). Note that we refine a slightly different version of MPQC-SWIA, thereby circumventing the need for the post-quantum Fully Homomorphic Encryption (FHE) assumption needed in \cite{ACC+21}.

MPQC-SWIA guarantees that whenever an abort happens, $\cMPC$ will output a partition of parties with all honest parties staying in the same group.
Our approach for instantiating the setup is to run MPQC-SWIA hierarchically to prepare the states that the setup would generate. The hierarchical MPQC-SWIA maintains a grouping between parties, where all parties are initially in the same group. Each group will try to run MPQC-SWIA by themselves, and a group breaks into two whenever MPQC-SWIA fails. At some point, all parties must have succeeded in running MPQC-SWIA within their group (or they will continue running MPQC-SWIA within descent subgroups), so they can proceed to execute the online MPQC-PVIA protocol. By employing the security of MPQC-PVIA with preprocessing, it is guaranteed that either the honest parties obtain their outputs, or some malicious party in the group that contains all honest parties will be publicly identified.

\begin{figure}[H]
    \centering
    \caption{{\bf Hierarchical MPQC-SWIA} parties try to run the offline setup using MPQC-SWIA. Initially, $G$ contains all the parties. When the first MPQC-SWIA run by $G$ terminates with a failure, parties in $G$ separate into two groups $G_0$ and $G_1$, who run another MPQC-SWIA within their own group. In this figure, $G_{1}$ executes MPQC-SWIA successfully and obtains the setup output. They can proceed to execute the online MPQC-PVIA protocol.}
    \label{fig:Hierarchical MPQC-SWIA}
\scalebox{0.8}{\begin{tikzpicture}
\tikzset{grow'=right,level distance=32pt}
\tikzset{execute at begin node=\strut}
\tikzset{every tree node/.style={anchor=base west}}
\Tree [.$G$   [.$G_{0}$ [.$G_{00}$ ... ]
                 [.$G_{01}$ ... ] ]
             [.$G_{1}$ Output ] ]
\end{tikzpicture}}
\end{figure}
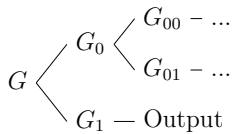

\subsection{Best-of-Both-Worlds Security}

One advantage of our protocol design is its flexibility to provide best-of-both-worlds security. That is, we construct an MPQC protocol that simultaneously achieves full security when there are at most $\thres < \frac{n}{2}$ corruptions and satisfies security with publicly verifiable identifiable abort against at most $n-1-\thres$ corrupted parties.

Prior to this work, the honest-majority and the dishonest-majority worlds were once separated because of a tension between sharing and extracting quantum information. We elaborate on it as follows. MPQC protocols that obtain full security in an honest majority setting \cite{BCG06} are based on verifiable quantum secret sharing (VQSS). In these protocols, each party individually creates VQSS of their input and distributes the shares across parties. The problem is that the secret shares sent between malicious parties are private information. Once the number of corrupted parties reaches one-half, the simulator cannot extract the adversary's input from the available secret shares. This is also why current MPQC protocols designed for a dishonest majority need every quantum message to be transmitted through all parties: the simulator can extract inputs when the quantum message passes through an honest party. As a result, protocols against a dishonest majority cannot divide a piece of quantum information across multiple parties, and a single malicious party is sufficient to destroy the information subjected to the computation.

Our solution to this tension is to utilize the offline-online structure of our protocol and incorporate quantum error correction codes (QECC). First, our offline setup\footnote{Similar to the previous subsection, this setup can be instantiated using MPQC-SWIA.} prepares QECC codewords on EPR pairs and distributes the codewords evenly across parties. Afterward, the parties can perform distributed computation over QECC codewords in the online protocol. 
In this protocol, the honestly generated QECC codewords facilitate the sharing of quantum information. Moreover, to extract quantum information even in the presence of a malicious majority, the setup can entangle a trapdoor with the states prepared for the parties and use the trapdoor to extract online inputs. We see that the offline setup acts as a vital piece of machinery that allows information extraction while preparing for the online distributed computation.

Our BoBW-MPQC protocol is reminiscent of the classical BoBW-cMPC protocols \cite{BoBW-pos,BoBW-neg,BoBW-1/p}. A key difference is that the classical protocols need to broadcast secret sharings and invoke the ideal functionality on the inputs multiple times, both of which are infeasible in MPQC due to no-cloning. Our protocol does not follow the same pattern, and we achieve the same goal in the merit of quantum teleportation. 

\section{Preliminary}\label{sec:prelim}

Let $[n] = \{1, \cdots, n\}$. We denote by $A_{[n]}$ the tuple $(A_1,\cdots,A_n)$. Uniform sampling from a set $S$ is denoted by $s\from S$.
A function $f \colon \bbN \to [0,1]$ is called negligible, if for every polynomial $\poly(\cdot)$ and all sufficiently large $n$, it holds that $f(n)<|{1}/{\poly(n)}|$. We use $\negl(\cdot)$ to denote an unspecified negligible function. 

Quantum states are written in lowercase Greek alphabets, \eg $\rho,\sigma$. Quantum operations are written in uppercase Latin alphabets, \eg $U, V$.
We write $\onreg{\rho}{M}$ and $\onreg{U}{M}$ to specify that $\rho$ is stored in register $\gray{M}$ and $U$ operates on register $\gray{M}$. The notation $(\rho, \sigma)$ denotes a state on two registers that may be entangled. The letters \qpt stands for quantum polynomial time.

Fix a prime $p$. A qupit in pure state $\ket{\phi}$ is a unit vector in the $p$-dimensional Hilbert space $\bbC^{p}$ and can be identified with the density operator $\mixednospace{\ket{\phi}} := \ketbra{\phi}$. The set of $n$-qupit mixed  states, denoted $\Den{n}$, consists of positive semi-definite operators on $\bbC^{p^n}$ with trace $1$. We sometimes identify a mixed state $\rho$ with its purification, which is a pure state $\ket{\phi}$ such that $\mixednospace{\ket{\phi}}$ has partial trace $\rho$. We also consider sub-normalized mixed states, which are positive semi-definite operators with trace at most $1$. We identify a distribution $\{\rho_j\}$ of sub-normalized states with the state $\E_j \rho_j$. Two sequences of sub-normalized states $\rho(n), \sigma(n) \in \Den{\poly(n)}$ are said to be statistically indistinguishable, denoted $\rho \approx \sigma$, if they have trace distance $\Tr|\rho(n)-\sigma(n)| = \negl(n) \tr(\rho(n))$. 


\subsection{Quantum Computation}

A quantum operation is a completely positive, trace preserving (CPTP) map acting on mixed states. Any such map can be represented as $\{A_j\}$ which maps a mixed state $\rho$ to the mixed state $\sum_j A_j \rho A_j^\dag$. Each $A_j$ defines a completely positive (CP) map $\rho \mapsto A_j \rho A_j^\dag$. For example, the measurement operator $\{\ketbra{j}\}_{j \in \bbZ_p}$ in the computation basis is a CPTP map, whereas each projector $\ketbra{j}$ is only a CP map. Every unitary operator defines a CPTP map.


Consider the phase $\omega = e^{2\pi i/p}$, the shift operator $\Xg: \ket{j} \mapsto \ket{j+1}$ and the clock operator $\Zg: \ket{j} \mapsto \omega^j \ket{j}$. Write 
$\Xg^{(x_1,\cdots,x_n)} \Zg^{(z_1,\cdots,z_n)} = \bigotimes_{j\in[n]}\Xg^{x_j}\Zg^{z_j}$ where each $x_j, z_j \in \bbZ_p$.
We define the Pauli basis $\PauliGroup^*_n = \{\Xg^x \Zg^z ~|~ x,z \in \bbZ_p^n\}$, which is a basis for the space of linear operators on $\bbC^{p^n}$. Decomposing a linear operator according to this basis is called the Pauli decomposition. For convenience, we identify the Pauli $P_a = \Xg^{x_a} \Zg^{z_a} \in \PauliGroup^*_n$ with the string $(z_a, x_a) = (z(P_a), x(P_a)) \in \bbZ_p^{2n}$. Define the Pauli group $\PauliGroup_n$ as $\{\omega^k \Xg^x \Zg^z ~|~ k \in \bbZ_p, ~x,z \in \bbZ_p^n\}$
and the Clifford group $\CliffordGroup_n$ as the normalizer of $\PauliGroup_n$ in the unitary group quotient by global phases. That is, a unitary $C \in \CliffordGroup_n$ if and only if for all $A \in \PauliGroup_n$, $CAC^\dag \in \PauliGroup_n$. Intuitively, it means that with a reasonable update of the Pauli gate, we can swap the order where a Clifford gate and a Pauli gate are applied.

The Clifford group is generated by the Fourier transform gate $\Hg: \ket{j} \mapsto \frac{1}{\sqrt{p}}\Sigma_k~ \omega^{jk} \ket{k}$, the phase gate $\Pg: \ket{j} \mapsto \omega^{j(j-1)/2} \ket{j}$ and the sum gate $\CX: \ket{j,k} \mapsto \ket{j,k+j}$ \cite{cliffordGenerator}. When $p=2$, the phase gate is defined as $\Pg: \ket{j} \mapsto i^{j}\ket{j}$ instead.
One can sample uniformly random Clifford gates in polynomial time \cite{qubitClifford,qupitClifford}.
We will write $\onreg{\CX_{(b_1,\cdots,b_n)}}{R_0,\cdots,R_n}$ as the abbreviation of $\onreg{\CX^{b_1}}{R_0,R_1} \cdots \onreg{\CX^{b_n}}{R_0,R_n}$.

Universal quantum computation can be carried out with Clifford gates and $\Tg$ gates, where $\Tg: \ket{j} \mapsto e^{\frac{2\pi i \eta_j}{p^2}}\ket{j}$ with $\eta_j= p\binom{j}{3}-j\binom{p}{3}+\binom{p+1}{4}$ \cite{Distill}. When $p=2$, the $\Tg$ gate is defined as $\Tg: \ket{j} \mapsto e^{\frac{\pi i j}{4}}\ket{j}$ instead. Although $\Tg$ gate is not in the Clifford group, it can be applied using classically controlled Clifford operations with the help of the $\Tg$ state $\ket{\Tg}=\Tg\ket{+}$, where $\ket{+}=\frac{1}{\sqrt{p}}\sum_{j=0}^{p-1} \ket{j}$. $\Tg$ states can be purified from noisy ones using classically controlled Clifford gates \cite{BK05,Distill}. 
We note that the phase gate $\Sg$ can also be applied using classically controlled $\Xg, \Zg, \CX$ gates with the help of the state $\ket{\Sg} = \Sg\ket{+}$.

\subsection{Quantum One-Time Pad}
\begin{dfn} A quantum one-time pad (QOTP) with key $P \in \PauliGroup_n$ is a symmetric-key encryption scheme that consists of the following two algorithms.
\begin{itemize}
    \item Encryption:  $\mathsf{QOPT.Enc}_P(\rho):= P \rho P^\dag$.
    \item Decryption: $\mathsf{QOPT.Dec}_P(\rho):= P^\dag \rho P$.
\end{itemize}
\end{dfn}
It is well known that the ciphertext under QOTP is maximally mixed:
\begin{lemma}[Pauli Twirl]
\label{lemma:qotp-security}
For every state $\onreg{\ket{\phi}}{M,N} = \sum_{u} \onreg{\ket{u}}{M} \ot \onreg{\ket{\phi_u}}{N}$ it holds that $$\E_{P \randsample \PauliGroup_n} \mixed{ \onreg{P}{M} \onreg{\ket{\phi}}{M,N} } = \onreg{\bigg(
\E_{r \randsample \bbZ_p^n}  \mixed{ \ket{r} } \bigg)}{M}  \ot \onreg{\bigg( \sum_u \mixed{\ket{\phi_u}} \bigg)}{N}$$
\end{lemma}

The same result holds when $P$ is randomly sampled from the Clifford group $\CliffordGroup_n$. Moreover, it is well known that QOTP can split a quantum attack into a probabilistic combination of Pauli attacks. This work considers a specific scenario where an untrusted party measures a state which is protected under QOTP. We formulate the following lemma, which shows that any attack would be equivalent to a probabilistic combination of Pauli attacks that cause different shifts. We prove the lemma in \cref{appendix:proof-of-lemma}.


\begin{lemma}[Pauli Twirl with Measurement] \label{lemma:qotp-measure}
Let $\onreg{\ket{\phi}}{M,N} = \sum_{u \in \bbZ_p^n} \onreg{\ket{u}}{M} \ot \onreg{\ket{\phi_u}}{N}$ be a state and $v \in \bbZ_p^n$ be the target measurement result. For any attack $\onreg{A}{M,N} = \sum_{Q \in \PauliGroup^*_n} \left( \onreg{Q}{M} \ot \onreg{A_Q}{N}\right)$ applied on the QOTP-protected state, it holds that
\begin{align*}
    &\E_{P \from \PauliGroup_n} \mixed{\onreg{\ketbra{v + x(P)}}{M} \onreg{A}{M,N} \onreg{P}{M} \onreg{\ket{\phi}}{M,N}}\\
    =& \E_{r \from \bbZ_p^n} \sum_{u} \mixed{ \bigg( \sum_{x(Q)=v-u} Q \ot A_Q \bigg) \big(\ket{r} \ot \ket{\phi_u} \big) } 
\end{align*}
\end{lemma}


\subsection{Quantum Authentication Code}
\label{section:AuthCode}
Quantum authentication code detects whether unauthorized alterations have been made to the data. When alternation is detected, the algorithm will output a rejection symbol $\bot$.

\begin{dfn}[Quantum Authentication Code, \cite{AuthCode}]
    \label{dfn:auth_code} A quantum authentication code consists of three algorithms. The key generation algorithm $\Gen$ takes in the security parameter $1^\secparam$ and the message size $1^\ell$ and outputs a random secret key $\sk$. The encoding algorithm $\Enc$ maps a secret key $\sk$ and a quantum message on $\gray{M}$ to a quantum ciphertext on $\gray{MT}$. The decoding algorithm $\Dec$ maps a secret key $\sk$ and a quantum ciphertext on $\gray{MT}$ to a quantum message $\gray{M}$. These algorithms should satisfy the following properties.
    \begin{itemize}[leftmargin=*]
        \setlength{\itemsep}{1pt}
        \item Completeness: For every secret key $\sk$, it holds that $\Dec_\sk \circ \Enc_\sk = \identity$.
        \item Security: For any quantum map ${\adv}$, there exists two CP maps ${\adv}_{\acc}$ and ${\adv}_{\rej}$ such that ${\adv}_{\acc} + {\adv}_{\rej}$ is trace preserving and that for any (possibly entangled) states ${\rho}, {\rho_{\aux}}$,
    {\large
    \begin{align*}
        \left\{ \scriptstyle{
            (\rho', \rho'_{\aux})
        }
        \given \substack{
            \sk \from \Gen(1^\secparam, 1^\ell)\\
            \sigma \from \Enc(\sk, \rho)\\
            (\sigma', \rho'_{\aux}) \from \adv(\sigma, \rho_{\aux})\\
            \rho' \from \Dec(\sk, \sigma')
        }
        \right\} \underset{\scriptscriptstyle{\negl(\secparam)}}{\approx}
        \bigg(\scriptstyle{
             \big({\rho},\; {\adv}_{\acc}(\rho_{\aux}) \big) + \big({\ketbra{\bot}},\; {\adv}_{\rej}(\rho_{\aux}) \big)
        }\bigg)
    \end{align*}
    }
    \end{itemize}
\end{dfn}

Here, we recall the Clifford authentication code from \cite{CliffordCodes}. The key generation algorithm outputs a uniformly random Clifford gate $\onreg{E}{M,T}$. The encoding procedure augments the message state $\onreg{\rho}{M}$ with traps $\onreg{\ket{0}^{\ot \numtraps}{}}{T}$ and applies $\onreg{E}{M,T}$. The decoding procedure applies $E^\dag$ followed by measuring the register $\gray{T}$ in the computational basis. If the measurement results are not all zero, the content of $\gray{M}$ is replaced with $\ket{\bot}$. The Clifford authentication code satisfies \cref{dfn:auth_code}. The following lemma is crucial to its proof, and we will use the lemma directly later on.

\begin{lemma}[Pauli Partitioning by Clifford, \cite{CliffordCodes,PauliPartition}]
\label{lemma:clifford}
For every Pauli operators $Q, Q' \in \PauliGroup_n$ that do not lie in $\{ \omega^k I ~|~ k \in \bbZ_p \}$, it holds that
$$
\underset{C \from \CliffordGroup_n}{\Pr}\left[C^\dag Q C = Q'\right] = \negl(n).
$$
\end{lemma}
The Clifford code also supports homomorphic computation for any Clifford operation. Consider a Clifford-code ciphertext $\Enc_E(\onreg{\rho}{M})$ with secret key $\sk = {E}$. To perform a Clifford gate $G$ on $\rho$, it suffices to update the secret key as $\sk' = {E} {G^{\dag}}$. This works because we have
$\Enc_\sk(\rho) = E({\rho}, {\ket{0}^{\ot \secparam}}) = {E} {G^{\dag}} ({G\rho}, {\ket{0}^{\ot \secparam}}) = \Enc_{\sk'}(G(\rho))$.

\subsection{Quantum Error-Correction Code}

Quantum error correction code can protect quantum states from errors as long as the number of errors is limited. In this work, it suffices to consider erasure errors.

\begin{dfn}[Quantum Error Correction Code] A $[[\Qnum,k]]_p$ quantum error correction code consists of two algorithms. The encoding algorithm $\QEnc: \Den{k}\to\Den{\Qnum}$ encodes a $k$-qupit message into a $\Qnum$-qupit codeword. The decoding algorithm $\QDec:\Den{\Qnum}\times \{0,1\}^\Qnum \to\Den{k}$ takes a modified codeword and its location of errors and outputs a $k$-qupit message. A quantum error correction code is said to correct $\thres$ erasure errors, if for any $\rho \in \Den{k}$ and any quantum channel $\onreg{\Delta}{R}$ acting on $|\gray{R}| < \thres$ qupits, it holds that
$$
\QDec\left(\onreg{\Delta}{R} \QEnc(\rho), \identity_R \right) = \rho
$$
where $\identity_R$ specifies the locations of $\gray{R}$ among the $\Qnum$ qupits.
\end{dfn}

To arrive at best-of-both-worlds security for any threshold $\thres < \frac{n}{2}$, we can use the quantum polynomial code of \cite{aharonov1997fault}, which also satisfies other desirable properties.


\begin{lemma}[Polynomial Code, \cite{aharonov1997fault}]
\label{lemma:qecc}
For every $\thres < \frac{n}{2}$ and prime $p > \Qnum$, there exists a $[[\Qnum,1]]_p$ quantum error correction code that corrects $\thres$ erasure errors with the following additional properties:
\begin{itemize}
    \item Syntax: The encoding algorithm applies a Clifford gate to input $\rho$ and ancilla $\ket{0}^{\ot (\Qnum-1)}$. We will denote the Clifford gate for encoding as $\QECC$.
    \item Transversal Measurement: The decoding algorithm commutes with qupit-wise measurement in the computational basis.
    \item Fault-Tolerant Computation: 
        $\Xg, \Zg, \CX, \Hg$ gates and their inverses can be applied to the underlying message $\rho$ by locally applying some of these gates (and measurements) to the individual components of the codeword (using ancillas).
\end{itemize}
\end{lemma}

A state injection technique (\cref{appendix:state-injection}) shows that the $\Sg$ gate can be performed through $\Xg, \Zg, \CX^{-1}$ gates and measurements in the computational basis using ancillas. Combining with the transversal measurement and fault-tolerant computation stated above, the entire Clifford group $\CliffordGroup_n = \langle \Sg, \Hg, \CX \rangle$ can be applied fault-tolerantly using ancillas under the polynomial code of \cref{lemma:qecc}.



\subsection{Quantum Teleportation}\label{prelim:teleport}

Quantum teleportation allows parties to transmit quantum messages using only classical communication and pre-shared quantum states. Below, an EPR pair $(\onreg{e_S}{S_1,\cdots,S_n}, \onreg{e_R}{R_1,\cdots,R_n})$ of length $n$ stands for the state $\bigotimes_{j \in [n]} \onreg{\ket{\Phi^{+}}}{S_j, R_j}$ where $\ket{\Phi^+} = \frac{1}{\sqrt{p}} \Sigma_{j=0}^{p-1} \ket{j, j}$. 

\begin{dfn}[Quantum Teleportation Without Measurement]
Let $(e_S, e_R)$ be an EPR pair of length $n$ pre-shared between a sender holding input $\psi \in \Den{n}$ and a receiver. Quantum teleportation consists of two algorithms. We will also abbreviate $\TPSend$ as $\TP$.
\begin{itemize}
    \item  $\TPSend(\onreg{\psi}{M}, \onreg{e_S}{S})$ applies $\onreg{\Hg}{M} \onreg{\CX^\dag}{M,S}$ to $(\onreg{\psi}{M}, \onreg{e_S}{S})$ and outputs registers $\gray{M}, \gray{S}$.
    \item  $\TPReceive(z,x,\onreg{e_R}{R})$ applies $({\Xg}^x {\Zg}^z)^\dag$ to $\onreg{e_R}{R}$ and outputs register $\gray{R}$.
\end{itemize}
\end{dfn}

When we speak of teleporting a state $\onreg{\psi}{M}$ via register $\gray{S}$, we mean to apply $\onreg{\TPSend}{M,S}$, measure $(\gray{M}, \gray{S})$ in the computational basis and interpret the measurement result $(z,x)$ as the Pauli $\Xg^x \Zg^z$.
The following lemma states that the teleportation result $\Xg^x \Zg^z$ can help the receiver recover the original quantum message $\psi$. A proof can be found in \cref{appendix:proof-of-lemma}.

\begin{lemma}
\label{lemma:teleport}
Let $(\psi, \tau)$ be a purified state independent of $(e_S,e_R)$. Then
$$
\left(\TPSend(\onreg{\psi}{M},\onreg{e_S}{S}), \onreg{e_R}{R}, \onreg{\tau}{N}\right) 
= \frac{1}{p^{n}}~\sum_{x,z \in \bbZ_p^n}~ \onreg{\ket{z}}{M} \ot  \onreg{\ket{x}}{S} \ot \left(\onreg{\left(\Xg^x \Zg^z\right) \psi}{R}, \onreg{\tau}{N}\right)
$$
\end{lemma}

\section{Model and Definition}
We focus on interactive protocols between $n$ parties $\Pc_1,\cdots,\Pc_n$ with quantum computational power. They can communicate using pairwise authenticated quantum channels and a broadcast channel for classical messages. We work in the synchronous communication model where the protocol proceeds in rounds, and each message will certainly arrive at the end of each round. In addition, we consider the presence of a protocol observer $\Observer$ who passively receives and records classical information from the broadcast channel all the time. The adversary $\adv$ can statically corrupt a set $I \subset \{\Pc_1, \cdots, \Pc_n\}$ of up to $n-1$ parties. 

The quantum computation to be performed is modeled as a quantum circuit $C$, which takes $n$ parts of quantum inputs and produces $n$ parts of quantum outputs. Without loss of generality, we assume that the corresponding inputs and outputs have equal size. We always apply \cite{BK05,Distill} to convert $\circuit$ into the following format, incurring only a polynomial growth in description size. The ancilla $\phi_{\anc}$ consists of $\ket{0},\ket{\Sg},\ket{\Tg}$ states, and the circuit operates on a total of $\ell_{\total} = \sum_{i} \ell_i + d$ qupits.
\begin{idealmodel}{Specification of Quantum Circuit $\circuit$}
\label{spec:circuit}
\begin{enumerate}[leftmargin=*]
    \setlength{\itemsep}{0pt}
    \item Take input registers $\gray{R_1},\cdots,\gray{R_n}$ where $|\gray{R_i}|=\ell_i$.
    \item Initialize register $\gray{N} = (\gray{N_1},\cdots,\gray{N_{d}})$ as some ancilla (magic) state $\phi_{\anc}$ where $ |\gray{N_i}|=1$.
    \item For $k=d,\cdots,1$, perform the following computation, denoted as $\circuit[k]$:
    \begin{enumerate}
        \setlength{\itemsep}{0pt}
        \item Measure $\gray{N_k}$ in the computational basis and obtain a pit $b_k$.
        \item Compute a classical circuit $f_k(b_k,\cdots,b_d)$ that outputs a Clifford gate $G_{k-1}$. 
        \item Apply $G_{k-1} \in \CliffordGroup_{\ell_1 +\cdots + \ell_n + k-1}$ on registers $(\gray{R_1,\cdots,R_n,N_{1},\cdots,N_{k-1}})$.
    \end{enumerate}
    \item Output registers $\gray{R_1},\cdots,\gray{R_n}$.
\end{enumerate}    
\end{idealmodel}

We will also consider evaluating a quantum circuit in a fault-tolerant manner. Under the quantum error-correction code of \cref{lemma:qecc}, the Clifford group can be applied fault-tolerantly with ancillas. Hence, a quantum circuit can be written in the above form, with  each $G_{k-1}$ additionally admits a fault-tolerant expression $(G_{k-1}^{(1)}, \cdots, G_{k-1}^{(n)})$ consisting of Clifford gates such that
$$
(G_{k-1}^{(1)} \ot \cdots \ot G_{k-1}^{(n)}) \circ \QEnc = \QEnc \circ G_{k-1}
$$

\subsection{The Ideal World of BoBW-MPQC-PVIA}\label{sec:MPQC-def}

A multi-party quantum computation protocol is defined using the real vs. ideal paradigm. In the ideal world, the parties delegate the computation $C$ to a trusted party $\trustp$. The only way for the corrupted parties to interrupt the delegation is to ask $\trustp$ to abort, in which case $\trustp$ publicly announces their identities. The ideal world of \bobw multi-party quantum computation \pvia is formally defined as follows. We denote its joint output distribution as $\ideal^{\MPQC}_{\adv_I(\rho_{\aux})}(1^{\secparam}, \thres, C, \rho_1, \cdots, \rho_n)$.

\begin{idealmodel}{$\ideal^{\MPQC}$: Best-of-Both-Worlds Multi-party Quantum Computation\\\qquad\qquad Secure with Publicly Verifiable Identifiable Abort}
\begin{description}
\item[Common input:]\quad\\
The security parameter $1^{\secparam}$, fault-tolerance threshold $\thres$ and quantum circuit $C$.
\item[Input:]\quad\\
$\Pc_i$ holds input $\rho_i$. $\adv_I$ holds input $\rho_{\aux}$ and controls parties in $I$.

\item[$\trustp$ receives inputs and performs computation:]\quad\\
Each party $\Pc_i$ sends some $\tilde{\rho}_i$ as input to $\trustp$. Honest parties choose $\tilde{\rho}_i = \rho_i$.\\
$\trustp$ computes $(\rho'_1, \cdots, \rho'_n) \from C(\tilde{\rho}_1, \cdots, \tilde{\rho}_n)$.

\item[$\trustp$ sends back outputs:]\quad\\
$\trustp$ sends $\rho'_i$ to all $\Pc_i \in I$.\\
$\Pc_i \in I$ can send $\abort$ message to $\trustp$. Let $J$ be the set of parties who indeed do so.\\
If $|J| > \thres$, $\trustp$ publicly aborts to $J$.\\
If $|J|\le\thres$, $\trustp$ sends $\rho'_i$ to all $\Pc_i \not \in I$.

\item[Output:]\quad\\ Honest parties output whatever output received from $\trustp$.\\
The observer $\Observer$ outputs whatever public information received from $\trustp$.\\
The adversary $\adv_I$ outputs a function of his view.
\end{description}
\end{idealmodel}

For a protocol $\Pi$, we denote by $\real^{\Pi}_{\adv_I(\rho_{\aux})}(1^{\secparam}, \rho_1, \cdots, \rho_n)$ the joint output distribution of the honest parties, the observer, and the adversary at the end of protocol $\Pi$ when executed by $\Pc_i(\rho_i)$ in the presence of an adversary $\adv_I(\rho_{\aux})$ corrupting parties in $I$. For a protocol $\Pi$ with a trusted setup $\Sigma$, we define $\real^{\Pi \circ \Sigma}_{\adv_I(\rho_{\aux})} (1^{\secparam}, {\rho_1, \cdots, \rho_n})$ similarly with $\Sigma$ being executed by a trusted party prior to $\Pi$.

\begin{dfn}\label{dfn:mpqcbobw}
We say that a protocol $\Pi$ is a \bobw multi-party quantum computation \pvia (BoBW-MPQC-PVIA) of threshold $\thres$ over a circuit $\circuit$, if for every $|I| \le n-1-\thres$ and every non-uniform (\qpt) adversary $\adv_I$ corrupting parties in $I$, there is a non-uniform (\qpt) simulator $\Sim_I$ corrupting parties in $I$, such that for any quantum inputs $\rho_i \in \Den{\numinputs_i}$, $i\in [n]$,
$$ \real^{\Pi}_{\adv_I(\rho_{\aux})}(1^{\secparam}, \rho_1, \cdots, \rho_n) \approx \ideal^{\MPQC}_{\Sim_I(\rho_{\aux})}(1^{\secparam}, \thres, \circuit, \rho_1, \cdots, \rho_n)$$
If the protocol $\Pi$ has a trusted setup $\Sigma$, the indistinguishability requirement is replaced with
$$ \real^{\Pi \circ \Sigma}_{\adv_I(\rho_{\aux})}(1^{\secparam}, \rho_1, \cdots, \rho_n) \approx \ideal^{\MPQC}_{\Sim_I(\rho_{\aux})}(1^{\secparam}, \thres, \circuit, \rho_1, \cdots, \rho_n)$$
\end{dfn}

\begin{dfn}\label{dfn:mpqcpvia}
We say that $\Pi$ is a multi-party quantum computation \pvia (MPQC-PVIA) over a circuit $\circuit$ if \cref{dfn:mpqcbobw} holds for $\thres = 0$.
\end{dfn}

\subsection{(Preprocessing) MPC-Hybrid Model}\label{sec:MPC-hybrid}

Following \cite{DNS12,DGJ+20,ACC+21,BCKM21}, we assume an ideal functionality $\cMPC$ for \emph{reactive}\footnote{It can be equipped with an internal state that may be taken into account when it is called next time.} classical multiparty computation within our MPQC protocol.
The ideal world of classical MPC is similar to that of MPQC defined in section \cref{sec:MPQC-def}, but allows only classical messages and classical computation. In additional to producing $n$ private outputs, the classical computation is allowed to generate an additional output which the ideal functionality publicly outputs if there is no abort. In our presentation, we will simply view $\cMPC$ as a trusted classical party. We refer this setting as the MPC-hybrid model. The preprocessing MPC-hybrid model extends the MPC-hybrid model by allowing an input-independent trusted setup to be executed prior to the actual protocol.

One can instantiate the MPC ideal functionality using a post-quantum \bobw MPC protocol \pvia and publicly verifiable output. A concrete construction of such a protocol is to plug the MPC protocol of \cite{BOS+20} into the compiler of \cite{BoBW-pos} to attain \bobw security, and then apply \cite{Unr10}’s lifting theorem to obtain post-quantum security.

\section{Auditable Quantum Authentication (AQA)}\label{Sec:AQA}

This section presents a new primitive called \emph{Auditable Quantum Authentication} (AQA) that lets a sender send quantum messages to a receiver and be accountable for his sending action. AQA is designed to identify the malicious sender only, while the receiving behavior is automatically guaranteed by successfully passing the test. In contrast to traditional quantum authentication codes like the Clifford or Trap codes, which necessitate the receiver to verify the checking bits, AQA adopts a different approach. It obliges the sender to generate the proof that should be verified by the auditor.

\begin{dfn}[Auditable Quantum Authentication]
\label{dfn:aqa}
An auditable quantum authentication scheme consists of the following five algorithms:
\begin{itemize}
\setlength{\itemsep}{1pt}
    \item $\Setup(1^\secparam, 1^\ell) \rightarrow (\sk, \phi_S, \phi_R)$ takes as input the security parameter $\secparam$ and the message length $\ell$ and outputs a classical secret key $\sk$, a quantum sending state $\phi_S$ and a quantum receiving state $\phi_R$.
    \item $\Enc(\sk, \rho) \rightarrow \sigma$ takes as input a classical secret key $\sk$, a quantum message state $\rho \in \Den{\ell}$ and outputs a quantum authenticated state $\sigma$.
    \item $\Send(\sigma, \phi_S) \rightarrow \pf$ takes as input a quantum authenticated state $\sigma$, a quantum sending state $\phi_S$ and outputs a classical proof $\pf$.
    \item $\Audit(\sk, \pf) \rightarrow \dk$ is a \emph{classical} algorithm that takes as input a secret key $\sk$, a proof $\pf$ and outputs a decryption key $\dk$. When the proof is invalid, $\dk$ will be set as $\bot$.
    \item $\Recv(\dk, \phi_R) \rightarrow \rho'$ takes as input a classical decryption key $\dk$, a quantum receiving state $\phi_R$ and outputs a quantum message state $\rho'$. When $\dk = \bot$, $\rho'$ will be set as $\bot$.
\end{itemize}
These algorithms should satisfy the following properties:
\begin{itemize}
\setlength{\itemsep}{1pt}
    \item Sender completeness: For every quantum message state $\rho \in \Den{\ell}$, it holds that
    {\large
        \begin{align*}
            \Pr\left[ \scriptstyle{
                \dk \neq \bot
            }
            \given \substack{
                (\sk, \phi_S, \phi_R) \from \Setup(1^\secparam, 1^\ell)\\
                \sigma \from \Enc(\sk, \rho)\\
                \pf \from \Send(\sigma, \phi_S)\\
                \dk \from \Audit(\sk, \pf)
            }
            \right] = 1
        \end{align*}
    }
    \item Receiver security: There exists algorithms $\widetilde{\Setup}, \widetilde{\Enc}, \widetilde{\Audit}$ such that $\Recv \circ \widetilde{\Enc} = \identity$ and that for every completely positive map $\adv$ and every possibly entangled quantum states $\rho, \rho_{\aux}$, it holds that
    {\large
        \begin{align*}
            \left\{ \scriptstyle{
                (\dk, \phi_A)
            }
            \given \substack{
                (\sk, \phi_S, \phi_R) \from \Setup(1^\secparam, 1^\ell)\\
                \sigma \from \Enc(\sk, \rho)\\
                (\pf, \phi_A) \from \adv(\sigma, \phi_S, \phi_R, \rho_{\aux})\\
                \dk \from \Audit(\sk, \pf)
            }
            \right\} \underset{\scriptscriptstyle{\negl(\secparam)}}{\approx} \left\{ \scriptstyle{
                (\dk, \phi_A)
            }
            \given \substack{
                (\sk, \phi_S, \sigma) \from \widetilde{\Setup}(1^\secparam, 1^\ell)\\
                (\dk, \phi_R) \from \widetilde{\Enc}(\rho)\\
                (\pf, \phi_A) \from \adv(\sigma, \phi_S, \phi_R, \rho_{\aux})\\
                \dk \from \bot \text{ if } \bot \from \widetilde{\Audit}(\sk, \pf)
            }
            \right\}
        \end{align*}
    }
\end{itemize}

\end{dfn}

Sender completeness guarantees that the honest sender always passes the audit. Receiver security is defined through the indistinguishability of two kinds of executions, implying that whatever property is satisfied by the right-hand side will also hold up to a negligible error for the left-hand side. 

In particular, receiver security captures the following properties. 
First, it guarantees that the sender's inputs $(\sigma, \phi_S)$ are as if they can be generated independently of $\rho$, and hence contain no information about the message $\rho$. Similarly, the receiver's input $\phi_R$ is as if it already encodes $\rho$. 
Second, $\dk$ is the only information required for the receiver to recover $\rho$ from $\phi_R$, and the honest receiver always obtains the true message given that the audit accepts.
Third, all the adversarial sender can do, even if the sender and the receiver collude, is to completely destroy $\dk$ at the cost of making the audit output $\bot$ at the same time.

\begin{figure}[H]
    \centering
    \caption{AQA real world (left) and ideal world (right)}
    \label{fig:AQArealideal}

\begin{tikzcd}[row sep=small, column sep=small]
                          &  & \Setup \arrow[rrd, "\phi_R"] \arrow[lld, "\phi_S"'] &  &          &                           &  & \widetilde{\Setup} \arrow[lld, "{\phi_S, \sigma}"']                 &  &          \\
\Sender \arrow[rrd, "\pf"'] &  & \Enc(\rho) \arrow[ll, "\sigma"]                     &  & \Receiver & \Sender \arrow[rrd, "\pf"'] &  & \widetilde{\Enc}(\rho) \arrow[rr, "\phi_R"] \arrow[d, "\dk", dotted] &  & \Receiver \\
                          &  & \Audit \arrow[rru, "\dk"'] \arrow[d, "?"]            &  &          &                           &  & \widetilde{\Audit} \arrow[rru, "\dk"'] \arrow[d, "?"]                &  &          \\
                          &  & \bot                                               &  &          &                           &  & \bot                                                               &  &         
\end{tikzcd}
\end{figure}
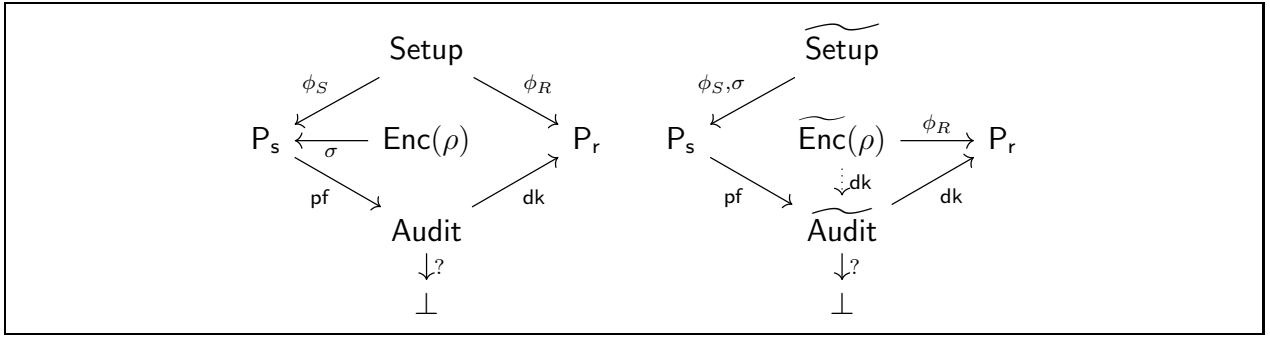
\subsection{Construction}

\begin{Construction}{Clifford-Form $\AQA$}
\label{algo:AQA}
\vspace{-6mm}
\begin{itemize}[leftmargin=*]
\setlength{\itemsep}{1pt}
    \item $\Setup(1^\secparam, 1^\ell):$
    \begin{enumerate}
    \setlength{\itemsep}{1pt}
        \item Sample random Clifford $F \randsample \CliffordGroup_{\ell+\numtraps}$ and Pauli $P_M, P_{S} \randsample \PauliGroup_{\ell+ \numtraps}, P_{R} \randsample \PauliGroup_{\ell}, P_{C} \randsample \PauliGroup_{\numtraps}$.
        \item Prepare EPR pairs on registers $(\gray{\hat{S}}, (\gray{\hat{R}}, \gray{\hat{C}}))$ with $|\gray{\hat{S}}| = \ell+ \numtraps$, $|\gray{\hat{R}}| = \ell $, $|\gray{\hat{C}}| = \numtraps$.
        \item Apply $\onreg{P_{S}}{\hat{S}} \onreg{P_{R}}{\hat{R}} \onreg{P_{C}}{\hat{C}} \onreg{F^\dag}{\hat{R}, \hat{C}}$. Name the resulting state as $\onreg{\phi_S}{\hat{S},\hat{C}}$ and $\onreg{\phi_R}{\hat{R}}$.
        \item Output $(\sk, \phi_S, \phi_R)$ where $\sk = (F, P_M, P_{S}, P_{C}, P_{R})$.
    \end{enumerate}
    \item $\Enc(\sk, \rho)$: Parse $\sk = (F, P_M, P_{S}, P_{C}, P_{R})$ and output $\sigma = P_M F(\rho \ot 0^\numtraps)F^\dag P_M^\dag$.
    \item $\Send(\onreg{\sigma}{\hat{M}}, \onreg{\phi_S}{\hat{S},\hat{C}})$:
    \begin{enumerate}
    \setlength{\itemsep}{1pt}
        \item Teleport $\onreg{\sigma}{\hat{M}}$ via $\gray{\hat{S}}$ and obtain the teleportation result $\hat{P}$.
        \item Measure $\gray{\hat{C}}$ in the computational basis and obtain the measurement result $\hat{c}$.
        \item Output $\pf = (\hat{P}, \hat{c})$.
    \end{enumerate}
    \item $\Audit(\sk,\pf)$:
    \begin{enumerate}
    \setlength{\itemsep}{1pt}
        \item Parse $\sk = (F, P_M, P_{S}, P_{C}, P_{R})$ and $\pf = (\hat{P}, \hat{c})$.
        \item Compute the quantum one-time pad $P_{M, S} := \TP (P_{M}{\ot}P_{S}) \TP^{\dag}$. Set the decoded teleportation result $\hat{P}'$ as the Pauli with string representation $\big(z(\hat{P}), x(\hat{P})\big) \oplus x(P_{M,S})$.
        \item Split the twirled Pauli $F^\dag \hat{P}' F \in \PauliGroup_{\ell+\numtraps}$ as $\hat{P}'_R, \hat{P}'_C$ that act on $\ell, \numtraps$ qupits respectively.
        \item 
        If $\hat{c} \neq  x( \hat{P}'_C) \oplus x(P_C)$, output $\dk = \bot$. Otherwise, output $\dk =  P_R \hat{P}'_R$. 
    \end{enumerate}
    \item $\Recv(\dk, \phi_R):$ If $\dk \neq \bot$, parse $\dk$ as a Pauli gate and output $\rho' = \dk^\dag (\phi_R) \dk$.
\end{itemize}

In light of the homomorphic property of the Clifford code (\cref{section:AuthCode}), we additionally define $\EncGate$ which extracts the encoding Clifford gate from the secret key. It will become useful in applications that make use of homomorphic computation.
\begin{itemize}[leftmargin=*]
    \item $\EncGate(\sk):$ Parse $\sk = (F, P_M, P_{S}, P_{C}, P_{R})$ and output $P_M F$.
\end{itemize}

\end{Construction}

\subsection{Security}

\begin{theorem}
\label{thm:aqa}
    \emph{Construction \ref{algo:AQA}} is an Auditable Quantum Authentication scheme.
\end{theorem}

\begin{proof} 
We take the following steps. First, we analyze the state $(\dk, \phi_A)$ that results from executing $(\Setup, \Enc, \adv, \Audit)$ in a row. Second, we prove sender completeness by plugging in $\adv = \Send$. Third, we show that the distribution of $(\dk, \phi_A)$ generated above is indistinguishable from a simpler state. Last, we construct $\widetilde{\Setup}, \widetilde{\Enc}, \widetilde{\Audit}$ that satisfy the requirements of receiver security.
\\

\noindent \underline{Step 1}.
Without loss of generality, we assume that $\adv$ has the same output length in $\phi_A$ as the input length. We can also assume that $\adv(\onreg{\sigma}{\hat{M}}, \onreg{\phi_S}{\hat{S},\hat{C}},\onreg{\phi_R}{\hat{R}}, \onreg{\rho_{\aux}}{\hat{W}})$ produces the classical proof $\pf$ by measuring the registers $(\gray{\hat{M}},\gray{\hat{S}},\gray{\hat{C}})$ in the computational basis. Moreover, we can assume that the CP map $\adv$ takes the form $\tau \mapsto A \tau A^\dag$ because every CP map can be decomposed into a sum of such operators and indistinguishability of subnormalized states extends under addition. We denote the Pauli decomposition of $A' = A \;\onreg{\TP^\dag{}}{\hat{M},\hat{S}}$ as $\sum_{Q \in \PauliGroup^*} \onreg{Q}{\hat{M},\hat{S},\hat{C}} \ot \onreg{A'_Q}{\hat{R} ,\hat{W}}$.

Let $(\hat{\EPRS},(\hat{\EPRR}, \hat{\EPRC} ))$ be the EPR pair prepared during $\Setup$. By defintion, the execution of $\Setup$ and $\Enc$
yields a random classical key $\sk = (F, P_M, P_S, P_C, P_R)$ and quantum state
$$
    \of{\onreg{\sigma}{\hat{M}}, \onreg{\phi_S}{\hat{S},\hat{C}},\onreg{\phi_R}{\hat{R}}}
    = \mixed{\of{ \onreg{P_{M} F \left({\rho}, {\ket{0}^{\ot\numtraps}}\right)}{\hat{M}}, \onreg{P_{S} \hat{\EPRS}}{\hat{S}}, \onreg{\left(P_{R}{\ot}  P_{C}\right)F^\dag \left(\hat{\EPRR}, \hat{\EPRC}\right)}{\hat{R}, \hat{C}}}}
$$
After the execution of $\adv$, the joint state is
\begin{align*}
    &\E_{\sk} \mixed{A \left( \onreg{P_{M} F \left({\rho}, {\ket{0}^{\ot \numtraps}}\right)}{\hat{M}}, \onreg{P_{S} \hat{\EPRS}}{\hat{S}}, \onreg{\left(P_{R}{\ot}  P_{C}\right)F^\dag \left(\hat{\EPRR}, \hat{\EPRC}\right)}{\hat{R}, \hat{C}}, \onreg{\rho_{\aux}}{\hat{W}} \right) \ot \ket{\sk}}\\
    =& \E_{\sk} \mixed{A' \left( \TP\left( \onreg{P_{M} F \left({\rho}, {\ket{0}^{\ot \numtraps}}\right)}{\hat{M}}, \onreg{P_{S} \hat{\EPRS}}{\hat{S}}\right), \onreg{\left(P_{R}{\ot}  P_{C}\right)F^\dag \left(\hat{\EPRR}, \hat{\EPRC}\right)}{\hat{R}, \hat{C}}, \onreg{\rho_{\aux}}{\hat{W}} \right) \ot \ket{\sk}}\\
    =& \E_{\sk} \mixed{A' \left( P_{M,S}\TP\left( \onreg{F \left({\rho}, {\ket{0}^{\ot \numtraps}}\right)}{\hat{M}}, \onreg{ \hat{\EPRS}}{\hat{S}}\right), \onreg{\left(P_{R}{\ot}  P_{C}\right)F^\dag \left(\hat{\EPRR}, \hat{\EPRC}\right)}{\hat{R}, \hat{C}}, \onreg{\rho_{\aux}}{\hat{W}} \right) \ot \ket{\sk}}\\
    =& \E_{\sk} \mixed{A' \left(  \sum_{P} \frac{1}{p^{\ell + \numtraps}} \left(\onreg{P_{M, S} \ket{z(P), x(P)}}{\hat{M}, \hat{S}}, \onreg{\left(P_{R}{\ot} P_{C}\right)F^\dag P F \left({\rho}, {\ket{0}^{\ot \numtraps}}\right)}{\hat{R}, \hat{C}}, \onreg{\rho_{\aux}}{\hat{W}} \right)\right) \ot \ket{\sk}}
\end{align*}
The first and second equalities follows from the definitions of $A'$ and $P_{M,S}$, and the last equality is by quantum teleportation (lemma \ref{lemma:teleport}). Let us define the linear functions 
\begin{align*}
    L_{F}(P) &:= \left(z(P), x(P), x(F^\dag P F)_{[\ell+1:\ell+\numtraps]}\right) \in \bbZ_p^{2\ell+ 3\numtraps}\\
    K_{F}(P) &:= \left(F^\dag P F\right)_{[1:\ell]} \in \PauliGroup_{\ell}
\end{align*}
for every Clifford operator $F$. The joint state can be simplified as
\begin{align*}
    \E_{\sk} \mixed{A' \left(  \sum_{P} \frac{1}{p^{\ell + \numtraps}} \left(\onreg{(P_{M, S} \ot P_C) \ket{L_{F}(P)}}{\hat{M}, \hat{S}, \hat{C}}, \onreg{P_{R}K_{F}(P) \left({\rho}\right)}{\hat{R}}, \onreg{\rho_{\aux}}{\hat{W}} \right)\right) \ot \ket{\sk}}
\end{align*}
The next step is to apply $\Audit$, which checks whether  the value stored in $(\gray{\hat{M}},\gray{\hat{S}},\gray{\hat{C}})$ is equal to $L_F(\hat{P}) \oplus x(P_{M,S} \ot P_C)$ for some $\hat{P} \in \PauliGroup_{\ell+\numtraps}$. If there is such a $\hat{P}$, then $\Audit$ outputs $\dk = P_R K_{F}(\hat{P})$; otherwise, it outputs $\dk = \bot$. 
To analyze the resulting state post-selected on finding $\hat{P}$, we can apply Pauli twirl with target measurement result $ L_{F}(\hat{P})$ (lemma \ref{lemma:qotp-measure}) using the Pauli decomposition of $A'$. We obtain
{\small
\begin{align*}
    &\E_{r, F, P_R}\; \sum_{P} \mixed{ \sum_{x(Q) = L_{F}(\hat{P} / P)} \frac{1}{p^{\ell + \numtraps}} Q \ket{r}
    \ot A'_Q \bigg(P_{R} K_{F}(P) \left({\rho}\right), \rho_{\aux} \bigg) \ot \ket{P_{R} K_{F}(\hat{P})}}\\
    \overset{_{P_R \from P_R K_F(P)
    }}{=}&\E_{r, F, P_{{R}}}  \sum_{P} \mixed{ \sum_{x(Q) = L_{F}(\hat{P} / P)} \frac{1}{p^{\ell + \numtraps}} Q \ket{r}
    \ot A'_Q \bigg(P_{R} \left({\rho}\right), \rho_{\aux} \bigg) \ot \ket{P_{R} K_{F}(\hat{P} / P)}}\\
    \overset{_{\quad P \from \hat{P}/P \quad
    }}{=}& \frac{1}{p^{2(\ell + \numtraps)}} \E_{r, F, P_{{R}}}  \sum_{P} \mixed{ \sum_{x(Q) = L_{F}(P)} Q \ket{r}
    \ot A'_Q \bigg(P_{R} \left({\rho}\right), \rho_{\aux} \bigg) \ot \ket{P_{R} K_{F}(P)}}
\end{align*}
}
Summing the post-selected states corresponding to every $\hat{P} \in \PauliGroup_{\ell+\numtraps}$, we obtain the state conditioned that $\Audit$ accepts.
\begin{align}
\label{state:AQA:real-acc}
    \E_{r, F, P_{{R}}} \sum_{P} \mixed{ \sum_{x(Q) = L_{F}(P)} Q \onreg{\ket{r}}{\hat{M},\hat{S},\hat{C}}
    \ot A'_Q \bigg(\onreg{P_{R} \left({\rho}\right)}{\hat{R}}, \onreg{\rho_{\aux}}{\hat{W}} \bigg) \ot \ket{P_{R} K_{F}(P)}}
\end{align} 
Similarly, the state conditioned that $\Audit$ rejects is
\begin{align}
\label{state:AQA:real-rej}
    \E_{r, F, P_{{R}}} \sum_{x \not\in \Range(L_{F})} \mixed{ \sum_{x(Q) =x} Q \onreg{\ket{r}}{\hat{M},\hat{S},\hat{C}}
    \ot A'_Q \bigg(\onreg{P_{R} \left({\rho}\right)}{\hat{R}}, \onreg{\rho_{\aux}}{\hat{W}} \bigg) \ot \ket{\bot}}
\end{align}

\noindent \underline{Step 2}.
To see sender completeness, we take $\adv = \Send$, which applies $\onreg{\TP}{\hat{M},\hat{S}}$ followed by measuring $(\gray{\hat{M}},\gray{\hat{S}},\gray{\hat{C}})$ in the computational basis. The induced $A' = A \; {\TP^\dag}$ is a measurement in the computational basis, and the Pauli decomposition of $A'$ involves only terms with $x(Q) = 0 \in \Range(L_F)$. For such $A'$, expression (\ref{state:AQA:real-rej}) shows that $\Audit$ never outputs $\bot$.
\\

\noindent \underline{Step 3}. 
We claim that $(\ref{state:AQA:real-acc})+(\ref{state:AQA:real-rej})$ is statistically indistinguishable to $(\ref{state:AQA:ideal-acc})+(\ref{state:AQA:ideal-rej})$.
\begin{align}
    \E_{r, P_{{R}}} \mixed{ \sum_{x(Q) = 0} Q \onreg{\ket{r}}{\hat{M},\hat{S},\hat{C}}
    \ot A'_Q \bigg(\onreg{P_{R} \left({\rho}\right)}{\hat{R}}, \onreg{\rho_{\aux}}{\hat{W}} \bigg) \ot \ket{P_{R}}}
    \label{state:AQA:ideal-acc}
    \\
    \E_{r, P_{{R}}} \sum_{x \neq 0} \mixed{ 
        \sum_{x(Q) = x} Q \onreg{\ket{r}}{\hat{M},\hat{S},\hat{C}}
        \ot A'_Q \bigg(\onreg{P_{R} \left({\rho}\right)}{\hat{R}}, \onreg{\rho_{\aux}}{\hat{W}} \bigg)
        \ot \ket{\bot}
    }
    \label{state:AQA:ideal-rej}
\end{align}
The trace distance between (\ref{state:AQA:real-acc}) and (\ref{state:AQA:ideal-acc}) is
\begin{equation}
\begin{aligned}[b]
    & \tr \left| \E_{F, r, P_{R}} \sum_{P \neq \Ig} \mixed{\sum_{x(Q)=L_{F}(P)} Q \ket{r} \ot A'_Q \left(P_{R}\big(\rho\big), \rho_{\aux} \right) \ot \ket{P_{R} K_{F}(P)} } \right| \\
    =& \tr \left| \sum_{x \neq 0} \sum_{P \neq \Ig} \E_{F, r, P_{R}} 1_{L_{F}(P) = x} \mixed{\sum_{x(Q)=x} Q \ket{r} \ot A'_Q \left(P_{R}\big(\rho\big), \rho_{\aux} \right) \ot \ket{P_{R} K_{F}(P)} } \right|\\
    \le& \sum_{x \neq 0} \sum_{P \neq \Ig}  \E_{F, r, P_{R}} 1_{L_{F}(P) = x} \text{ } \tr \left| \mixed{\sum_{x(Q)=x} Q \ket{r} \ot A'_Q \left(P_{R}\big(\rho\big), \rho_{\aux} \right) \ot \ket{P_{R} K_{F}(P)} } \right|\\
    =& \sum_{x \neq 0} \E_{r, P_{R}} \left( \sum_{P \neq \Ig} \Pr_{F} \left[L_{F}(P) = x\right] \right) \tr \left( \mixednospace{\sum_{x(Q)=x} Q \ket{r} \ot A'_Q \left(P_{R}\big(\rho\big), \rho_{\aux} \right) } \right)\\
    =& \sum_{x \neq 0} \E_{r,P_R} \Pr\left[{{x \in \Range(L_F)}}\right]\; \tr \left( \mixednospace{\sum_{x(Q)=x} Q \ket{r} \ot A'_Q \left(P_{R}\big(\rho\big), \rho_{\aux} \right) } \right)
    \label{state:AQA:difference}
\end{aligned}
\end{equation}
where the third line is from triangle inequality, the fourth line is because $\mixednospace{\cdot}$ is positive, and the last line follows from the observation that each $x$ has at most one $P$ such that $L_F(P)=x$. Similarly, the trace distance between (\ref{state:AQA:real-rej}) and (\ref{state:AQA:ideal-rej}) is
\begin{align*}
    \E_{r,P_R} \sum_{x \neq 0} \left(1-\Pr[x \not\in \Range(L_F)]\right) \tr \left( \mixednospace{\sum_{x(Q)=x} Q \ket{r} \ot A'_Q \left(P_{R}\big(\rho\big), \rho_{\aux} \right) } \right)
\end{align*}
which is equal to (\ref{state:AQA:difference}). By triangle inequality, the trace distance between $(\ref{state:AQA:real-acc})+(\ref{state:AQA:real-rej})$ and $(\ref{state:AQA:ideal-acc})+(\ref{state:AQA:ideal-rej})$ is upper bounded by two times (\ref{state:AQA:difference}), which is
\begin{align*}
    & \sum_{x \neq 0} \E_{r,P_R} 2\Pr\left[{{x \in \Range(L_F)}}\right]\; \tr \left( \mixednospace{\sum_{x(Q)=x} Q \ket{r} \ot A'_Q \left(P_{R}\big(\rho\big), \rho_{\aux} \right) } \right)\\
    \le& \negl(\numtraps) \text{ } \sum_{x \neq 0} \E_{s, P_{R}} \tr \left( \mixednospace{\sum_{x(Q)=x} Q \ket{r} \ot A'_Q \left(P_{R}\big(\rho\big), \rho_{\aux} \right) } \right)\\
    =& \negl(\numtraps) \text{ } \tr \left( \ref{state:AQA:ideal-rej} \right) \;\le\; \negl(\numtraps) \text{ } \tr \left( (\ref{state:AQA:ideal-acc})+(\ref{state:AQA:ideal-rej}) \right).
\end{align*}
The second line follows from the Pauli partitioning by Clifford (lemma $\ref{lemma:clifford}$). The third line holds by the linearity of trace and the fact that (\ref{state:AQA:ideal-acc}), (\ref{state:AQA:ideal-rej}) are orthogonal. This establishes the statistical indistinguishability between $(\ref{state:AQA:real-acc})+(\ref{state:AQA:real-rej})$ and $(\ref{state:AQA:ideal-acc})+(\ref{state:AQA:ideal-rej})$.\\

\noindent \underline{Step 4}. Finally, we construct $\widetilde{\Setup}, \widetilde{\Enc}, \widetilde{\Audit}$ and prove receiver security.
\\

\noindent
\fbox{\parbox{\textwidth}{
    $\widetilde{\Setup}(1^\secparam, 1^\ell)$:\\
    1. Sample a random string $\sk \gets \bbZ_p^{2\ell+3\lambda}$.\\
    2. Output $(\sk, \phi_S, \sigma)$, where $(\onreg{\sigma}{\hat{M}}, \onreg{\phi_S}{\hat{S},\hat{C}})$ is the result of applying $\onreg{\TP^\dag}{\hat{M},\hat{S}}$ to $\onreg{\ket{\sk}}{\hat{M},\hat{S},\hat{C}}$.
}}
\fbox{\parbox{0.49\textwidth}{
    $\widetilde{\Enc}(\rho)$:\\
    1. Sample a random Pauli $\dk \from \PauliGroup_{\ell}$.\\
    2. Output $(\dk, \sigma)$ where $\sigma = \dk (\rho) \dk^\dag$.
}}
\fbox{\parbox{0.49\textwidth}{
    $\widetilde{\Audit}(\sk,\pf)$:\\
    1. Parse $\pf = (\hat{P}, \hat{c})$.\\
    2. Output $\bot$ if $(z(\hat{P}),x(\hat{P}),\hat{c})\neq \sk$.
}}
\vspace{2mm}

It is direct to see that $\Recv \circ \widetilde{\Enc} = \identity$. We now analyze the state $(\dk, \phi_A)$ that results from executing $(\widetilde{\Setup}, \widetilde{\Enc}, \adv)$ in a row and replacing $\dk \from \bot$ if $\bot \from \widetilde{\Audit}$. The execution of $\widetilde{\Setup}$ and $\widetilde{\Enc}$ yields random classical keys $\sk=r,\; \dk=P$ and quantum state
$$
    \of{\onreg{\sigma}{\hat{M}}, \onreg{\phi_S}{\hat{S},\hat{C}},\onreg{\phi_R}{\hat{R}}}
    = \mixed{\of{ \onreg{\TP^\dag}{\hat{M},\hat{S}} \onreg{\ket{r}}{\hat{M},\hat{S},\hat{C}}, \onreg{P \rho}{\hat{R}} }}
$$
After the execution of $\adv$, the joint state is
\begin{align*}
    &\E_{r, P} \mixed{A \of{ \onreg{\TP^\dag}{\hat{M},\hat{S}} \onreg{\ket{r}}{\hat{M},\hat{S},\hat{C}}, \onreg{P \rho}{\hat{R}}, \onreg{\rho_{\aux}}{\hat{W}} } \ot \ket{r} \ot \ket{P} }\\
    =&\E_{r, P} \mixed{A' \of{ \onreg{\ket{r}}{\hat{M},\hat{S},\hat{C}}, \onreg{P \rho}{\hat{R}}, \onreg{\rho_{\aux}}{\hat{W}} } \ot \ket{r} \ot \ket{P} }
\end{align*}
The next step is to apply $\widetilde{\Audit}$, which corresponds to the projection $\onreg{\ketbra{r}}{\hat{M},\hat{S},\hat{C}}$. The state conditioned that $\widetilde{\Audit}$ accepts can be analyzed through Pauli twirl with target measurement result $0$ (lemma \ref{lemma:qotp-measure}), which yields exactly (\ref{state:AQA:real-acc}). Similarly, the state conditioned that $\widetilde{\Audit}$ rejects is exactly (\ref{state:AQA:real-rej}). Hence, $(\dk,\phi_A)$ generated from $(\Setup, \Enc, \adv, \Audit)$ and from $(\widetilde{\Setup}, \widetilde{\Enc}, \adv, \widetilde{\Audit})$ are indistinguishable, which establishes receiver security.

\end{proof}

\section{MPQC-PVIA with Trusted Setup}
\label{sec:MPQC-PVIA}

Here, we present our MPQC protocol with a trusted setup. We make use of the Clifford-form $\AQA$ developed in the previous section together with $\cMPC$ to achieve PVIA security. For simplicity, we work with qubits \ie $p=2$ here. The protocol is divided into two parts: \begin{enumerate}
    \item An offline setup: a setup $\Sigma^\PVIA$ prepares EPR pairs $(\EPRS_i,\EPRR_i)$ of length $\ell_i$ and distributes the sending side $\EPRS_i$ to party $\Pc_i$. Next, the trusted setup encodes all of the receiving sides $\EPRR_1,\cdots,\EPRR_n$ and the ancilla $\phi_{\anc}$ into a single ciphertext $\sigma$, which is sent to the server. This server can be any participant in the MPQC game; for simplicity, we can assume it's the first party.
    The above quantum states will later be utilized in the input encoding stage. The trusted setup then executes $\AQA.\Setup$ to obtain portals for the output delivery stage. Finally, the trusted setup transmits information related to the secret keys to $\cMPC$.
    \item  An online phase: every party acts as a client who teleports their input to the server. Directed by $\cMPC$, the server evaluates the circuit on the ciphertext $\sigma$. Finally, the server returns the outputs to all clients using AQA.
\end{enumerate}  


\begin{protocol}{$(\Sigma^\PVIA, \Pi^{\PVIA})$ for MPQC-PVIA with Trusted Setup}\label{proto:pvia}{\bf Common Input:} A quantum circuit $\circuit$ in the format of \cref{spec:circuit}.\vspace{3mm}

{\bf Trusted Setup $\Sigma^\PVIA$:}
\begin{enumerate}
    \setlength{\itemsep}{1pt}
    \item Prepare EPR pairs $(\onreg{\EPRS_i}{S_i}, \onreg{\EPRR_i}{R_i})$ with $|\gray{S_i}| = |\gray{R_i}| = \ell_i$.
    \item Initialize the ancilla register $\gray{N} = (\gray{N_1},\cdots,\gray{N_d})$ as $\phi_{\anc}$.
    \item Initialize the trap register $\gray{T} = (\gray{T_1},\cdots,\gray{T_{n+d}})$ as $\ket{0}^{\ot (n+d)\numtraps}$.
    \item Sample $(\sk_i, \phi_{S,i}, \phi_{R,i}) \from \AQA.\Setup(1^{\numtraps}, 1^{\ell_i})$.
    \item Sample random Clifford $E \randsample \CliffordGroup_{\ell_{\total}+(n+d)\numtraps}$ and apply $\onreg{E}{R_1,\cdots,R_n,N,T}$ with result $\sigma$.
    \item Send $(\onreg{\sigma}{\gray{R_1},\cdots,\gray{R_n},\gray{N},\gray{T}}, \phi_{S,1}, \cdots, \phi_{S,n})$ to server and send $(\onreg{\EPRS_i}{S_i}, \phi_{R,i})$ to client $i$.
    \item Send the secrets $(E, \sk_1, \cdots, \sk_n)$ to $\cMPC$.
\end{enumerate}
{\bf Online Input:} Client $i$ receives $\rho_i \in \Den{\ell_i}$.\vspace{3mm}

{\bf Protocol $\Pi^\PVIA$:}
\vspace{2mm}\par
    \quad \underline{Input Encoding:}
    \begin{enumerate}
    \setlength{\itemsep}{1pt}
        \item Client $i$ teleports $\rho_i$ via $\onreg{\EPRS_i}{S_i}$ and sends the teleportation result $P_i$ to $\cMPC$.
        \item $\cMPC$ sets the key $E_d = \onreg{E}{R_1,\cdots,R_n,N,T} \onreg{P_1}{R_1} \cdots \onreg{P_n}{R_n}$.
    \end{enumerate}
    \quad \underline{Computation:}
    \begin{enumerate}
    \setlength{\itemsep}{1pt}
    \setcounter{enumi}{2}
        \item For $k=d,\cdots,1$:
        \vspace{-2mm}
        \begin{enumerate}
        \setlength{\itemsep}{1pt}
            \item $\cMPC$ sends the gate $V_k = (\onreg{E'_{k-1}}{{R_{[n]},N_{[k-1]},T_{[n+k-1]}}} \ot \onreg{P'_k  \CX_{c_k}}{N_k,T_{n+k}}) \onreg{E^\dag_k}{{R_{[n]},N_{[k]},T_{[n+k]}}}$ to the server using a random Clifford $E'_{k-1}\from \CliffordGroup_{\Sigma_i \ell_i + (k-1) +(n+k-1)\numtraps}$, a random Pauli $P'_k\from \PauliGroup^*_{1+\numtraps}$, and a random string $c_k\from \zo^{\numtraps}$.
            \item Server applies $V_k$ to registers $(\gray{R_{[n]}},\gray{N_{[k]}},\gray{T_{[n+k]}})$, measures $(\gray{N_{k}},\gray{T_{n+k}})$ in the computational basis and sends the measurement outcome $r_k \in \zo^{1+\numtraps}$ to $\cMPC$.
            \item $\cMPC$ sets $b_k\in\zo$ as the solution to $r_k \oplus x(P'_k) = b_k (1, c_k)$ if there is a solution.\\ Otherwise, $\cMPC$ publicly outputs the server as malicious and aborts.
            \item $\cMPC$ computes $\onreg{G_{k-1}}{R_{[n]},N_{[k-1]}} = f_k(b_k,\cdots,b_d)$ and sets the key $E_{k-1} = E'_{k-1} G_{k-1}^\dag$.
        \end{enumerate}
    \end{enumerate}
    \quad \underline{Output Delivery:}
    \begin{enumerate}
    \setlength{\itemsep}{1pt}
    \setcounter{enumi}{3}
        \item $\cMPC$ sends server $V' = \onreg{\AQA.\EncGate(\sk_1)}{R_1, T_1} \cdots \onreg{\AQA.\EncGate(\sk_n)}{R_n, T_n} \onreg{E_0^\dag}{R_1, \cdots, R_n, T_1, \cdots, T_n}$.
        \item Server applies $\onreg{V'}{R_1, \cdots, R_n, T_1, \cdots, T_n}$ and obtains $(\onreg{\hat{\sigma}_1}{\hat{M}_1}, \cdots, \onreg{\hat{\sigma}_n}{\hat{M}_n})$ where  $\gray{\hat{M}_i} := (\gray{R_i},\gray{T_i})$.
        \item Server computes $\pf_i \from \AQA.\Send(\hat{\sigma}_i, \phi_{S,i})$ and sends the result to $\cMPC$.
        \item $\cMPC$ computes $\dk_i \from \AQA.\Audit(\sk_i, \pf_i)$. If $\dk_i = \bot$, $\cMPC$ publicly outputs the server as malicious and aborts. Otherwise, $\cMPC$ sends $\dk_i$ to client $i$.
        \item Client $i$ outputs $\rho'_i \from \AQA.\Recv(\dk_i, \phi_{R,i})$.
    \end{enumerate}
\end{protocol}

\subsection{Security}
\begin{thm}\label{thm:pvia}
    $(\Sigma^{\PVIA}, \Pi^{\PVIA})$ is a multi-party quantum computation \pvia with trusted setup in the MPC-hybrid model as defined in \cref{dfn:mpqcpvia}.
    \ie For every non-uniform (\qpt) adversary $\adv$ corrupting party set $I$ with $|I| \le n-1$, there is a non-uniform (\qpt) adversary $\Sim_\adv$ corrupting $I$, such that for any (possibly entangled) states $\rho_1,\cdots \rho_n, \rho_{\aux}$,
\begin{align*}
    \{\real^{\Pi^{\PVIA} \circ \Sigma^{\PVIA}}_{\adv(\rho_{\aux})}(1^{\secparam}, \circuit, \rho_1,\cdots,\rho_n) \} \approx \{\ideal^{\MPQC}_{\Sim_{\adv}(\rho_{\aux})}(1^{\secparam}, 0, \circuit, \rho_1,\cdots,\rho_n) \}
\end{align*}
\end{thm}

\begin{proof}
Consider the following hybrid worlds modified from the real protocol gradually. We describe each hybrid in terms of the changes made to the previous hybrid.

\begin{itemize}[leftmargin=*]
\item $\hybrid_1$: Introduce a trusted party $\trustp^{\hyb}$ who executes both the setup $\Sigma^\PVIA$ and $\cMPC$.

\item $\hybrid_2$:
\begin{itemize}
    \item $\Sigma^\PVIA$ step $1$: $\trustp^{\hyb}$ prepares EPR pairs on registers $(\gray{S_i}, \gray{\tilde{R}_i}), (\gray{\tilde{S}_i}, \gray{R_i})$ with $|\gray{S_i}|=|\gray{\tilde{R}_i}|=|\gray{\tilde{S}_i}|=|\gray{R_i}|=\ell_i$ and keeps $\{(\gray{\tilde{R}_i}, \gray{\tilde{S}_i})\}$. Let $\tilde{\EPRR}_i$ be the content of $\gray{\tilde{R}_i}$.
    \item $\Sigma^\PVIA$ step $2$: $\trustp^{\hyb}$ prepares EPR pairs on $(\gray{D},\gray{N})$ of length $d$ and keeps $\gray{D}$.
    \item $\Pi^\PVIA$ step $2$: $\trustp^{\hyb}$ extracts the input $\onreg{\tilde{\rho}_i}{\tilde{R}_i} = P_i^\dag \onreg{\tilde{\EPRR}_i}{\tilde{R}_i}$ and prepares ancilla $\onreg{\phi_{\anc}}{\tilde{N}}$. Next, it teleports $\onreg{\tilde{\rho}_i}{\tilde{R}_i}, \onreg{\phi_{\anc}}{\tilde{N}}$ back to the server via registers $\gray{\tilde{S}_i}, \gray{D}$ respectively. If the teleportation results are $\tilde{P}_i, \tilde{P}_{\anc}$, it sets $E_d = \onreg{E}{R_1,\cdots,R_n,N,T} \onreg{\tilde{P}_1}{R_1} \cdots \onreg{\tilde{P}_n}{R_n} \onreg{\tilde{P}_{\anc}}{N}$.
\end{itemize}

\item $\hybrid_{3+d-k}$, where $k = d,\cdots,1$:
\begin{itemize}
    \item $\Sigma^\PVIA$ step $5$: For $h=d,\cdots,k$, $\trustp^{\hyb}$ resets $(\gray{N_{h}},\gray{T_{n+h}})$ as random string $\tilde{r}_{h}\randsample \zo^{\numtraps+1}$,
    samples random Clifford $V_h \randsample \CliffordGroup_{\ell^{\total}+(n+h)\numtraps}, \tilde{E}_{k-1} \randsample \CliffordGroup_{\ell^{\total}+(n+k-1)\numtraps}$ and sets $E = \onreg{V^\dag_d}{R_{[n]}, N_{[d]}, T_{[n+d]}} \cdots \onreg{V^\dag_k}{R_{[n]}, N_{[k]}, T_{[n+k]}} \onreg{\tilde{E}_{k-1}}{R_{[n]}, N_{[k-1]}, T_{[n+k-1]}}$.
    \item $\Pi^\PVIA$ step $2$: $\trustp^{\hyb}$ computes $\circuit[k] \cdots \circuit[d]\left(\onreg{\tilde{\rho}_1}{\tilde{R}_1}, \cdots, \onreg{\tilde{\rho}_n}{\tilde{R}_n}, \onreg{\phi_{\anc}}{\tilde{N}}\right)$. Next, it teleports the states in registers $\gray{\tilde{R}_{i}}$ and $\gray{\tilde{N}'_{[k-1]}}$ back to the server. If the teleportation results are $\tilde{P}_{i},\tilde{P}_{{\anc}}$, it sets $E_{k-1} = \onreg{\tilde{E}_{k-1}}{R_1,\cdots,R_n,,N_{[k-1]},T_{[n+k-1]}} \onreg{\tilde{P}_1}{R_1}\cdots \onreg{\tilde{P}_n}{R_n} \onreg{\tilde{P}_{{\anc}}}{N_{[k-1]}}$.
    \item $\Pi^\PVIA$ step $3$ iteration $h$ for $h = d,\cdots, k$: $\trustp^{\hyb}$ sends $V_h$ to the server and receives $r_h$ in return. If $r_h \neq \tilde{r}_h$, $\trustp^{\hyb}$ publicly outputs the server as malicious and aborts.
\end{itemize}

\item $\hybrid_{3+d}$: 
\begin{itemize}
    \item $\Sigma^\PVIA$ step $4$: $\trustp^{\hyb}$ samples $(\sk_i, \phi_{S,i}, \sigma_i) \from \AQA.\widetilde{\Setup}(1^\secparam, 1^{\ell_i})$, uses the content $\EPRR_i$ of register $\gray{R_i}$ as $\phi_{R,i}$, and reassigns register $(\gray{R_i}, \gray{T_i})$ as $\sigma_i$.
    \item $\Pi^\PVIA$ step $2$: $\trustp^{\hyb}$ computes the circuit $\circuit$
    with input $(\onreg{\tilde{\rho}_1}{\tilde{R}_1}, \cdots \onreg{\tilde{\rho}_n}{\tilde{R}_n})$, teleports the output via register $\gray{\tilde{S}_i}$, and sets $\dk_i$ as the teleportation result $\tilde{P}_i$.
    \item $\Pi^\PVIA$ step $5$: $\trustp^{\hyb}$ sets $V' = \tilde{E}_0^\dag$.
    \item $\Pi^\PVIA$ step $7$: $\trustp^{\hyb}$ resets $\dk_i = \bot $ if $ \bot \from \AQA.\widetilde{\Audit}(\sk_i, \pf_i)$.
\end{itemize}

\end{itemize}

The last hybrid world is equivalent to the ideal world with the following simulator.

\begin{simulator}{$\Sim^{\PVIA}_{\adv(\rho_{\aux})}$ for MPQC-PVIA with Trusted Setup}\label{sim:pvia}
\begin{enumerate}
    \setlength{\itemsep}{1pt}
    \item Fake setup (if client $i$ is corrupted):\begin{enumerate}
        \item Prepare EPR pairs on registers $(\gray{S_i}, \gray{\tilde{R}_i})$ with $|\gray{S_i}| = |\gray{\tilde{R}_i}| = \ell_i$.
        \item Prepare EPR pairs on registers $({\gray{\tilde{S}_i}}, {\gray{\hat{R}_i}})$ with  $|\gray{\tilde{S}_i}| = |\gray{\hat{R}_i}| = \ell_i$.
        \item Send $(\gray{S_i}, \gray{\hat{R}_i})$ to $\adv$.
    \end{enumerate}
    \item Fake setup (if server is corrupted):
    \begin{enumerate}
        \item Sample $(\sk_i, \phi_{S,i}, \onreg{\sigma_i}{R_i,T_i}) \from \AQA.\widetilde{\Setup}(1^\secparam, 1^{\ell_i})$ for $i \in [n]$.
        \item Initialize registers $(\gray{N_k},\gray{T_{n+k}})$ with random $\tilde{r}_k \randsample \zo^{\numtraps+1}$ for $k \in [d]$.
        \item Sample random Clifford gates $V_k \randsample \CliffordGroup_{\ell^{\total}+(n+k)\numtraps}$ for $k = 0,1,\cdots,d$.
        \item Apply $\onreg{V_d^\dag}{R_{[n]},N_{[d]},T_{[n+d]}} \cdots \onreg{V_1^\dag}{R_{[n]},N_{[1]},T_{[n+1]}} \onreg{V_0^\dag}{{R_{[n]},T_{[n]}}}$ with resulting state $\sigma$.
        \item Send $(\onreg{\sigma}{R_1,\cdots,R_n,N,T},\phi_{S,1},\cdots,\phi_{S,n})$ to $\adv$.
    \end{enumerate}
    \item Input extraction: Receive $P_i$ from $\adv$ and extract the input $\onreg{\tilde{\rho_i}}{\tilde{R}_i}=\onreg{P_i^\dag \tilde{\EPRR}_i}{\tilde{R}_i}$.
    \item Invoke the ideal functionality: Send $\tilde{\rho_i}$ to $\trustp$ and receive the output $\tilde{\rho}'_i$ from $\trustp$.
    \item Check the abort decision (if server is corrupted): \begin{enumerate}
        \item For $k=d,\cdots,1$, send $V_k$ to $\adv$ and receive $r_k$ in return.\\
        If $r_k \neq \tilde{r}_k$, send $\abort$ to $\trustp$ in the name of the server.
        \item Send $V_0$ to $\adv$.
        \item For $i \in [n]$, receive $\pf_i$ from $\adv$.\\
        If $\bot \from \AQA.\widetilde{\Audit}(\sk_i,\pf_i)$, send $\abort$ to $\trustp$ in the name of the server.
        \end{enumerate}
    \item Output delivery: Teleport $\tilde{\rho}'_i$ via $\gray{\tilde{S_i}}$, obtain its result $\tilde{P}_i$ and send $\dk_i = \tilde{P}_i$ to $\adv$ if the server did not send $\abort$ in the previous step.
    \item Output $\adv$'s output.
\end{enumerate}
\end{simulator}

To prove \cref{thm:pvia}, it suffices to show indistinguishability between consecutive hybrids. When comparing consecutive hybrids, we will view the execution of the adversary and the protocol until a certain step as a quantum operation, which is a sum of CP maps of the form $\tau \mapsto A \tau A^\dag$. Then, it suffices to show indistinguishability under every such CP map because indistinguishability of subnormalized states extend under addition. 

Note that the indistinguishability holds whenever the adversary aborts the classical MPC. The reason is that the abort prevents $\cMPC$ from producing information related to the Pauli or Clifford keys, so the quantum states are all maximally mixed according to Pauli twirl and Clifford twirl (lemma \ref{lemma:qotp-security}). Hence, we will assume that the adversary does not abort $\cMPC$ from now on.
Below, we sometimes abbreviate parts of the state that are currently irrelevant with dots and omit writing the expectation over currently irrelevant keys.

\begin{itemize}[leftmargin=*] \label{hybrid:analysis}
    \item $\real = \hybrid_1$: This is because $\Sigma^{\PVIA}, \cMPC$ and $\trustp^{\hyb}$ are all trusted executions.
    \item $\hybrid_1 = \hybrid_2$: Consider executing the protocol until the end of step $1$. Suppose the CP map $A$ projects  $(\gray{M_{[n]}},\gray{S_{[n]}})$ to $\ket{P_{[n]}}$ in step $1$. The state of $\hybrid_1$ after protocol step $2$ is
    \begin{align*}
        & \E_{E \from \CliffordGroup}\mixed{ A\left( E\left(\onreg{\EPRR_{[n]}}{R_{[n]}},\onreg{\phi_{\anc}}{N},\onreg{\ket{0}^{\ot (n+d)\numtraps}}{T}\right),\cdots\right) \ot \ket{E_d}} \\
        =& \E_{E_d \from \CliffordGroup}\mixed{ A\left( E_d\left(\onreg{P_{[n]}^\dag \EPRR_{[n]}}{R_{[n]}},\onreg{\phi_{\anc}}{N},\onreg{\ket{0}^{\ot (n+d)\numtraps}}{T}\right),\cdots\right) \ot \ket{E_d}}
    \end{align*}
    In $\hybrid_2$, we can change the order of $A$ and the teleportation made by $\trustp^{\hyb}$ because these operators act on disjoint registers. By quantum teleportation, the state of $\hybrid_2$ after protocol step $2$ is as follows, which is the same as in $\hybrid_1$.
    \begin{align*}
        & \E_{E \from \CliffordGroup}\mixed{ A\left( E\left(\onreg{\tilde{P}_{[n]} P_{[n]}^\dag \tilde{\EPRR}_{[n]}}{R_{[n]}}, \onreg{ \tilde{P}_{\anc} \phi_{\anc}}{N},\onreg{\ket{0}^{\ot (n+d)\numtraps}}{T}\right),\cdots\right) \ot \ket{E_d}} \\
        =& \E_{E_d \from \CliffordGroup}\mixed{ A\left( E_d\left(\onreg{P_{[n]}^\dag \tilde{\EPRR}_{[n]}}{R_{[n]}},\onreg{\phi_{\anc}}{N},\onreg{\ket{0}^{\ot (n+d)\numtraps}}{T}\right),\cdots\right) \ot \ket{E_d}}
    \end{align*}
    
    \item $\hybrid_{2+d-k} \approx \hybrid_{3+d-k}$ for $k=d,\cdots,1$:
    Consider executing the protocol until the end of step $3$ iteration $k\text{+}1$. Suppose the CP map $A$ projects  $(\gray{M_{[n]}},\gray{S_{[n]}})$ to $\ket{P_{[n]}}$ in step $1$ and projects $\gray{N_d},\cdots,\gray{N_{k+1}}$ to $b_d,\cdots,b_{k+1}$ in step $3$ for iterations $d$ to $k+1$. Let us denote the partial computation result $C[k+1]\cdots C[d] \big(P_{[n]}^\dag \tilde{\EPRR}_{[n]}, \phi_{\anc}\big)$ as $\onreg{\tilde{\rho}_{(k)}{}}{\tilde{R}_{[n]},\tilde{N}_{[k]}} = \sum_{b}\onreg{\tilde{\rho}_{(k,b)}{}}{\tilde{R}_{[n]},\tilde{N}_{[k-1]}}\ot \onreg{\ket{b}}{\tilde{N}_{k}}$. In $\hybrid_{2+d-k}$, the state at protocol step $3(a)$ for iteration $k$ is 
    {\small
    \begin{align*}
        & \E_{E_k, E'_{k\text{-}1}, P'_k, c_k} \mixed{A\left( E_k\left(\onreg{\tilde{\rho}_{(k)}{}}{R_{[n]},N_{[k]}}, \onreg{\ket{0}^{\ot (n+k)\numtraps}}{T_{[n+k]}} \right),\cdots\right)\ot \ket{P'_k,c_k}}\\
        =& \E_{V_k, E'_{k\text{-}1}, P'_k, c_k} \mixed{A\left( V_k^\dag\left( \sum_b E'_{k-1}\big(\tilde{\rho}_{(k,b){}}, \ket{0}^{\ot (n+k-1)\numtraps}\big)\ot P'_k \CX_{c_k}\big(\ket{b}, \ket{0}^{\ot \numtraps}\big)\right),\cdots\right) \ot \ket{P'_k,c_k}}
    \end{align*}
    }The equality follows from the definition of $V_k$. Since the protocol gives the attacker access to $\ket{V_k}$ in step $3(a)$, we can merge $V_k^\dag$ and $A$ into a CP map $A'$ that operates also on $\ket{V_k}$. Also, we can simplify $\CX_{c_k}(b, \ket{0}^{\ot \numtraps}) = \ket{b(1,c_k)}$. Thus, the state is eqaul to
    {\small
    \begin{align*}
        \E_{E'_{k\text{-}1}, P'_k, c_k} \mixed{A'\left( \sum_b E'_{k-1}\big(\tilde{\rho}_{(k,b)}, \ket{0}^{\ot (n+k-1)\numtraps}\big)\ot \onreg{P'_k \ket{b(1,c_k)}}{N_k,T_{n+k}},\cdots\right) \ot \ket{P'_k,c_k}}
    \end{align*}
    }Afterwards, protocol step $3(c)$ projects registers $(\gray{N_k}, \gray{T_{n+k}})$ to $\ket{x(P'_k) + b_k(1,c_k)}$ for $b_k \in \zo$. The two cases are similar, and we demonstrate using the case $b_k = 0$. To analyze, we apply Pauli twirl with target measurement result $(0,0^{\numtraps})$ (lemma \ref{lemma:qotp-measure}) and use the Pauli decomposition
    $A' = \sum_{Q \in \PauliGroup^*_{\numtraps+1}} \onreg{Q}{N_{k},T_{n+k}} \ot A'_Q$. The state after the projection of step $3(c)$ for obtaining solution $b_k=0$ is
    {\small
    \begin{align*}
        & \E_{E'_{k\text{-}1}, P'_k, c_k} \mixed{\onreg{\ketbra{x(P'_k)}}{N_k, T_{n+k}} A'\left( \sum_b \onreg{P'_k \ket{b(1,c_k)}}{N_k,T_{n+k}} \ot E'_{k-1}\big(\tilde{\rho}_{(k,b)}, \ket{0}^{\ot (n+k-1)\numtraps}\big) ,\cdots\right)}\\
        =& \E_{E'_{k\text{-}1},\tilde{r}_k, c_k} \biggl( \mixed{\sum_{x(Q)=(0,0^\numtraps)} Q \ket{\tilde{r}_k} \ot A'_Q \left(E'_{k-1}\big(\tilde{\rho}_{(k,0)}, \ket{0}^{\ot (n+k-1)\numtraps}\big),\cdots\right)} \\
        & \qquad\qquad + \mixed{\sum_{x(Q)=(1,c_k)} Q \ket{\tilde{r}_k} \ot A'_Q \left( E'_{k-1}\big(\tilde{\rho}_{(k,1)}, \ket{0}^{\ot (n+k-1)\numtraps}\big),\cdots\right)} \biggl)\\
        \approx&\text{ } \E_{E'_{k-1}, \tilde{r}_k} \mixed{\sum_{x(Q)=(0,0^\numtraps)} Q \ket{\tilde{r}_k} \ot A'_Q \left(E'_{k-1}\big(\tilde{\rho}_{(k,0)}, \ket{0}^{\ot (n+k-1)\numtraps}\big),\cdots\right)}
    \end{align*}
        
    }The indistinguishability follows because each $c_k$ occurs with negligible probability. After setting the key $E_{k-1} = E'_{k-1} G_{k-1}^\dag$ in protocol step $3(k)$, the state in $\hybrid_{2+d-k}$ becomes
    {\small 
    \begin{align*}
        & \E_{E_{k-1}, \tilde{r}_k} \mixed{\sum_{x(Q)=(0,0^\numtraps)} Q\left(\ket{\tilde{r}_k}\right) \ot  A'_Q\left(E_{k-1}\left(G_{k-1}\big(\tilde{\rho}_{(k,0)}\big), \ket{0}^{\ot (n+k-1)\numtraps}\right),\cdots\right)}\\
        =& \E_{E_{k-1}, \tilde{r}_k} \mixed{\sum_{x(Q)=(0,0^\numtraps)} Q\left(\ket{\tilde{r}_k}\right) \ot  A'_Q\left(E_{k-1}\left(\tilde{\rho}_{(k-1)}, \ket{0}^{\ot (n+k-1)\numtraps}\right),\cdots\right)}
    \end{align*}
    }In $\hybrid_{3+d-k}$, the state of acceptance after projecting to $b_k=0$ is
    {\small
    \begin{align*}
    &\E_{E_{k-1}, \tilde{r}_k} \mixed{\onreg{\ketbra{\tilde{r}_k}}{N_{k}, T_{n+k}} A'\left(\onreg{\ket{\tilde{r}_k}}{N_k, T_{n+k}}, E_{k-1}\left(\tilde{\rho}_{(k-1)}, \ket{0}^{\ot (n+k-1)\numtraps}\right),\cdots\right)}\\
    =& \E_{E_{k-1}, \tilde{r}_k} \mixed{\sum_Q \ketbra{\tilde{r}_k} Q \ket{\tilde{r}_k} \ot A'_Q \left(E_{k-1}\left(\tilde{\rho}_{(k-1)}, \ket{0}^{\ot (n+k-1)\numtraps}\right),\cdots\right)}\\
    =& \E_{E_{k-1}, \tilde{r}_k} \mixed{ \sum_{x(Q)=(0,0^\numtraps)} Q\left(\ket{\tilde{r}_k}\right) \ot A'_Q \left(E_{k-1}\left(\tilde{\rho}_{(k-1)}, \ket{0}^{\ot (n+k-1)\numtraps}\right),\cdots\right)}
    \end{align*}
    }Thus, the states in $\hybrid_{2+d-k}$ and $\hybrid_{3+d-k}$ are indistinguishable following the projections for both $b_k \in \zo$. When both projections fail, $\trustp^{\hyb}$ aborts in both $\hybrid_{2+d-k}$ and $\hybrid_{3+d-k}$ by publicly outputting the server $\Pc_1$ as malicious and will not use the Clifford keys anymore. By Clifford twirl (lemma \ref{lemma:qotp-security}), the states in both $\hybrid_{2+d-k}$ and $\hybrid_{3+d-k}$ will be indistinguishable to 
    {\small
    \begin{align*}
        & \E_{\tilde{r}_k, \tilde{r}_k^*} \mixed{ \sum_{x(Q)\neq(0,0^\numtraps)} \onreg{Q\left(\ket{\tilde{r}_k}\right)}{N_k,T_{n+k}} \ot A'_Q \left(\onreg{\ket{\tilde{r}^*}}{N_{[k-1]}, T_{[n+k-1]}},\cdots\right) \ot \ket{\bot_{\Pc_1}}}
    \end{align*}}

    \item $\hybrid_{2+d} \approx \hybrid_{3+d}$: Consider executing the protocol until the end of step $6$. Let $(\tilde{\rho}'_1,\cdots, \tilde{\rho}'_n)$ be the quantum output of $\circuit$. The state of $\hybrid_{2+d}$ at step $6$ is
    \begin{align*}
        & \E_{E_0,\sk_{i}} \mixed{A \of{ \onreg{E_0}{R_1,\cdots,R_n,T_1,\cdots,T_n} \of{ \onreg{\tilde{\rho}'_{i}{}}{R_{i}}, \onreg{\ket{0}^{\ot \numtraps}}{T_i} } , \ket{V'},\phi_{S,i}, \phi_{R,i}, \rho_{\aux} } \ot \ket{\sk_i}}\\
        =& \E_{V', \sk_i} \mixed{A \of{ \onreg{V'^{\dag}}{R_1,\cdots,R_n,T_1,\cdots,T_n} \of{ \onreg{\AQA.\EncGate(\sk_i) \of{\tilde{\rho}'_{i}, \ket{0}^{\ot \numtraps}}}{R_i, T_i} } , \ket{V'}, \phi_{S,i}, \phi_{R,i}, \rho_{\aux} } \ot \ket{\sk_i} }\\
        =& \E_{V', \sk_i} \mixed{A \of{ \onreg{V'^{\dag}}{R_1,\cdots,R_n,T_1,\cdots,T_n} \of{ \onreg{\AQA.\Enc \of{\sk_i, \tilde{\rho}'_{i}}}{R_i,T_i} } , \ket{V'}, \phi_{S,i}, \phi_{R,i}, \rho_{\aux} } \ot \ket{\sk_i} }
    \end{align*}
    where the first equality follows from the definition of $V'$ in $\hybrid_{2+d}$ and the second follows from the definition of $\AQA.\Enc$. After protocol step $7$, the state of $\hybrid_{2+d}$ can be generated under the process $(\AQA.\Setup, \AQA.\Enc, A', \Audit)$ where $A'$ is a CP map that merges $A$ and $V'^\dag$. On the other hand, the state of $\hybrid_{3+d}$ at step $6$ is 
    \begin{align*}
        & \E_{\tilde{E}_0,\sk_{i}, \tilde{P}_i} \mixed{A \of{ \onreg{\tilde{E}_0}{R_1,\cdots,R_n,T_1,\cdots,T_n} \of{ \onreg{\sigma_i}{R_i,T_i} } , \ket{V'}, \phi_{S,i}, \tilde{P}_i(\tilde{\rho}'_i), \rho_{\aux} } \ot \ket{\tilde{P}_i} \ot \ket{\sk_i}}\\
        =& \E_{V',\sk_{i}, \dk_i} \mixed{A \of{ \onreg{V'^\dag}{R_1,\cdots,R_n,T_1,\cdots,T_n} \of{ \onreg{\sigma_i}{R_i,T_i} } , \ket{V'}, \phi_{S,i}, \dk_i(\tilde{\rho}'_i), \rho_{\aux} } \ot \ket{\dk_i} \ot \ket{\sk_i}}
    \end{align*}
    where the equality follows from the definition of $V'$ and $\dk_i$ in $\hybrid_{3+d}$. Here, $\sk_i, \phi_{S,i}, \sigma_i$ are generated from $\AQA.\widetilde{\Setup}$, whereas $\dk_i$ and the encoding of $\tilde{\rho}'_i$ can be generated from $\AQA.\widetilde{\Enc}$. After protocol step $7$, the state of $\hybrid_{3+d}$ can be generated under the process $(\AQA.\widetilde{\Setup}, \AQA.\widetilde{\Enc}, A', \widetilde{\Audit})$. By the receiver security of $\AQA$ (\cref{thm:aqa}), these two states are statistically indistinguishable.
    
    \item $\hybrid_{3+d} = \ideal$: Except for the computation of $\circuit$, the execution of $\trustp^{\hyb}$ in $\hybrid_{3+d}$ consists of $n$ independent parts, each interacting with only one party. Thus, we can collect the parts that interact with the corrupted parties as a simulator $\Sim$, who also interacts with the ideal functionality of MPQC to compute $\circuit$. By quantum teleportation and \cref{thm:aqa}, the honest parties and their corresponding parts of $\trustp^{\hyb}$ indeed send the honest inputs to the ideal functionality, never send $\abort$ to the ideal functionality, and always output the honest outputs. Hence, $\hybrid_{3+d}$ is identical to $\ideal$.
\end{itemize}
\vspace{-1em}
\end{proof}

\section{BoBW-MPQC-PVIA with Trusted Setup}
In this section, we construct a best-of-both-worlds multi-party quantum computation protocol secure with publicly verifiable identifiable abort (BoBW-MPQC-PVIA). The protocol is similar to the MPQC-PVIA protocol but with $n$ servers instead. $\Pc_j$ will be assigned as server $j$. The protocol is divided into two parts: \begin{enumerate}
    \item An offline setup: a setup $\Sigma^\BoBW$ prepares EPR pairs $(\EPRS_i,\EPRR_i)$ of length $\ell_i$ and distributes the sending side $\EPRS_i$ to party $\Pc_i$. Next, the setup uses a QECC scheme to encode every $\EPRR_i$ into $\{\EPRR_i^{(j)}\}_{j\in [n]}$ and the ancilla $\phi_{\anc}$ into $\{\phi_{\anc}^{(j)}\}_{j \in [n]}$. Afterwards, the setup encrypts the $j$-th part of the QECC codewords into a Clifford ciphertext and send it to server $j$. The setup then executes $\AQA.\Setup$ to prepare for the output delivery stage. Finally, the setup sends the secret keys to $\cMPC$.
    \item An online phase: every party acts as a client who teleports their input to the servers. Directed by $\cMPC$, the servers evaluate the circuit in a fault-tolerant manner. Each server $j$ operates only on the ciphertext that encrypts the $j$-th part of the QECC codewords. Finally, the servers returns the outputs to the clients using AQA.
\end{enumerate}  

Let $\thres < \frac{n}{2}$ and let QECC be a $[[n,1]]_p$ polynomial code that corrects $\thres$ erasure errors as in \cref{lemma:qecc}. Recall that $\QECC$ stands for the Clifford gate for QECC encoding. We present the protocol formally as follows:

\begin{protocol}{$(\Sigma^{\BoBW}, \Pi^{\BoBW})$ for BoBW-MPQC-PVIA with Trusted Setup}\label{proto:bobw}{\bf Common Input:} A threshold $\thres$ and a quantum circuit $\circuit$ in the format of \cref{spec:circuit}.
\vspace{3mm}

{\bf Trusted Setup $\Sigma^\BoBW$:}
\begin{enumerate}
    \setlength{\itemsep}{1pt}
    \item Prepare EPR pairs $(\onreg{\EPRS_i}{S_i}, {\EPRR_i})$ of length $\ell_i$.
    Compute $(\onreg{\EPRR^{(1)}_{i}}{R_{1,i}},\cdots,\onreg{\EPRR^{(n)}_{i}}{R_{n,i}})\from \QEnc({\EPRR_i})$. 
    \item Initialize the ancilla $(\onreg{\phi^{(1)}_{{\anc}}}{N_{1}},\cdots,\onreg{\phi^{(n)}_{{\anc}}}{N_{n}})\from \QEnc({\phi_{\anc}})$ where $\gray{N_j}=(\gray{N_{j,1}},\cdots,\gray{N_{j,d}})$.
    \item Initialize the trap registers $\gray{T_j} = (\gray{T_{j,1}},\cdots,\gray{T_{j,n+d}})$ as $\ket{0}^{\ot (n+d)\numtraps}$.
    \item Sample $(\sk_{j \to i},\; \phi_{S,{j \to i}},\; \phi_{R,{j \to i}}) \from \AQA.\Setup(1^{\numtraps}, 1^{\ell_i})$.
    \item Sample random Clifford $E_{j} \randsample \CliffordGroup_{\ell_{\total}+(n+d)\numtraps}$ and apply $\onreg{E_j}{R_{j,1},\cdots,R_{j,n},N_{j},T_j}$ with result $\sigma_j$.
    \item Send $(\onreg{\EPRS_i}{S_i}, \onreg{\sigma_i}{\gray{R_{i,1}},\cdots,\gray{R_{i,n}},\gray{N_i},\gray{T_i}}, \phi_{S,i\to 1}, \cdots, \phi_{S,i \to n}, \onreg{\phi_{R,1\to i}}{\hat{R}_{1,i}}, \cdots, \onreg{\phi_{R,n\to i}}{\hat{R}_{n,i}})$ to party $\Pc_i$.
    \item Send the secrets $\{E_j, \sk_{j \to i}\}_{i,j \in [n]}$ and the corruption list $J = \emptyset$ to $\cMPC$.
\end{enumerate}
{\bf Online Input:} Party $\Pc_i$ receives $\onreg{\rho_i}{M_i} \in \Den{\ell_i}$.\vspace{3mm}

{\bf Protocol $\Pi^\BoBW$:}
\vspace{2mm}\par
    \quad \underline{Input Encoding:}  
    \begin{enumerate}
    \setlength{\itemsep}{1pt}
        \item Client $i$ teleports $\rho_i$ via $\onreg{\EPRS_i}{S_i}$ and sends the teleportation result $P_i$ to $\cMPC$.
        \item $\cMPC$ computes the effective teleportation Pauli $P_{j,i} \in \PauliGroup_{\ell_i}$ for each server $j$ where $(P_{1,i} \ot \cdots \ot P_{n,i}) = \QECC\; P_i\; \QECC^\dag$. Accordingly to the Pauli gates $(P_{j,1},\cdots, P_{j,n})$, $\cMPC$ sets the key $E_{j,d} = \onreg{E_j}{R_{j,1},\cdots,R_{j,n},N_j,T_j} \onreg{P_{j,i}}{R_{j,1}} \cdots \onreg{P_{j,n}}{R_{j,n}}$.
    \end{enumerate}
    \quad \underline{Computation:}
    \begin{enumerate}
    \setlength{\itemsep}{1pt}
    \setcounter{enumi}{2}
        \item For $k=d,\cdots,1$ and for each server $j$:
        \begin{enumerate}
        \setlength{\itemsep}{1pt}
            \item $\cMPC$ sends $V_{j,k} = (\onreg{E'_{j,k-1}}{{R_{j,[n]},N_{j,[k-1]},T_{j,[n+k-1]}}} \ot \onreg{P'_{j,k}  \CX_{c_{j,k}}}{N_{j,k},T_{j,n+k}}) \onreg{E^\dag_{j,k}}{{R_{j,[n]},N_{j,[k]},T_{j,[n+k]}}}$ to server $j$ using a random Clifford $E'_{j,k-1}\from \CliffordGroup_{\Sigma_i \ell_i + (k-1) +(n+k-1)\numtraps}$, a random Pauli $P'_{j,k}\from \PauliGroup^*_{1+\numtraps}$, and random $c_{j,k}\from \bbZ_p^{\numtraps}$.
            \item Server $j$ applies $V_{j,k}$ to registers $(\gray{R_{j,[n]}},\gray{N_{j,[k]}},\gray{T_{j,[n+k]}})$, measures $(\gray{N_{j,k}},\gray{T_{j,n+k}})$ in the computational basis and sends the measurement outcome $r_{j,k} \in \bbZ_p^{1+\numtraps}$ to $\cMPC$.
            \item $\cMPC$ sets $b_{j,k}\in \bbZ_p$ as the solution to $r_{j,k} \oplus x(P'_{j,k}) = b_{j,k} (1, c_{j,k})$ if there is a solution. Otherwise, $\cMPC$ adds server $j$ into the corruption list $J$ and will not interact with this server anymore.
            \item If $|J|>\thres$, $\cMPC$ publicly outputs $J$ as malicious and aborts.
            \item $\cMPC$ computes the decoded measurement result $b_k \from \QDec((b_{1,k},\cdots,b_{n,k}),J)$, the next gate $G_{k-1} = f_{k}(b_1,\cdots,b_k)$, and its fault-tolerant version $(G_{k-1}^{(1)},\cdots,G_{k-1}^{(n)})$. Accordingly, $\cMPC$ sets the key $E_{j,k-1} = E'_{j,k-1} G_{k-1}^{(j)\dag}$.
        \end{enumerate}
    \end{enumerate}
    \quad \underline{Output Delivery:}
    \begin{enumerate}
    \setlength{\itemsep}{1pt}
    \setcounter{enumi}{3}
        \item $\cMPC$ sends the gate $V'_j = \onreg{\AQA.\EncGate(\sk_{j \to 1})}{R_{j,1}, T_{j,1}} \cdots \onreg{\AQA.\EncGate(\sk_{j \to n})}{R_{j,n}, T_{j,n}} \onreg{E_{j,0}^\dag}{R_{j,[n]}, T_{j,[n]}}$ to server $j$.
        \item Server $j$ applies $\onreg{V'_j}{R_{j,1}, \cdots, R_{j,n}, T_{j,1}, \cdots, T_{j,n}}$ and obtains $(\onreg{\hat{\sigma}_{j,1}}{\hat{M}_{j,1}}, \cdots, \onreg{\hat{\sigma}_{j,n}}{\hat{M}_{j,n}})$ where $\gray{\hat{M}_{j,i}} := (\gray{R_{j,i}},\gray{T_{j,i}})$.
        \item Server $j$ computes $\pf_{j,i} \from \AQA.\Send(\hat{\sigma}_{j,i}, \phi_{S,j \to i})$ and sends the result to $\cMPC$.
        \item $\cMPC$ computes $\dk_{j,i} \from \AQA.\Audit(\sk_{j \to i}, \pf_{j,i})$. If $\dk_{j,i} = \bot$, $\cMPC$ adds server $j$ into the corruption list $J$. If $|J|>\thres$, $\cMPC$ publicly outputs $J$ as malicious and aborts.
        \item $\cMPC$ sends $\dk_{j,i}$ to client $i$.
        \item Client $i$ computes $\rho'_{j,i} \from \AQA.\Recv(\dk_{j,i}, \phi_{R,j \to i})$ and outputs $\rho'_i \from \QDec((\rho'_{1,i}, \cdots \rho'_{n,i}), J)$.
    \end{enumerate}

\end{protocol}

\subsection{Security}
\begin{theorem}\label{thm:bobw}
    When parameterized by $\thres < \frac{n}{2}$, $(\Sigma^{\BoBW}, \Pi^{\BoBW})$ is a \bobw multi-party quantum computation \pvia of threshold $\thres$ with trusted setup in the MPC-hybrid model as defined in \cref{dfn:mpqcbobw}.
    \ie For every non-uniform (\qpt) adversary $\adv$ corrupting party set $I$ with $|I| \le n-\thres-1$, there is a non-uniform (\qpt) adversary $\Sim_\adv$ corrupting $I$, such that for any (possibly entangled) states $\rho_1,\cdots \rho_n, \rho_{\aux}$,
    \begin{align*}
        \{\real^{\Pi^{\BoBW} \circ \Sigma^{\BoBW}}_{\adv(\rho_{\aux})}(1^{\secparam}, \thres, \circuit, \rho_1,\cdots,\rho_n) \} \approx \{\ideal^{\MPQC}_{\Sim_{\adv}(\rho_{\aux})}(1^{\secparam}, \thres, \circuit, \rho_1,\cdots,\rho_n) \}
    \end{align*}
\end{theorem}

The proof of \cref{thm:bobw} bears resemblance to that of \cref{thm:pvia}, with some additional analysis owing to the use of quantum error correction codes.

\begin{proof}
Consider the following hybrid worlds modified from the real protocol gradually. We describe each hybrid in terms of the changes made to the previous hybrid.

\begin{itemize}[leftmargin=*]
\item $\hybrid_1$: Introduce a trusted party $\trustp^{\hyb}$ who executes both the setup $\Sigma^{\BoBW}$ and $\cMPC$.

\item $\hybrid_2$:
\begin{itemize}
    \item $\Sigma^\BoBW$ step $1$: $\trustp^{\hyb}$ prepares EPR pairs on $(\gray{S_i}, \gray{\tilde{R}_i}), (\gray{\tilde{S}_{j,i}}, \gray{R_{j,i}})$ with $|\gray{S_i}| = |\gray{\tilde{R}_i}| = |\gray{\tilde{S}_{j,i}}| = |\gray{R_{j,i}}| = \ell_i$ and keeps $\{(\gray{\tilde{R}_i}, \gray{\tilde{S}_{j,i}})\}$. Let $\tilde{\EPRR}_i$ be the content of $\gray{\tilde{R}_i}$.
    \item $\Sigma^\BoBW$ step $2$: $\trustp^{\hyb}$ prepares EPR pairs on $(\gray{D_j},\gray{N_j})$ of length $d$ and keeps $\gray{D_j}$.
    \item $\Pi^{\BoBW}$ step $2$: $\trustp^{\hyb}$ extracts the input $\onreg{\tilde{\rho}_i}{\tilde{R}_i} = P_i^\dag \onreg{\tilde{\EPRR}_i}{\tilde{R}_i}$, performs $(\tilde{\rho}^{(1)}_{i}, \cdots \tilde{\rho}^{(n)}_{i}) \from \QEnc({\tilde{\rho}_i})$, and performs $(\phi^{(1)}_{{\anc}}, \cdots \phi^{(n)}_{{\anc}}) \from \QEnc({\phi_{\anc}})$. 
    Next, it teleports ${\tilde{\rho}^{(j)}_{i}}, {\phi^{(j)}_{\anc}}$ back to the server $j$ via registers $\gray{\tilde{S}_{j,i}}, \gray{D_j}$ respectively. If the teleportation results are $\tilde{P}_{j,i}, \tilde{P}_{j,\anc}$, it sets $E_{j,d} = \onreg{E_j}{R_{j,1},\cdots,R_{j,n},N_j,T_j} \onreg{\tilde{P}_{j,1}}{R_{j,1}} \cdots \onreg{\tilde{P}_{j,n}}{R_{j,n}} \onreg{\tilde{P}_{j,{\anc}}}{N_j}$.
\end{itemize}

\item $\hybrid_{3+d-k}$, where $k = d,\cdots,1$:
\begin{itemize}
    \item $\Sigma^{\BoBW}$ step $5$: For $h=d,\cdots,k$ and for every $j$, $\trustp^{\hyb}$ resets $(\gray{N_{j,h}},\gray{T_{j,n+h}})$ as a random string $\tilde{r}_{j,h}\randsample \bbZ_p^{\numtraps+1}$, samples Clifford $V_{j,h} \randsample \CliffordGroup_{\ell^{\total}+(n+h)\numtraps}, \tilde{E}_{j,k-1} \randsample \CliffordGroup_{\ell^{\total}+(n+k-1)\numtraps}$ and sets $E_j = \onreg{V^\dag_{j,d}}{R'_{j,[n]}, N'_{j,[d]}, T'_{j,[n+d]}} \cdots \onreg{V^\dag_{j,k}}{R'_{j,[n]}, N'_{j,[k]}, T'_{j,[n+k]}} \onreg{\tilde{E}_{j,k-1}}{R'_{j,[n]}, N'_{j,[k-1]}, T'_{j,[n+k-1]}}$.
    \item $\Pi^\BoBW$ step $2$: $\trustp^{\hyb}$ computes $\onreg{\tilde{\rho}_{(k)}{}}{\tilde{R}_{[n]}, \tilde{N}_{[k-1]}} \from \circuit[k] \cdots \circuit[d]\left(\onreg{\tilde{\rho}_1}{\tilde{R}_1}, \cdots, \onreg{\tilde{\rho}_n}{\tilde{R}_n}, \onreg{\phi_{\anc}}{\tilde{N}}\right)$, performs $(\tilde{\rho}^{(1)}_{(k)},\cdots,\tilde{\rho}^{(n)}_{(k)}) \from \QEnc(\tilde{\rho}_{(k)})$, and teleports $\tilde{\rho}^{(j)}_{(k)}$ back to server $j$. If the teleportation result is $\tilde{P}_j$, it sets $E_{j,k} = \onreg{E_j}{R_{j,[n]},N_{j,[k-1]},T_{j,[n+k-1]}} \onreg{\tilde{P}_j}{R_{j,[n]}, N_{j,[k-1]}}$.
    \item $\Pi^\BoBW$ step $3$ iteration $h$ for $h = d,\cdots, k$: $\trustp^{\hyb}$ sends $V_{j,h}$ to server $j$ and receives $r_{j,h}$ in return. If $r_{j,h} \neq \tilde{r}_{j,h}$, $\trustp^{\hyb}$ adds server $j$ to the corruption list $J$ and will not interact with this server anymore. If $|J|>\thres$, $\trustp^{\hyb}$ publicly outputs $J$ as malicious and aborts.
\end{itemize}

\item $\hybrid_{3+d}$:
\begin{itemize}
    \item $\Sigma^\BoBW$ step $4$: $\trustp^{\hyb}$ samples $(\sk_{j \to i}, \phi_{S,j \to i}, \sigma_{j,i}) \from \AQA.\widetilde{\Setup}(1^\secparam, 1^{\ell_i})$, uses the content $\EPRR_{j,i}$ of register $\gray{R_{j,i}}$ as $\phi_{R,j \to i}$, and reassigns register $(\gray{R_{j,i}}, \gray{T_{j,i}})$ as $\sigma_{j,i}$.
    \item $\Pi^\BoBW$ step $2$: $\trustp^{\hyb}$ computes the circuit $\circuit$
    with input $(\onreg{\tilde{\rho}_1}{\tilde{R}_1}, \cdots \onreg{\tilde{\rho}_n}{\tilde{R}_n})$, obtains the quantum output $({\tilde{\rho}'_1}, \cdots {\tilde{\rho}'_n})$ of $\circuit$, performs $({\tilde{\rho}'^{(1)}_i}, \cdots, {\tilde{\rho}'^{(n)}_i}) \from \QEnc({\tilde{\rho}'_i})$, teleports $\tilde{\rho}'^{(j)}_i$ via register $\gray{\tilde{S}_{j,i}}$, and sets $\dk_{j,i}$ as the teleportation result $\tilde{P}_{j,i}$.
    \item $\Pi^\BoBW$ step $5$: $\trustp^{\hyb}$ sets $V'_j = \tilde{E}_{j,0}^\dag$.
    \item $\Pi^\BoBW$ step $7$: $\trustp^{\hyb}$ resets $\dk_{j,i} = \bot $ if $ \bot \from \AQA.\widetilde{\Audit}(\sk_{j \to i}, \pf_{j,i})$.
\end{itemize}
\end{itemize}

The last hybrid is equivalent to the ideal world with the following simulator.

\begin{simulator}{$\Sim^{\BoBW}_{\adv(\rho_{\aux})}$ for BoBW-MPQC-PVIA with Trusted Setup}\label{sim:bobw}
\begin{enumerate}
    \setlength{\itemsep}{1pt}
    \item Fake setup (if client $i$ is corrupted):\begin{enumerate}
        \item Prepare EPR pairs on registers $(\gray{S_i}, \gray{\tilde{R}_{i}})$ with $|\gray{S_i}| = |\gray{\tilde{R}_{i}}| = \ell_i$.
        \item Prepare EPR pairs on registers $({\gray{\tilde{S}_{j,i}}}, {\gray{\hat{R}_{j,i}}})$ with  $|\gray{\tilde{S}_{j,i}}| = |\gray{\hat{R}_{j,i}}| = \ell_{i}$.
        \item Send $(\gray{S_{i}}, \gray{\hat{R}_{j,i}})$ to $\adv$.
    \end{enumerate}
    \item Fake setup (if server $j$ is corrupted):
    \begin{enumerate}
        \item Sample $(\sk_{j\to i},\phi_{S,j\to i}, \onreg{\sigma_{j,i}}{R_{j,i}, T_{j,i}}) \from \AQA.\widetilde{\Setup}(1^\secparam, 1^{\ell_i})$ for $i \in [n]$.
        \item Initialize registers $(\gray{N_{j,k}},\gray{T_{j,n+k}})$ with random $\tilde{r}_{j,k} \randsample \bbZ_p^{\numtraps+1}$ for $k \in [d]$.
        \item Sample random Clifford gates $V_{j,k} \randsample \CliffordGroup_{\ell^{\total}+(n+k)\numtraps}$ for $k = 0,1,\cdots,d$.
        \item Apply $\onreg{V_{j,d}^\dag}{R_{j,[n]}, N_{j,[d]}, T_{j,[n+d]}} \cdots \onreg{V_{j,1}^\dag}{R_{j,[n]}, N_{j,[1]}, T_{j,[n+1]}}  \onreg{V_{j,0}^\dag}{R_{j,[n]}, T_{j,[n]}}$ with resulting state $\sigma_j$.
        \item Send $(\onreg{\sigma_j}{R_{j,1},\cdots,R_{j,n},N_j,T_j},\phi_{S,j\to 1},\cdots,\phi_{S,j \to n})$ to $\adv$.
    \end{enumerate}
    \item Input extraction: Receive $P_i$ from $\adv$ and extract the input $\onreg{\tilde{\rho_i}}{\tilde{R}_i}=\onreg{P_i^\dag \EPRR_i}{\tilde{R}_i}$.
    \item Invoke the ideal functionality: Send $\tilde{\rho_i}$ to $\trustp$ and receive the output $\tilde{\rho}'_i$ from $\trustp$.
    \item Check the abort decision (if server $j$ is corrupted): \begin{enumerate}
        \item For $k=d,\cdots,1$, send $V_{j,k}$ to $\adv$ and receive $r_{j,k}$ in return.\\
        If $r_{j,k} \neq \tilde{r}_{j,k}$, send $\abort$ to $\trustp$ in the name of server $j$.
        \item Send $V_{j,0}$ to $\adv$.
        \item For $i \in [n]$, receive $\pf_{j,i}$ from $\adv$.\\
        If $\bot \from \AQA.\widetilde{\Audit}(\sk_{j,i},\pf_{j,i})$, send $\abort$ to $\trustp$ in the name of server $j$.
        \end{enumerate}
    \item Output delivery: Perform $(\tilde{\rho}'^{(1)}_{i},\cdots,\tilde{\rho}'^{(n)}_{i}) \from \QEnc(\tilde{\rho}'_i)$, teleport $\tilde{\rho}'^{(j)}_{i}$ via $\gray{\tilde{S}_{j,i}}$, and obtain its teleportation result $\tilde{P}_{j,i}$. Set $\dk_{j,i} = \tilde{P}_{j,i}$ if server $j$ did not send $\abort$ in the previous step; otherwise, set $\dk_{j,i} = \bot$. Send $\dk_{j,i}$ to $\adv$. 
    \item Output $\adv$'s output.
\end{enumerate}
\end{simulator}
To prove our theorem, it suffices to show indistinguishability between consecutive hybrids. In the following, $\QEnc_j$ will represent the $j$-th part of the output of $\QEnc$.
\begin{itemize}[leftmargin=*]
    \item $\real = \hybrid_1$: This is because $\Sigma^{\BoBW}, \cMPC$ and $\trustp^{\hyb}$ are all trusted executions.
    \item $\hybrid_1 = \hybrid_2$: Consider executing the protocol until the end of step $1$. Suppose the CP map $A$ projects  $(\gray{M_{[n]}},\gray{S_{[n]}})$ to $\ket{P_{[n]}}$ in step $1$. The state of $\hybrid_1$ after protocol step $2$ is
    \begin{align*}
        & \E_{E_{j} \from \CliffordGroup}\mixed{ A\left( E_j\left( \onreg{\QEnc_{}\left( {\EPRR_{[n]}},{\phi_{{\anc}}}\right)}{R_{j,[n]}, N_j}, \onreg{\ket{0}^{\ot (n+d)\numtraps}}{T_j}\right),\cdots\right) \ot \ket{E_{j,d}}} \\
        =& \E_{E_{j,d} \from \CliffordGroup}\mixed{ A\left( E_{j,d}\left( \onreg{P_{j,[n]}^\dag}{R_{j,[n]}} \onreg{\QEnc_{j}\left( {\EPRR_{[n]}},{\phi_{{\anc}}}\right)}{R_{j,[n]}, N_j}, \onreg{\ket{0}^{\ot (n+d)\numtraps}}{T_j}\right),\cdots\right) \ot \ket{E_{j,d}}} \\
        =& \E_{E_{j,d} \from \CliffordGroup}\mixed{ A\left( E_{j,d}\left( \onreg{\QEnc_{j}\left( {P_{[n]}^\dag \EPRR_{[n]}},{\phi_{{\anc}}}\right)}{R_{j,[n]}, N_j} , \onreg{\ket{0}^{\ot (n+d)\numtraps}}{T_j}\right),\cdots\right) \ot \ket{E_{j,d}}} 
    \end{align*}
    where the first equality is by the definition of $E_{j,d}$ and the second equality follows from $(P_{1,[n]}\ot \cdots \ot P_{n,[n]})^\dag\; \QECC = \QECC\; P_{[n]}^\dag$.
    In $\hybrid_2$, we can change the order of $A$ and the teleportation made by $\trustp^{\hyb}$ because these operators act on disjoint registers. By quantum teleportation, the state of $\hybrid_2$ after protocol step $2$ is as follows, which is the same as in $\hybrid_1$.
    \begin{align*}
        & \E_{E_j \from \CliffordGroup}\mixed{ A\left( E_j\left(\onreg{\tilde{P}_{j,[n]} \tilde{\rho}^{(j)}_{[n]}}{R_{j,[n]}}, \onreg{ \tilde{P}_{j,\anc} \phi^{(j)}_{\anc}}{N_j},\onreg{\ket{0}^{\ot (n+d)\numtraps}}{T_j}\right),\cdots\right) \ot \ket{E_{j,d}}} \\
        =& \E_{E_{j,d} \from \CliffordGroup}\mixed{ A\left( E_{j,d}\left(\onreg{\tilde{\rho}^{(j)}_{[n]}{}}{R_{j,[n]}}, \onreg{ \phi^{(j)}_{\anc}}{N_j},\onreg{\ket{0}^{\ot (n+d)\numtraps}}{T_j}\right),\cdots\right) \ot \ket{E_{j,d}}} \\
        =& \E_{E_{j,d} \from \CliffordGroup}\mixed{ A\left( E_{j,d}\left( \onreg{\QEnc_{j}\left( {P_{[n]}^\dag \tilde{\EPRR}_{[n]}},{\phi_{{\anc}}}\right)}{R_{j,[n]}, N_j} , \onreg{\ket{0}^{\ot (n+d)\numtraps}}{T_j}\right),\cdots\right) \ot \ket{E_{j,d}}} 
    \end{align*}
    
    \item $\hybrid_{2+d-k} = \hybrid_{3+d-k}$: This is similar to \ref{hybrid:analysis} except here, we allow multiple servers $j \in [n]$.
    Consider executing the protocol until the end of step $3$ iteration $k\text{+}1$. Suppose the CP map $A$ projects  $(\gray{M_{j,[n]}},\gray{S_{j,[n]}})$ to $\ket{P_{j,[n]}}$ in step $1$ and obtains $b_d,\cdots,b_{k+1}$ in step $3$ for iterations $d$ to $k\text{+}1$. The partial computation result $C[k+1]\cdots C[d] \big(P_{[n]}^\dag \tilde{\EPRR}_{[n]}, \phi_{\anc}\big)$ is denoted as $\onreg{\tilde{\rho}_{(k)}{}}{\tilde{R}_{[n]},\tilde{N}_{[k]}} = \sum_{u \in \bbZ_p}\onreg{\tilde{\rho}_{(k,u)}{}}{\tilde{R}_{[n]},\tilde{N}_{[k-1]}}\ot \onreg{\ket{u}}{\tilde{N}_{k}}$. 
    Let us write $\QEnc_j(\tilde{\rho}_{(k)})$ as $\onreg{\tilde{\rho}^{(j)}_{(k)}{}}{\tilde{R}_{j,[n]},\tilde{N}_{j,[k]}} = \sum_{u_j \in \bbZ_p} \onreg{\tilde{\rho}_{(k,j,u_j)}{}}{\tilde{R}_{j,[n]},\tilde{N}_{j,[k-1]}} \ot \onreg{\ket{u_j}}{\tilde{N}_{j,k}}$. In $\hybrid_{2+d-k}$, the state at protocol step $3(a)$ for iteration $k$ is 
    {\small
    \begin{align*}
        & \E
        \mixed{A\left( E_{j,k}\left(\onreg{\tilde{\rho}^{(j)}_{(k)}{}}{R_{j,[n]},N_{j,[k]}}, \onreg{\ket{0}^{\ot (n+j)\numtraps}}{T_{j,[n+k]}} \right),\cdots\right)\ot \ket{P'_{j,k},c_{j,k}}}\\
        =& \E
        \mixed{A\left( V_{j,k}^\dag\left( \sum_{u_j \in \bbZ_p} E'_{j,k-1}\big(\tilde{\rho}_{(k,j,u_j)}, \ket{0}^{\ot (n+j-1)\numtraps}\big)\ot P'_{j,k} \CX_{c_{j,k}}\big(\ket{u_j}, \ket{0}^{\ot \numtraps}\big)\right),\cdots\right) \ot \ket{P'_{j,k},c_{j,k}}}
    \end{align*}
    }The equality holds by the definition of $V_{j,k}$. Since the protocol gives the attacker access to $\ket{V_{j,k}}$ in step $3(a)$, we can merge $V_{j,k}^\dag$ and $A$ into a CP map $A'$ that operates also on $\ket{V_{j,k}}$. Also, we can simplify $\CX_{c_{j,k}}(\ket{u_j}, \ket{0}^{\ot \lambda}) = \ket{u_j (1,c_{j,k})}$. Thus, the state is equal to
    {\small
    \begin{align*}
        \E
        \mixed{A'\left( \sum_{u_j \in \bbZ_p} E'_{j,k-1}\big(\tilde{\rho}_{(k,j,u_j)}, \ket{0}^{\ot (n+k-1)\numtraps}\big)\ot \onreg{P'_{j,k} \ket{u_j (1,c_{j,k})}}{N_{j,k},T_{j,n+k}},\cdots\right) \ot \ket{P'_{j,k},c_{j,k}}}
    \end{align*}
    }Afterwards, protocol step $3(c)$ projects registers $(\gray{N_{j,k}}, \gray{T_{j,n+k}})$ to $\ket{x(P'_{j,k})+b_{j,k}(1,c_{j,k})}$ for all possible values of $b_{j,k}$. 
    To analyze, we apply Pauli twirl with target measurement result $b_{j,k}(1,c_{j,k})$ (lemma \ref{lemma:qotp-measure}) and use the Pauli decomposition $A' = \sum_{{Q_j}\in \PauliGroup^*_{\numtraps+1}} \onreg{Q_j}{N_{j,k},T_{j,n+k}} \ot A'_{Q_j}$.
    The state after the projection (denoted $\measurement_{b_{j,k}}$) of step $3(c)$ for obtaining solution $b_{j,k}$ is
    {\small
    \begin{align*}
        & \E
        \mixed{\onreg{\measurement_{b_{j,k}}}{N_{j,k}, T_{j,n+k}} A'\left( \sum_{u_j \in \bbZ_p} \onreg{P'_{j,k} \ket{u_j(1, c_{j,k})}}{N_{j,k},T_{j,n+k}} \ot E'_{j,k-1}\big(\tilde{\rho}_{(k,j,u_j)}, \ket{0}^{\ot (n+k-1)\numtraps}\big) ,\cdots\right)}\\
        =& \E
        \;\;\sum_{u_j}
        \mixed{\sum_{x(Q_j)=(b_{j,k} - u_j)(1,c_{j,k})} {Q_j}\ket{\tilde{r}_{j,k}} \ot A'_{Q_j}\left(E'_{j,k-1}\big(\tilde{\rho}_{(k,j,u_j)}, \ket{0}^{\ot (n+k-1)\numtraps}\big),\cdots\right)}
    \end{align*}
    }The state at step $3(c)$ conditioned that there is a solution for $b_{j,k}$ is
    {\small
    \begin{align*}
        \E_{c_{j,k}, \tilde{r}_{j,k}, E'_{j,k\text{-}1}} \sum_{b_{j,k}} \sum_{u_j}
        \mixed{\sum_{x(Q_j)=(b_{j,k} - u_j)(1,c_{j,k})} {Q_j}\ket{\tilde{r}_{j,k}} \ot A'_{Q_j}\left(E'_{j,k-1}\big(\tilde{\rho}_{(k,j,u_j)}, \ket{0}^{\ot (n+k-1)\numtraps}\big),\cdots\right) \ot \ket{b_{j,k}} }
    \end{align*}
    }where the last register is the solution stored by $\trustp^{\hyb}$. Observe that each $c_{j,k}$ occurs with negligible probability, so the sum of summands for $u_j \neq b_{j,k}$ have negligible trace norm when averaged over $c_{j,k}$. Hence, the above state is indistinguishable to the following state where only $u_j = b_{j,k}$ is left. Note that this state is as if the last qupit of $\tilde{\rho}^{(j)}_{(k)}$ is being measured in the computational basis and only $\trustp^{\hyb}$ knows the measurement result:
    {\small \begin{align}
        \E_{\tilde{r}_{j,k}, E'_{j,k\text{-}1}} \sum_{b_{j,k}} \mixed{\sum_{x(Q_j)=0} {Q_j}\ket{\tilde{r}_{j,k}} \ot A'_{Q_j}\left(E'_{j,k-1}\big(\tilde{\rho}_{(k,j,b_{j,k})}, \ket{0}^{\ot (n+k-1)\numtraps}\big),\cdots\right) \ot \ket{b_{j,k}} }
        \label{eq:bobw-measure-success}
    \end{align}
    }The state at step $3(c)$ conditioned that there is no solution for $b_{j,k}$ is indistinguishable, using a similar argument and the Clifford twirl (\cref{lemma:qotp-security}), to the following state:
    {\small \begin{align}
        \E_{\tilde{r}^*_{j,k}} \mixed{\sum_{x(Q_j) \neq 0} ({Q_j}\ot A'_{Q_j}) \left(\onreg{\ket{\tilde{r}^*_{j,k}}}{R_{j,[n]},N_{j,[k]},T_{j,[n+k]}}, \cdots\right) \ot \ket{1_{j\in J}}}
        \label{eq:bobw-measure-failure}
    \end{align}
    }where $1_{j\in J}$ is a flag that indicates server $j$ is in the corruption list $J$. Next, we analyze the state at step $3(d)$. If the corruption list has size $|J|> \thres$, then $J$ will be publicly announced and $E'_{j,k-1}$ will be discarded. By the Clifford twirl (\cref{lemma:qotp-security}), the residual joint state consists of a maximally mixed state and the information of $J$. 
    
    If $|J|\le \thres$, $\trustp^{\hyb}$ decodes $b_k \from \QDec((b_{1,k},\cdots,b_{n,k}), J)$. Since QECC tolerates $\thres$ erasure errors, the values of $\{b_{j,k}\}_{j\in J}$ does not affect the QECC decoding result. The transversal measurement property of QECC shows that decoding the measurement results $b_{j,k}$ of $\gray{\tilde{N}_{j,k}}$ is the same as obtaining the measurement result of $\gray{\tilde{N}_k}$. The state obtained from these two procedures on registers $(\gray{R_{j,[n]}},\gray{N_{j,[k-1]}})$ are equal, which gives $\tilde{\rho}_{(k,j,b_{j,k})} = \QEnc_j(\tilde{\rho}_{k,b_k})$. By plugging the equality to \cref{eq:bobw-measure-success}, we see that the state held by server $j \not \in J$ at step $3(d)$ is
    \begin{align*}
        \sum_{x(Q_j)=0} Q_j \ket{\tilde{r}_{j,k}} \ot A'_{Q_j} \left( E'_{j,k-1}\big(\QEnc_j \left(\tilde{\rho}_{(k,b_k)}\right), \ket{0}^{\ot (n+k-1)\numtraps}\big),\cdots \right)
    \end{align*}
    Finally, protocol step $3(e)$ sets $E_{j,k-1} = E'_{j,k-1} G_{k-1}^{(j)\dag}$. The state for $j \not \in J$ now becomes
    {\small 
    \begin{align}
        & \sum_{x(Q_j)=0} Q_j \ket{\tilde{r}_{j,k}} \ot A'_{Q_j} \left( E_{j,k-1}\big( G_{k-1}^{(j)} \QEnc_j \left(\tilde{\rho}_{(k,b_k)}\right), \ket{0}^{\ot (n+k-1)\numtraps}\big),\cdots \right)\\
        =& \sum_{x(Q_j)=0} Q_j \ket{\tilde{r}_{j,k}} \ot A'_{Q_j} \left( E_{j,k-1}\big( \QEnc_j \left(G_{k-1} (\tilde{\rho}_{(k,b_k)})\right), \ket{0}^{\ot (n+k-1)\numtraps}\big),\cdots \right)\\
        =& \sum_{x(Q_j)=0} Q_j \ket{\tilde{r}_{j,k}} \ot A'_{Q_j} \left( E_{j,k-1}\big( \QEnc_j \left(\tilde{\rho}_{(k-1)}\right), \ket{0}^{\ot (n+k-1)\numtraps}\big),\cdots \right)
        \label{eq:bobw-measure-final-inside}
    \end{align}
    }where the first equality follows from $(G_{k-1}^{(1)}, \cdots, G_{k-1}^{(n)})$ being the fault-tolerant version of $G_{k-1}$ and the second equality arises because $G_{k-1}(\tilde{\rho}_{(k,b_k)})$ is the result of performing the $k$-th iteration of computation on $\tilde{\rho}_{(k)}$. We have now described all the cases in $\hybrid_{2+d-k}$.
    
    In $\hybrid_{3+d-k}$, the state after projecting to the acceptance condition $r_{j,k}=\tilde{r}_{j,k}$ is
    {\small
    \begin{align*}
    &\E_{\tilde{r}_{j,k}, E_{j}} \mixed{\onreg{\ketbra{\tilde{r}_{j,k}}}{N_{j}, T_{j,n+k}} A'\left(\onreg{\ket{\tilde{r}_{j,k}}}{N_{j,k}, T_{j,n+k}}, E_{j}\left(\eps_{(k)}, \ket{0}^{\ot (n+k-1)\numtraps}\right),\cdots\right)}\\
    =& \E_{\tilde{r}_{j,k}, E_{j}} \mixed{ \sum_{x(Q_j)=0} Q\left(\ket{\tilde{r}_{j,k}}\right) \ot A'_{Q_j}\left(E_{j}\left(\eps_{(k)}, \ket{0}^{\ot (n+k-1)\numtraps}\right),\cdots\right)}
    \end{align*}
    }where $\eps_{(k)}$ is the half EPR pairs prepared on $\gray{R_{j,[n]}}, \gray{N_{j,[k-1]}}$ and the equality follows from \cref{lemma:qotp-measure}. Next, $\trustp^{\hyb}$ teleports $\QEnc_j(\tilde{\rho}_{(k-1)})$ and fixes its teleportation Pauli to $E_j$ which yields $E_{j,k}$. This gives the state
    {\small
    \begin{align}
        \E_{\tilde{r}_{j,k}, E_{j,k-1}} \sum_{u_j} \mixed{ \sum_{x(Q_j)=0} Q\left(\ket{\tilde{r}_{j,k}}\right) \ot A'_{Q_j}\left(E_{j,k-1}\left(\QEnc_j(\tilde{\rho}_{(k-1)}), \ket{0}^{\ot (n+k-1)\numtraps}\right),\cdots\right)}
        \label{eq:bobw-measure-success-ideal}
    \end{align}
    }The state after projecting to the rejection condition $r_{j,k} \neq \tilde{r}_{j,k}$ would be exactly the same as \cref{eq:bobw-measure-failure}, using a similar argument and the Clifford twirl (\cref{lemma:qotp-security}). Next, if $|J|> \thres$, then $E_{j,k-1}$ will be discarded. By Clifford twirl, the residual joint state is same as in $\hybrid_{2+d-k}$. If $|J|\le \thres$, the state described in \cref{eq:bobw-measure-success-ideal} is the same as the state described in \cref{eq:bobw-measure-final-inside}. Thus, $\hybrid_{2+d-k}$ and $\hybrid_{3+d-k}$ are indistinguishable.
    
    \item $\hybrid_{2+d} = \hybrid_{3+d}$: Consider executing the protocol until the end of step $6$. Let $(\tilde{\rho}'_1,\cdots, \tilde{\rho}'_n)$ be the quantum output of $\circuit$ and $\tilde{\rho}'^{(j)}_i = \QEnc_j(\tilde{\rho}'_i)$. The state of $\hybrid_{2+d}$ at step $6$ is
    {\small
    \begin{align*}
        & \E_{E_{j,0},\sk_{j \to i}} \mixed{A \of{ \onreg{E_{j,0}}{R_{j,[n]},T_{j,[n]}} \of{ \onreg{\tilde{\rho}'^{(j)}_{i}{}}{R_{j,i}}, \onreg{\ket{0}^{\ot \numtraps}}{T_{j,i}} } , \ket{V'_j},\phi_{S,j\to i}, \phi_{R,j\to i}, \cdots} \ot \ket{\sk_{j \to i}}}\\
        =& \E_{V'_j, \sk_{j \to i}} \mixed{A \of{ \onreg{V'^{\dag}_j}{R_{j,[n]},T_{j,[n]}} \of{ \onreg{\AQA.\EncGate(\sk_{j \to i}) \of{\tilde{\rho}'^{(j)}_{i}, \ket{0}^{\ot \numtraps}}}{R_i, T_i} } , \ket{V'}, \phi_{S,i}, \phi_{R,i}, \cdots } \ot \ket{\sk_{j \to i}} }\\
        =& \E_{V'_j, \sk_{j \to i}} \mixed{A \of{ \onreg{V'^{\dag}_j}{R_{j,[n]},T_{j,[n]}} \of{ \onreg{\AQA.\Enc \of{\sk_{j \to i}, \tilde{\rho}'^{(j)}_{i}}}{R_i,T_i} } , \ket{V'}, \phi_{S,i}, \phi_{R,i}, \cdots } \ot \ket{\sk_{j \to i}} }
    \end{align*}
    }where the first equality follows from the definition of $V'$ in $\hybrid_{2+d}$ and the second follows from the definition of $\AQA.\Enc$. After protocol step $7$, the state of $\hybrid_{2+d}$ can be generated under the process $(\AQA.\Setup, \AQA.\Enc, A', \Audit)$ where $A'$ is a CP map that merges $A$ and $V'^\dag$. On the other hand, the state of $\hybrid_{3+d}$ at step $6$ is 
    {\small
    \begin{align*}
        & \E_{\tilde{E}_0,\sk_{j \to i}, \tilde{P}_{j,i}} \mixed{A \of{ \onreg{\tilde{E}_0}{R_{j,[n]},T_{j,[n]}} \of{ \onreg{\sigma_{j,i}}{R_{j,i},T_{j,i}} } , \ket{V'_j}, \phi_{S,j\to i}, \tilde{P}_{j,i}(\tilde{\rho}'^{(j)}_{i}), \cdots } \ot \ket{\tilde{P}_{j,i}} \ot \ket{\sk_{j \to i}}}\\
        =& \E_{V'_j,\sk_{j \to i}, \dk_{j,i}} \mixed{A \of{ \onreg{V'^\dag_j}{R_{j,[n]},T_{j,[n]}} \of{ \onreg{\sigma_{j,i}}{R_{j,i},T_{j,i}} } , \ket{V'_j}, \phi_{S,j\to i}, \dk_{j,i}(\tilde{\rho}'^{(j)}_i), \cdots } \ot \ket{\dk_{j,i}} \ot \ket{\sk_{j\to i}}}
    \end{align*}
    }where the equality follows from the definition of $V'_j$ and $\dk_{j,i}$ in $\hybrid_{3+d}$. Here, $\sk_{j\to i}, \phi_{S,j\ to i}, \sigma_{j,i}$ are generated from $\AQA.\widetilde{\Setup}$, whereas $\dk_{j,i}$ and the encoding of $\tilde{\rho}'^{(j)}_i$ can be generated from $\AQA.\widetilde{\Enc}$. After protocol step $7$, the state of $\hybrid_{3+d}$ can be generated under the process $(\AQA.\widetilde{\Setup}, \AQA.\widetilde{\Enc}, A', \widetilde{\Audit})$. By the receiver security of $\AQA$ (\cref{thm:aqa}), these two states are statistically indistinguishable.
    
    \item $\hybrid_{3+d} = \ideal$: Except for the computation of $\circuit$, the execution of $\trustp^{\hyb}$ in $\hybrid_{3+d}$ consists of $n$ independent parts, each interacting with only one party. Thus, we can collect the parts that interact with the corrupted parties as a simulator $\Sim$, who also interacts with the ideal functionality of MPQC to compute $\circuit$. By quantum teleportation, the completeness and security of our AQA, and the recoverability of QECC, the honest parties and their corresponding parts of $\trustp^{\hyb}$ indeed send the honest inputs to the ideal functionality, never send $\abort$ to the ideal functionality, and always output the honest outputs. Hence, $\hybrid_{3+d}$ is identical to $\ideal$.
\end{itemize}
\vspace{-1em}
\end{proof}
\section{BoBW-MPQC-PVIA without Trusted Setup}\label{Sec:SWIA+PVIA}
So far, we have demonstrated an BoBW-MPQC-PVIA protocol in the preprocessing model with a trusted setup. Now, we will illustrate how to instantiate the preprocessing phase \emph{without} a trusted setup. Our approach involves using MPQC-SWIA  \eg \cite{ACC+21} to implement the preprocessing phase.  Note that \cite{ACC+21} requires a post-quantum fully homomorphic encryption assumption to achieve the security; however, we do not need this assumption when we want to compute circuits that have no inputs and apply only Clifford gates to $\ket{0}$'s and $\ket{\Tg}$'s. We defer the explicit MPQC-SWIA construction to \cref{sec:modified-swia}. We reformulate MPQC-SWIA as the following lemma. 
\begin{lemma}[MPQC-SWIA]\label{lem:ACC+21}
	There is a multi-party quantum computation $\Pi^{\SWIA}$ \; secure with identifiable abort in the MPC-hybrid model that computes quantum circuit $\circuit$, given that $\circuit$ takes no inputs and only applies Clifford to ancillary $\ket{0}$ and $\ket{\Tg}$ states. That is, for every non-uniform (\qpt) adversary $\adv$ corrupting party set $I$, there is a non-uniform (\qpt) adversary $\Sim^{\SWIA}_{\adv}$ corrupting $I$, such that
$$
    \{\real^{\Pi^{\SWIA}}_{\adv(\rho_{\aux})}(1^{\secparam}, \circuit)\} \approx \{\ideal^{\SWIA}_{\Sim^{\SWIA}_{\adv(\rho_{\aux})}}(1^{\secparam}, \circuit)\}
$$
\end{lemma}
\vspace{-3mm}
\begin{idealmodel}{$\ideal^{\SWIA}$: Multi-party Quantum Computation Secure with Identifiable Abort}
\label{ideal:swia}
\begin{description}
\item[Common input:]\quad\\
The security parameter $1^{\secparam}$ and a quantum circuit $C$ with no inputs.
\item[Input:]\quad\\
$\adv_I$ holds input $\rho_{\aux}$ and controls parties in $I$.

\item[Execution of $\trustp^{\SWIA}$:]\quad\\
$\trustp^{\SWIA}$ computes $(\rho_1',\rho_2',\cdots,\rho'_n, r'_{\cMPC}) \from C$, where $r'_{\cMPC}$ is classical.\\
$\trustp^{\SWIA}$ sends $\rho'_i$ to all $\Pc_i \in I$. Every $\Pc_i \in I$ can send $\abort$ message to $\trustp^{\SWIA}$.\\
Let $I_{\abort}$ be the set of parties who indeed send the $\abort$ message.\\ If $I_{\abort}$ is non-empty, $\trustp^{\SWIA}$ sends the partition $\{I_{\abort}, \Pc_{[n]}\backslash I_{\abort}\}$ to all parties.\\ 
Otherwise, $\trustp^{\SWIA}$ sends $\rho'_i$ to all $\Pc_i \not \in I$ and sends $r'_{\cMPC}$ to $\cMPC$.

\item[Output:]\quad\\ Honest parties output whatever output received from $\trustp^{\SWIA}$.\\
The adversary $\adv_I$ outputs a function of his view.
\end{description}
\end{idealmodel}

We say an execution succeeds if every party receives their part of the circuit output. Whenever the execution fails, all parties get to know how they have been partitioned into two groups. An honest party can infer that the group he does not belong to (\ie $I_{\abort}$) is the set of malicious parties who interfered the computation.

In contrast to \cite{ACC+21} where the partition information serves the purpose of SWIA, we utilize the partition information to design a preprocessing procedure that \emph{always} succeeds. Such a preprocessing procedure will be suitable for replacing the setup of MPQC-PVIA.

\subsection{Protocol}
First, we let the parties run MPQC-SWIA over the setup circuit $\Sigma^{\BoBW}$ that the trusted setup is supposed to run. If the MPQC-SWIA fails, the parties will be divided into two groups. Each group will then ignore other groups and run MPQC-SWIA independently. By iterating this process, each party will eventually find a group in which the MPQC-SWIA succeeds and receive an output of the setup circuit. This constitutes our preprocessing procedure.
Note that we do not publicly identify anyone during this stage. 

After obtaining the output of the setup circuit, every party can run the protocol $\Pi^{\BoBW}$ within their group to obtain their MPQC output. It is reasonable for each group to operate independently, as they view other groups as untrustworthy. Parties within a group can set the inputs of the parties outside the group as some default inputs such as $\ket{0}$. Moreover, MPQC-SWIA guarantees that the honest parties are always in the same group, so they will jointly compute their outputs. Our approach circumvents the false accusation problem encountered in \cite{ACC+21} because the parties in our protocol no longer accuse between groups. Instead, each group runs its own $\Pi^{\BoBW}$, which only aborts dishonest members.
Below is the formulation of our protocol. 

\begin{protocol}{$\Pi^{\MPQC}$ for BoBW-MPQC-PVIA}\label{proto:swia+pvia}
{\bf Input:}
\begin{enumerate}
    \setlength{\itemsep}{1pt}
    \item Everyone holds the threshold $\thres$ and the circuit description $C$.
    \item Party $\Pc_i$ holds private input $\rho_i \in \Den{\ell_i}$.
\end{enumerate}

{\bf Protocol:}
\begin{enumerate}
    \setlength{\itemsep}{1pt}
    \item Set $G = \{\{\Pc_1, \cdots, \Pc_n\}\}$ as the initial partition of parties (\ie no partition) and mark the set $\{\Pc_1,\cdots,\Pc_n\} \in G$ as unfinished.
    \item Repeat the following as long as $G$ contains a set $S$ that is marked as unfinished:
    \begin{enumerate}
        \setlength{\itemsep}{1pt}
        \item The parties in $S$ run $\Pi^{\SWIA}$ over the setup circuit $\Sigma^{\BoBW}(\thres, C_S)$, where $C_S$ is the circuit that prepares default inputs (\eg $\ket{0}$) for parties not in $S$ and runs $C$.
        \item If $\Pi^{\SWIA}$ succeeds or if $|S|=1$, parties in $S$ has obtained the output of the setup circuit. In this case, mark $S$ as finished.
        \item Otherwise, $\Pi^{\SWIA}$ instructs to partition $S$ into $\{S_0, S_1\}$ of $S$. In this case, replace $S$ with $S_0$ and $S_1$ in $G$, and mark both $S_0, S_1$ as unfinished.
    \end{enumerate}
    
    \item Run $\Pi^{\BoBW}(\thres, \circuit_S)$ within every set $S \in G$.
    \item Let $J$ be the union of all corruption lists output by all executions of $\Pi^{\BoBW}$. We note that every corruption list is public and only contains the parties that participate.
\end{enumerate}
{\bf Output:}
\begin{enumerate}
    \setlength{\itemsep}{1pt}
    \item If $|J|> \thres$, every party outputs $J$ as malicious and aborts.
    \item Otherwise, $\Pc_i$ outputs the result obtained from his execution of $\Pi^{\BoBW}$.
\end{enumerate}
\end{protocol}

\subsection{Security}
\begin{theorem}\label{thm:swia+pvia}
$\Pi^{\MPQC}$ is a \bobw multi-party quantum computation \pvia of threshold $\thres$ in the MPC-hybrid model.
\end{theorem}

\begin{proof}
We would like to show that the real world 
and the ideal world 
are indistinguishable using the following simulator. We regard $\adv$ as a stateful adversary throughout.
\begin{simulator}{$\Sim^{\MPQC}_{\adv(\rho_{\aux})}$ for BoBW-MPQC-PVIA}\label{sim:swia+pvia}
\begin{enumerate}
    \setlength{\itemsep}{1pt}
    \item Set $S = \{\Pc_1, \cdots, \Pc_n\}$ as the initial group that contains all the honest parties.
    \item Repeat the following as long as the inner $\ideal^{\SWIA}$ fails:
    \begin{enumerate}
        \setlength{\itemsep}{1pt}
        \item Run a simulated $\ideal^{\SWIA}$ that computes the first part of $\Sim^{\BoBW}_\adv(\circuit_S)$, which is a circuit that resembles the setup circuit $\Sigma^{\BoBW}(\circuit_S)$ but prepares different states.
        \item Run $\Sim^{\SWIA}_{\adv}$ who interacts with the simulated $\ideal^{\SWIA}$.
        \item Upon failure, let $I_{\abort}$ be the set of parties instructed by $\Sim^{\SWIA}_{\adv}$ to send $\abort$.
        \item Run $\adv$ to complete the steps in the protocol that involve only the malicious group $I_{\abort}$.
        \item Set $S$ as the updated group $S \backslash I_{\abort}$ that contains all the honest parties.
    \end{enumerate}
    \item For every corrupted party not in $S$, send their default input to the ideal functionality.
    \item Run the remaining part of $\Sim_{\adv}^{\BoBW}(\circuit_S)$, which interacts with the ideal functionality.
    \item Let $J$ be the union of all corrupted lists output by all executions of $\Pi^{\BoBW}$. Make all parties in $J$ send $\abort$ to the ideal functionality. 
    \item Output the output of $\adv$.
    \end{enumerate}
\end{simulator}

To prove indistinguishability, we introduce the following hybrids. We describe each hybrid in terms of the changes made to the previous hybrid, starting with $\hybrid_0 = \real$.
\begin{itemize}[leftmargin=*]
    \item $\hybrid_i$ for $i = 1,\cdots,n$: Replace the $i$-th real execution of $\Pi^\SWIA$ with an ideal execution of $\ideal^{\SWIA}$ that computes the corresponding setup circuit $\Sigma^{\BoBW}(\circuit_S)$.
    \item $\hybrid_{n+i}$ for $i = 1,\cdots,n$: If the $i$-th execution of $\ideal^\SWIA$ will involve all the honest parties, let the ideal execution compute the first part of the simulator $\Sim^{\BoBW}_\adv(\circuit_S)$ instead. When this execution succeeds, replace the corresponding real execution of $\Pi^{\BoBW}$ in protocol step $3$ with an ideal execution of $\ideal^{\MPQC}$ that involves only parties in $S$ and interacts with the remaining part of the simulator $\Sim^{\BoBW}_\adv(\circuit_S)$. Parties in $S$ still follows the output procedure of the protocol instead of directly outputting what they get from $\ideal^{\MPQC}$.
    \item $\hybrid_{2n+1}$: Replace the ideal execution of $\ideal^{\MPQC}(\circuit_S)$ that involves only a subset $S$ of parties with an ideal execution of $\ideal^{\MPQC}(\circuit)$ that involves all parties, where the parties not in $S$ always send default inputs. For every corruption list output by $\Pi^{\BoBW}$, let the malicious parties in the corruption list send $\abort$ to $\ideal^{\MPQC}$. Parties in $S$ directly outputs what they get from $\ideal^{\MPQC}$.
\end{itemize}
Since the sub-protocols $\Pi^{\SWIA}$ are executed sequentially among the protocol, lemma \ref{lem:ACC+21} implies that $\hybrid_{i-1}$ and $\hybrid_{i}$ are indistinguishable for $i=1,\cdots,n$. 
By theorem $\ref{thm:bobw}$, the hybrids $\hybrid_{n+i-1}$ and $\hybrid_{n+i}$ are indistinguishable for $i=1,\cdots,n$. We have $\hybrid_{2n}=\hybrid_{2n+1}$ by the definition of $\circuit_S$ and the observation that both hybrids make the same abort decisions. Moreover, if we merge all the simulators and all the ideal functionalities except for $\ideal^{\MPQC}$ in the last hybrid, we would get simulator \ref{sim:swia+pvia}. Thus, the last hybrid has the same output as running the ideal world with simulator $\ref{sim:swia+pvia}$. We therefore conclude that the real and ideal worlds are indistinguishable.

\end{proof}

\section{Discussion and Open Questions}
\label{Sec:discussion}
\paragraph{Alternative Forms of AQA}
In our paper, we construct a Clifford-form AQA. It is interesting to consider Trap-code or other forms of AQA. This opens up the feasibility of constructing MPQC or other protocols using different forms, whereas existing MPQC works in the dishonest majority setting follow only a rule of thumb along the line of the Clifford authentication code \cite{DNS12,DGJ+20,ACC+21,BCKM21}.

\paragraph{Applications of AQA}
Our AQA primitive paves the way for developing cryptographic protocols or primitives that offer public verifiability. For example, AQA may help construct publicly verifiable quantum fully homomorphic encryption (pvQFHE). However, one may need to seek a public-key version of AQA to reduce the extra round induced by the audit.

\paragraph{Constant-Round MPQC-PVIA.} Upon close examination of MPQC-PVIA, it becomes apparent that by incorporating a quantum garbled circuit from \cite{BCKM21} and a trusted setup, a constant-round MPQC-PVIA protocol can be achieved without difficulty. In fact, upon careful analysis, it is possible to condense the protocol to just three rounds. However, this requires a trusted setup, which is unfavorable in most cryptographic scenarios. Thus, constructing an instantiable constant-round setup that removes the trusted setup assumption would be a good direction.

\ifdefined\ShowAuthor
\section*{Acknowledgement}
The authors would like to thank Andrea Coladangelo for useful discussions. This research is supported by NSF CAREER award 2141536 and supported by NSTC QC project under Grant no. NSTC 111-2119-M-001-004.
\fi

\ifdefined\LLNCS
\bibliographystyle{splncs04}
\else
\bibliographystyle{alpha}
\fi
\bibliography{references}

\ifdefined\FullVersion
\appendix
\section{Proof of Lemmas}
\label{appendix:proof-of-lemma}

\begin{lemma}\label{lemma:Pauli-cross-term}
    Let $u \neq u'\in \bbZ_p^n$ and $Q,Q' \in \PauliGroup_n$ such that $x(Q)+u = x(Q')+u'$. Then
    $$
    \E_{P \from \PauliGroup_n} \left[ QP\ket{u}\bra{u'} P^\dag Q'^\dag \right] = 0
    $$
\end{lemma}
\begin{proof}
    Let $v = x(Q)+u = x(Q')+u'$. Then $\E_{P \from \PauliGroup_n} \left[QP\ket{u}\bra{u'}P^\dag Q'^\dag \right]$ is equal to
    \begin{align*}
        & \E_{x(P),z(P) \from \bbZ_p^n} \left[ \Xg^{x(Q)} \Zg^{z(Q)} \Xg^{x(P)} \Zg^{z(P)} \Xg^{u} \ketbra{0} \Xg^{-u'} \Zg^{-z(P)} \Xg^{-x(P)} \Zg^{-z(Q')} \Xg^{-x(Q')} \right]\\
        =& \E_{x(P),z(P) \from \bbZ_p^n} \left[ \omega^{(u-u')z(P)+u z(Q) - u' z(Q') + x(P)(z(Q)-z(Q'))} \ket{x(Q)+x(P)+u}\bra{x(Q')+x(P)+u'} \right]\\
        =& \E_{z(P) \from \bbZ_p^n} \left[ \omega^{(u-u')z(P)} \E_{x(P) \from \bbZ_p^n} \left[ \omega^{u z(Q) - u' z(Q') + x(P)(z(Q)-z(Q'))} \ket{x(P)+v}\bra{x(P)+v} \right] \right] = 0
    \end{align*}
    where the last equation follows from $u\neq u'$ and averaging over $z(P)$.
\end{proof}

\begin{lemma}[\cref{lemma:qotp-measure}]
Let $\onreg{\ket{\phi}}{M,N} = \sum_{u \in \bbZ_p^n} \onreg{\ket{u}}{M} \ot \onreg{\ket{\phi_u}}{N}$ be a state and $v \in \bbZ_p^n$ be the target measurement result. For any attack $\onreg{A}{M,N} = \sum_{Q \in \PauliGroup^*_n} \left( \onreg{Q}{M} \ot \onreg{A_Q}{N}\right)$ applied on the QOTP-protected state, it holds that
\begin{align*}
    &\E_{P \from \PauliGroup_n} \mixed{\onreg{\ketbra{v + x(P)}}{M} \onreg{A}{M,N} \onreg{P}{M} \onreg{\ket{\phi}}{M,N}}\\
    =& \E_{r \from \bbZ_p^n} \sum_{u} \mixed{ \bigg( \sum_{x(Q)=v-u} Q \ot A_Q \bigg) \big(\ket{r} \ot \ket{\phi_u} \big) } 
\end{align*}
\end{lemma}
\begin{proof}
\begin{align*}
    & \E_{P \from \PauliGroup_n} \mixed{\onreg{\ketbra{v + x(P)}}{M} \onreg{A}{M,N} \onreg{P}{M} \onreg{\ket{\phi}}{M,N}}
    \\=& \E_{P \from \PauliGroup_n} \mixed{\onreg{\ketbra{v + x(P)}}{M} \sum_{Q\in \PauliGroup_n^*} \sum_{u \in \bbZ_p^n} \left(\onreg{Q}{M} \ot \onreg{A_Q}{N}\right) \onreg{P}{M} \left(\onreg{\ket{u}}{M} \ot \onreg{\ket{\phi_u}}{N}\right)}
    \\=& \E_{P \from \PauliGroup_n} \mixed{\onreg{\ketbra{v + x(P)}}{M} \sum_{Q\in \PauliGroup_n^*} \sum_{u \in \bbZ_p^n} \left(\onreg{Q P \ket{u}}{M} \ot \onreg{A_Q \ket{\phi_u}}{N}\right)}
    \\=& \E_{P \from \PauliGroup_n} \mixed{ \sum_{u \in \bbZ_p^n} \sum_{x(Q) = v - u} \onreg{Q P \ket{u}}{M} \ot \onreg{A_Q \ket{\phi_u}}{N}}
    \\=& \E_{P \from \PauliGroup_n} \sum_{u \in \bbZ_p^n} \mixed{ \sum_{x(Q) = v - u} \onreg{Q P \ket{u}}{M} \ot \onreg{A_Q \ket{\phi_u}}{N}}
    \\=& \E_{r \from \bbZ_p^n} \sum_{u \in \bbZ_p^n} \mixed{ \sum_{x(Q) = v - u} \onreg{Q \ket{r}}{M} \ot \onreg{A_Q \ket{\phi_u}}{N}}
\end{align*}
The fourth line holds because $v+x(P) = x(Q)+x(P)+u$ if and only if $x(Q)=v-u$. The fifth line follows from expanding $\mixednospace{\cdot}$ and eliminating the cross terms for $u,u'\in\bbZ_p^n$ using \cref{lemma:Pauli-cross-term}. The last line follows from the Pauli twirl (\cref{lemma:qotp-security}).
\end{proof}

\begin{lemma}[\cref{lemma:teleport}]
\label{proof:lemma:teleport}
Let $(\psi, \tau)$ be a purified state independent of $(e_S,e_R)$. Then
$$
\left(\TPSend(\onreg{\psi}{M},\onreg{e_S}{S}), \onreg{e_R}{R}, \onreg{\tau}{N}\right) 
= \frac{1}{p^{n}}~\sum_{x,z \in \bbZ_p^n}~ \onreg{\ket{z}}{M} \ot  \onreg{\ket{x}}{S} \ot \left(\onreg{\left(\Xg^x \Zg^z\right) \psi}{R}, \onreg{\tau}{N}\right)
$$
\end{lemma}
\begin{proof}
Let $(\phi, \tau) = \sum_{u} \ket{u} \ot \ket{\tau_u}$. Then
\vspace{-2em}
\begin{multicols}{2}
\begin{align*}
&\onreg{\TP}{M,S} (\onreg{\psi}{M},\onreg{e_S}{S}, \onreg{e_R}{R}, \onreg{\tau}{N})\\
=& \onreg{\Hg}{M} \onreg{\CX^\dag}{M,S} (\onreg{\psi}{M},\onreg{e_S}{S}, \onreg{e_R}{R}, \onreg{\tau}{N}) \\
=& \onreg{\Hg}{M} \onreg{\CX^\dag}{M,S} \left(\frac{1}{\sqrt{p^{n}}}~\sum_{u,v}~ \ket{\onreg{u}{M},\onreg{v}{S},\onreg{v}{R}} \ot \onreg{\tau_u}{N}\right) \\
=& \onreg{\Hg}{M} \left(\frac{1}{\sqrt{p^{n}}}~\sum_{u,v}~ \ket{\onreg{u}{M},\onreg{v-u}{S},\onreg{v}{R}} \ot \onreg{\tau_u}{N}\right)
\end{align*}
\newline
\begin{align*}
=&\frac{1}{p^{n}}~\sum_{u,v,z}~ \omega^{\inner{z,u}} \ket{\onreg{z}{M},\onreg{v-u}{S},\onreg{v}{R}} \ot \onreg{\tau_u}{N} \\
=&\frac{1}{p^{n}}~\sum_{u,x,z}~ \omega^{\inner{z,u}} \ket{\onreg{z}{M},\onreg{x}{S},\onreg{x+u}{R}} \ot \onreg{\tau_u}{N} \\
=&\frac{1}{p^{n}}~\sum_{u,x,z}~ \onreg{\Xg^x\Zg^z}{R} \ket{\onreg{z}{M},\onreg{x}{S},\onreg{u}{R}} \ot \onreg{\tau_u}{N} \\
=&\frac{1}{p^{n}}~\sum_{x,z}~ \onreg{\ket{z}}{M} \ot \onreg{\ket{x}}{S} \ot \left( \onreg{(\Xg^x \Zg^z) \psi}{R} , \onreg{\tau}{N}\right)
\end{align*}
\end{multicols}
\end{proof}
\section{State Injection} \label{appendix:state-injection}
\paragraph{Phase state} $\ket{\Pg}=\Pg\ket{+}$\\
We can apply the phase gate $\Pg$ on a state $\ket{\psi}$ using $\CX^{-1}$, $\Xg$, $\Zg$ gates and $\ket{\Pg}$ state as follows:
$$
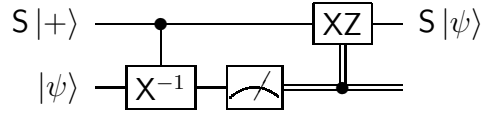
\Qcircuit @C=1em @R=.7em {
    \lstick{\Pg\ket{+}} & \ctrl{+1} & \qw    & \gate{\Xg\Zg} & \rstick{\Pg\ket\psi} \qw \\
    \lstick{\ket\psi} & \gate{\Xg^{-1}}     & \meter & \control\cw\cwx[-1] &  \rstick{} \cw
    }
$$
\begin{proof}
Write $\Pg \ket{+} = \frac{1}{\sqrt{p}}\sum_{j=0}^{p-1} \omega^{\frac{j(j-1)}{2}} \ket{j}$ and  $\ket{\psi}=\sum_{i=0}^{p-1}\alpha_i \ket{i}$. After applying $\CX^{-1}$, we get 
\begin{align*}
\CX^{-1}(\Pg \ket{+},\ket{\psi}) 
&\;=\;\; \CX^{-1}\left(\sum_{j=0}^{p-1} \omega^{\frac{j(j-1)}{2}} \ket{j},\sum_{i=0}^{p-1}\alpha_i \ket{i}\right) \\
&\;=\;\; \sum_{j=0}^{p-1}\sum_{i=0}^{p-1}\omega^{\frac{j(j-1)}{2}}\alpha_i\ket{j}\otimes\ket{i-j}\\
&\overset{_{k = i-j}}{=} \sum_{j=0}^{p-1}\sum_{k=0}^{p-1}\omega^{\frac{j(j-1)}{2}}\alpha_{j+k}\ket{j}\otimes\ket{k}
\end{align*}
where the subscript of $\alpha$ is taken modulo $p$. Then we measure the second register and obtain a measurement result $k'$. We now show by induction that applying $(\Xg\Zg)^{k'}$ to the first register produces $\Pg\ket{\psi}$. 
\begin{itemize}
    \item If $k'=0$, the state in the first register is $\sum_{j=0}^{p-1}\omega^{\frac{j(j-1)}{2}}\alpha_{j}\ket{j}=\Pg\ket{\psi}$. 
    \item If $k'>0$, the state in the first register is $\sum_{j=0}^{p-1} \omega^{\frac{j(j-1)}{2}}\alpha_{j+k'}\ket{j}$. Applying $\Xg \Zg$ to the state gives 
    $$\sum_{j=0}^{p-1} \omega^j \omega^{\frac{j(j-1)}{2}}\alpha_{j+k'}\ket{j+1}
    =\sum_{j=0}^{p-1} \omega^{\frac{(j+1)j}{2}}\alpha_{j+k'}\ket{j+1}
    =\sum_{j=1}^{p} \omega^{\frac{j(j-1)}{2}}\alpha_{j+k'-1}\ket{j}$$
    Since $\omega^d=1$, the term with $j=p$ in the last expression is equal to the term corresponding to $j=0$. Thus, the state is equivalent to $\sum_{j=0}^{p-1} \omega^{\frac{j(j-1)}{2}}\alpha_{j+k'-1}\ket{j},$
    which is what we would obtain if the measurement result were $k'-1$. Hence, applying $\Xg\Zg$ and the above argument $k'$ times would result in $\Pg\ket{\psi}$.
\end{itemize}
\end{proof}

\section{MPQC-SWIA}
\label{sec:modified-swia}

This section explains how we obtain the MPQC-SWIA protocol we need by utilizing two primitives proposed in \cite{ACC+21} to perform quantum computation that involves no inputs under fewer assumptions. In particular, our MPQC-SWIA protocol only assumes $\cMPC$ whereas \cite{ACC+21} assumes $\cMPC$ and a post-quantum fully homomorphic encryption (cFHE). First, we recall some primitives of \cite{ACC+21}.

\begin{itemize}
    \item Authentication Routing ($\AR$): The protocol takes as input the identity of two parties, one called the sender and the other called the receiver. The sender takes $k > n^2$ quantum packets $(\sigma_{1},\cdots,\sigma_{k})$ as input. At the end of the protocol, either the receiver obtains at least $k-n^2$ of the packets with known indices, or the protocol aborts and a partition of parties is known to everyone.The protocol directly generalizes to the qupit version using Clifford group for qupits and replacing $\GL(2,n)$ with $\GL(p,n)$.
    \item $\Tg$-Magic State Preparation ($\MSP$): The protocol takes parameters $1^\lambda, 1^n, 1^m$ as input. At the end of the protocol, either $\Pc_1$ receives $E(\ket{\Tg}^{\ot n}, \ket{0}^{\ot (m+n\lambda)})$ and $\cMPC$ receives $E$ for a random Clifford $E$, or the protocol aborts and a partition of parties is known to everyone.  The protocol directly generalizes to the qupit version using Clifford group for qupits and the magic state distillation of \cite{Distill}.
\end{itemize}
\begin{protocol}
{Our protocol $\Pi^{\SWIA}$ for MPQC-SWIA }\label{proto:swia}Common input: The parties hold the security parameter $1^\lambda$ and the circuit description, which samples a Clifford $G$ from a distribution $\mathbb{G}$ and applies it to the state $\ket{\Tg}^{\ot n} \ot \ket{0}^{\ot (m+n\lambda)}$.
    \begin{enumerate}
    \setlength{\itemsep}{1pt}
        \item Compute the following $n^3$ times. In the $k$-th iteration:
        \begin{itemize}[leftmargin=*]
        \setlength{\itemsep}{1pt}
            \item Call $\MSP$ so that $\Pc_1$ receives a state $E_k(\ket{\Tg}^{\ot n}, \ket{0}^{\ot (m+n\lambda)})$ and $\cMPC$ receives $E_k$.
            \item If $\MSP$ aborts, everyone receives a partition of parties and aborts.
            \item $\cMPC$ samples $G_k \from \mathbb{G}$ and sends the Clifford gate $V_k = G_k E_k^\dag$ to $\Pc_1$.
            \item $\Pc_1$ applies $V_k$ to $E_k(\ket{\Tg}^{\ot n}, \ket{0}^{\ot (m+n\lambda)})$ and obtains $(\sigma_{1,k}, \cdots, \sigma_{n,k})$.
        \end{itemize}
        \item $\Pc_1$ sends the states $\{\sigma_{i,k}\}_{k \in [n^3]}$ to $\Pc_i$ in the order of $i=2,\cdots,n$:
        \begin{itemize}[leftmargin=*]
        \setlength{\itemsep}{1pt}
            \item $\Pc_1$ runs $\AR$ to send $\{\sigma_{i,k}\}_{k \in K_{i-1}}$ to $\Pc_i$ where we set $K_1 = [n^3]$.
            \item If $\AR$ aborts, everyone receives a partition of parties and aborts.
            \item $\cMPC$ sets $K_{i}$ as the indices of the packets that are successfully sent through $\AR$.
        \end{itemize}
        \item $\cMPC$ sends the smallest index $k^*$ in $K_{n}$ to everyone.
        \item Party $\Pc_i$ outputs $\sigma_{i,k^*}$ and $\cMPC$ outputs $G_{k^*}$.
    \end{enumerate}

\end{protocol}
Initially, there are $n^3$ states prepared. Each execution of $\AR$ will either abort or drop up to $n^2$ packets. If $\Pc_1$ sends states to $\Pc_2,\cdots,\Pc_{n}$ with no abort, there will be no more than $n^2 (n-1) < n^3$ packet drops. Hence, the set $K_n$ will retain at least one element, which points to an iteration where all parties receive the circuit output successfully.
\fi

\end{document}